\documentclass[12pt]{article}
\usepackage[utf8]{inputenc}
\usepackage{amsmath, amsthm, bm, amssymb, hyperref, xcolor, caption, subcaption, graphicx, dsfont, multirow, booktabs, array,rotating,enumitem, mathtools, placeins}

\hypersetup{
	hidelinks,
    colorlinks = true,
    linkbordercolor = {white},
    citecolor = blue,
    urlcolor = blue,
    linkcolor = blue,
    linktoc = all % make both sections and subsections in ToC clickable
}

\usepackage[
  top=2cm,
  bottom=2cm,
  left=1in,
  right=1in,
  headheight=17pt, % as per the warning by fancyhdr
  includehead,includefoot,
  heightrounded, % to avoid spurious underfull messages
]{geometry} 

\usepackage{natbib}
\setcitestyle{authoryear}
\usepackage{grffile}

\author{Max Welz \\
  \href{mailto:max.welz@uzh.ch}{\texttt{max.welz@uzh.ch}} \\ Erasmus University Rotterdam \\ University of Zurich
  \and
  Patrick Mair \\
  \href{mailto:mair@fas.harvard.edu}{\texttt{mair@fas.harvard.edu}} \\ Harvard University
  \and
  Andreas Alfons \\
  \href{mailto:alfons@ese.eur.nl}{\texttt{alfons@ese.eur.nl}} \\ Erasmus University Rotterdam}
  
\date{}
    
\title{\textsc{Robust Estimation of Polychoric Correlation}\thanks{
This paper has been published in \emph{Psychometrika} at \url{https://doi.org/10.1017/psy.2025.10066}. This ArXiv version differs from the published version in layout, pagination, and typographic detail. Please cite the published version. 
This paper is based on Chapter~3 of MW's PhD thesis. We thank Patrick Groenen, Steffen Gr{\o}nneberg, and Jonas Moss for valuable feedback. We further thank the associate editor and three anonymous reviewers, whose comments led to a substantially improved paper.
This work was supported by a grant from the Dutch Research Council (NWO), research program Vidi (Project No. VI.Vidi.195.141).}}

\newcommand{\Poperator}{\mathbb{P}}

\newcommand{\var}[2]{\mathbb{V}\mathrm{ar}_{#1} \left[ #2 \right]}
\newcommand{\cor}[2]{\mathbb{C}\mathrm{or} \left[#1,\ #2\right]}
\newcommand{\corF}[3]{\mathbb{C}\mathrm{or}_{#1} \left[#2,\ #3\right]}

\renewcommand{\Pr}[2]{\Poperator_{#1} \left[ #2 \right]}
\newcommand{\CDF}[2]{\Phi_2\left(#1; #2 \right)}
\newcommand{\PDF}[2]{\phi_2\left(#1; #2 \right)}
\newcommand{\CDFuni}[1]{\Phi_1\left( #1 \right)}
\newcommand{\PDFuni}[1]{\phi_1\left( #1 \right)}
\renewcommand{\emph}[1]{\textit{#1}}
\newcommand{\X}{\mathcal{X}}
\newcommand{\Y}{\mathcal{Y}}

\newcommand{\pxy}[1]{p_{xy}(#1)}

\newcommand{\residual}[2]{\frac{#1}{\pxy{#2}}}
\newcommand{\funresidual}[3]{{#1}\Bigg(\residual{#2}{#3}\Bigg)}
\newcommand{\sxy}[1]{\boldsymbol{s}_{xy}(#1)}

\newcommand{\BW}[1]{\boldsymbol{W}(#1)}

\newcommand{\BOmega}{\bm{\Omega}}
\renewcommand{\vec}[1]{\boldsymbol{#1}}
\newcommand{\mat}[1]{\boldsymbol{#1}}
\newcommand{\matfun}[2]{\mat{#1}\left( #2 \right)}
\newcommand{\hatmatfun}[2]{\hat{\mat{#1}}_N\left( #2 \right)}
\newcommand{\matinv}[2]{\matfun{#1}{#2}^{-1}}
\newcommand{\hatmatinv}[2]{\hatmatfun{#1}{#2}^{-1}}

\newcommand{\Btheta}{\bm{\theta}}
\newcommand{\hatN}[1]{\hat{#1}_N}
\newcommand{\hatNMLE}[1]{\hatN{#1}^{\mathrm{\ MLE}}}
\newcommand{\rhohat}{\hatN{\rho}}
\newcommand{\rhohatt}{\rhohat^{(t)}}
\newcommand{\thetahat}{\hatN{\Btheta}}
\newcommand{\thetahatMLE}{\hatNMLE{\Btheta}}
\newcommand{\BTheta}{\bm{\Theta}}

\newcommand{\BSigma}[1]{\bm{\Sigma}\left({#1}\right)}

\newcommand{\gauss}{\textnormal{N}}

\newcommand{\I}[1]{\mathds{1}\left\{ #1 \right\}}
\newcommand{\wxy}[1]{\boldsymbol{w}_{#1}\left(\Btheta\right)}

\renewcommand{\d}{\textnormal{d}}

\renewcommand{\hat}[1]{\widehat{#1}}

\newcommand{\rhohatmean}{\hatN{\rho}^{\ \text{ave}}}
\newcommand{\SE}[1]{\textnormal{SE}\left(#1\right)}
\newcommand{\SEapprox}[1]{\textnormal{SE}^{\textnormal{approx}}\left(#1\right)}
\newcommand{\SEhat}[1]{\hat{\textnormal{SE}}\left(#1\right)}
\newcommand{\SEbarhat}[1]{\hat{\textnormal{SE}}^{\textnormal{ave}}\left(#1\right)}
\newcommand{\R}{\mathbb{R}}
\newcommand{\fhat}{\hat{f}_{N}}
\newcommand{\fhatxy}{\hat{f}_{N}(x,y)}

\newcommand{\convP}{\stackrel{\Poperator}{\longrightarrow}}
\newcommand{\convweak}{\stackrel{\mathrm{d}}{\longrightarrow}}

\newcommand{\feps}{f_{\varepsilon}}
\newcommand{\fepsfun}[1]{\feps \left(#1\right)}
\newcommand{\fepsxy}{\fepsfun{x,y}}
\newcommand{\Bfeps}{\boldsymbol{f}_\varepsilon}
\newcommand{\partialderivative}[2]{\frac{\partial #1 }{\partial #2}}
\newcommand{\partialderivativetwice}[2]{\frac{\partial^2 #1 }{\partial {#2}\partial {#2}^\top}}
\newcommand{\partialderivativetwiced}[3]{\frac{\partial^2 #1 }{\partial {#2}\partial{#3}}}
\newcommand{\partialderivativetwicesame}[2]{\frac{\partial^2 #1 }{\partial {#2}^2}}

\newcommand{\proglang}[1]{\textsf{#1}}
\newcommand{\pkg}[1]{\texttt{#1}}

\theoremstyle{plain}
\newtheorem{theorem}{\color{blue}Theorem}

\theoremstyle{definition}

\usepackage{apptools}
\AtAppendix{\counterwithin{lemma}{section}}
\AtAppendix{\counterwithin{proposition}{section}}
\AtAppendix{\counterwithin{corollary}{section}}

\begin{document}

\maketitle

\begin{abstract}
\noindent
Polychoric correlation is often an important building block in the analysis of rating data, particularly for structural equation models. However, the commonly employed maximum likelihood (ML) estimator is highly susceptible to misspecification of the polychoric correlation model, for instance through violations of latent normality assumptions. We propose a novel estimator that is designed to be robust against partial misspecification of the polychoric model, that is, when the model is misspecified for an unknown fraction of observations, such as careless respondents. To this end, the estimator minimizes a robust loss function based on the divergence between observed frequencies and theoretical frequencies implied by the polychoric model. In contrast to existing literature, our estimator makes no assumption on the type or degree of model misspecification. It furthermore generalizes ML estimation, is  consistent as well as asymptotically normally distributed, and comes at no additional computational cost. We demonstrate the robustness and practical usefulness of our estimator in simulation studies and an empirical application on a Big Five administration. In the latter, the polychoric correlation estimates of our estimator and ML differ substantially, which, after further inspection, is likely due to the presence of careless respondents that the estimator helps identify.
\end{abstract}


%%%%%%%%%%%%%%%%%%%%%%%%%%%%%%%%%%%%%%%%%%%%%%%%%%%
%%%%%%%%%%%%%%%%%%%%% MAIN TEXT %%%%%%%%%%%%%%%%%%%
%%%%%%%%%%%%%%%%%%%%%%%%%%%%%%%%%%%%%%%%%%%%%%%%%%%
\section{Introduction}
Ordinal data are ubiquitous in psychology and related fields. With such data, e.g., arising from responses to rating scales, it is often recommended to estimate correlation matrices through polychoric correlation coefficients \citep[e.g.,][]{foldnes2022,garrido2013,holgado2010}. 
The resulting polychoric correlation matrix is an important building block in subsequent multivariate models like factor analysis models and structural equation models (SEMs), as well as in exploratory methods like principal component analysis, multidimensional scaling, and clustering techniques \citep[see, e.g.,][for an overview]{mair2018}. 
An individual polychoric correlation coefficient is the population correlation between two underlying latent variables that are postulated to have generated the observed categorical data through an unobserved discretization process. Traditionally, it is assumed that the two latent variables are standard bivariate normally distributed \citep{pearson1901} to estimate the polychoric correlation coefficient from observed ordinal data. Estimation of this latent normality model, called the \emph{polychoric model}, is commonly carried out via maximum likelihood  \citep{olsson1979poly}. 
However, recent work has demonstrated that maximum likelihood (ML) estimation of polychoric correlation is highly sensitive to violations of the assumption of underlying normality. Violations of this assumption result in a misspecified polychoric model, which can lead to substantially biased estimates of its parameters and those of subsequent multivariate models \citep{gronneberg2022,foldnes2020polycor,foldnes2019identification,jin2017}, particularly SEMs using \emph{diagonally weighted least squares}  \citep{foldnes2022}, where the latter is based on weights derived under latent normality.

Motivated by the recent interest in non-robustness of ML, we study estimation of the polychoric model under a misspecification framework stemming from the robust statistics literature \citep[e.g.,][]{huber2009}. In this setup, which we call \emph{partial} misspecification here, the polychoric model is potentially misspecified for an unknown (and possibly zero-valued) fraction of observations. Heuristically, the model is misspecified such that the affected subset of observations contains little to no relevant information for the parameter of interest, the polychoric correlation coefficient. Examples of such uninformative observations include careless responses, misresponses, or responses due to item misunderstanding. Especially careless responding has been identified as a major threat to the validity of questionnaire-based research findings \citep[e.g.,][]{welz2024maxbias,huang2015ier,meade2012,crede2010,woods2006}. We demonstrate that already a small fraction of uninformative observations (such as careless respondents) can result in considerably biased ML estimates.

As a remedy and our main contribution, we propose a novel way to estimate the polychoric model which is robust to partial model misspecification. Essentially, the estimator poses the question ``What is the best fit that can be achieved with the polychoric model for (the majority of) the data at hand?'' The estimator compares the observed frequency of each contingency table cell with its expected frequency under the polychoric model, and automatically downweights cells whose observed frequencies cannot be fitted sufficiently well. As such,  
our estimator generalizes the ML estimator, but, in contrast to ML, does not crucially rely on correct specification of the model. Specifically, our estimator allows the model to be misspecified for an unknown fraction of uninformative responses in a sample, but makes \emph{no assumption} on the type, magnitude, or location of potential misspecification. The estimator is designed to identify such responses and to simultaneously reduce their influence so that the polychoric model can be accurately estimated from the remaining responses generated by latent normality. 
Conversely, if the polychoric model is correctly specified, that is, latent normality holds true for all observations, our estimator and ML estimation are asymptotically equivalent. As such, our proposed estimator can be thought of as a generalized ML estimator that is robust to potential partial model misspecification, due to, for instance (but not limited to) careless responding. We show that our robust estimator is consistent, asymptotically normal, and fully efficient under the polychoric model, while possessing similar asymptotic properties under misspecification, and it comes at no additional computational cost compared to ML.

The partial misspecification framework in this paper is fundamentally different to that considered in recent literature on misspecified polychoric models. In this literature \citep[e.g.,][]{lyhagen2023,gronneberg2022,foldnes2022,foldnes2020polycor,foldnes2019identification,jin2017}, the polychoric model is misspecified in the sense that \emph{all} (unobserved) realizations of the latent continuous variables come from a distribution that is nonnormal. Under this framework, which is also known as \emph{distributional misspecification}, the parameter of interest is the correlation coefficient of the latent nonnormal distribution, and all observations are informative for this parameter. While the distributional misspecification framework led to novel insights regarding (the lack of) robustness in ML estimation of polychoric correlation, the partial misspecification framework of this paper can provide complimentary insights regarding the effects of a fraction of uninformative observations in a sample (such as careless responses), which is our primary objective. 

Nevertheless, while our estimator is designed to be robust to partial misspecification caused by some uninformative responses, it can in some situations also provide a robustness gain under distributional misspecification. It turns out that if a nonnormal latent distribution differs from a normal distribution mostly in the tails, our estimator produces less biased estimates than ML because it can downweigh observations that are father from the center.

To enhance accessibility and adoption by empirical researchers, an implementation of our proposed methodology in \proglang{R} \citep{R} is freely available in the package \pkg{robcat} \citep[for ``ROBust CATegorical data analysis'';][]{robcat} on CRAN (the Comprehensive \proglang{R} Archive Network) at \url{https://CRAN.R-project.org/package=robcat}. Replication materials for all numerical results in this paper are provided on GitHub at \url{https://github.com/mwelz/robust-polycor-replication}.

This paper is structured as follows. 
We start with reviewing related literature (Section~\ref{sec:literature}) followed by the polychoric correlation model and ML estimation thereof (Section~\ref{sec:polycor}). 
Afterwards, we elaborate on the partial misspecification framework (Section~\ref{sec:model-misspecification}) and introduce our robust generalized ML estimator including its statistical properties (Section~\ref{sec:robust-polycor}).
These properties are then examined by a simulation study in which we vary the misspecification fraction systematically, and compare the result to the commonly employed standard ML estimator (Section~\ref{sec:simulation}). Subsequently, we demonstrate the practical usefulness in an empirical application on a Big Five administration \citep{goldberg1992} by \citet{arias2020}, where we find evidence of careless responding, manifesting in differences in polychoric correlation estimates of as much as~0.3 between our robust estimator and ML (Section~\ref{sec:application}). We then investigate the performance of the estimator under distributional misspecification (Section~\ref{sec:distributional-misspecification}) and conclude with a discussion of the results 
and avenues for further research (Section~\ref{sec:conclusion}).

\section{Related literature}\label{sec:literature}

ML estimation of polychoric correlations was originally believed to be fairly robust to slight to moderate distributional misspecification \citep{li2016,coenders1997,flora2004,maydeu2006}. This belief was based on simulations that generated data for nonnormal latent variables via the Vale-Maurelli (VM) method \citep{vale1983}, which were then discretized to ordinal data. However, \citet{gronneberg2019} show that the distribution of ordinal data generated in this way is indistinguishable from that of ordinal data stemming from discretizing normally distributed latent variables.\footnote{A key reason for this finding is that the VM method produces a latent vector whose distribution corresponds either exactly or near-exactly to a Gaussian copula \citep{foldnes2015} (except in regions where %the
certain polynomials used in the VM transformation are non-monotonous). It follows that the VM transformation  \emph{``inherits a Gaussian-like property, indicating that simulation studies based on the VM approach might give overly optimistic impressions of finite-sample properties of estimators with non-Gaussian data''} \citep[][p.~1078]{foldnes2015}. To simulate non-normality, one should instead strive for a copula with distinctively non-Gaussian features, such as tail dependence and asymmetry. This is exactly what %the
an alternative transformation method proposed by %of
\citet{gronneberg2017vita} does. For instance, generating data with the tail-dependent and tail-asymmetric Clayton copula leads to markedly different behavior of normality-based estimators than with the VM method \citep[e.g.,][]{foldnes2022}.}
In other words, simulation studies that ostensibly modeled nonnormality did in fact model normality. Simulating ordinal data in a way that ensures proper violations of latent normality \citep{gronneberg2017vita} reveals that polychoric correlation is in fact highly susceptible to distributional misspecification, resulting in possibly large biases \citep{foldnes2020polycor,foldnes2022,gronneberg2022,jin2017}. Consequently, it is recommended to test for the validity of the latent normality assumption, for instance by using the bootstrap test of \citet{foldnes2020polycor}.

Another source of model misspecification occurs when the polychoric model is only misspecified for an uninformative subset of a sample (partial misspecification), where, in the context of this paper, the term ``uninformative'' %means
refers to an absence of relevant information for polychoric correlation, %like
for instance in careless responses.  Careless responding \textit{``occurs when participants are not basing their response on the item content''}, for instance when a participant is \textit{``unmotivated to think about what the item is asking''} \citep{ward2023}. It has been shown to be a major threat to the validity of research results through a variety of psychometric issues, such as reduced scale reliability \citep{arias2020} and construct validity \citep{kam2015}, attenuated factor loadings, improper factor structure, and deteriorated model fit in factor analyses \citep{arias2020,huang2015detecting,woods2006}, as well as inflated type~I or type~II errors in hypothesis testing \citep{arias2020,huang2015ier,maniaci2014,mcgrath2010,woods2006}. Careless responding is widely prevalent \citep{ward2023,bowling2016,meade2012} with most estimates on its prevalence ranging from 10--15\% of study participants \citep{curran2016,huang2015ier,huang2012,meade2012}, while already a prevalence~5--10\% can jeopardize the validity of research findings  \citep{welz2024maxbias,arias2020,crede2010,woods2006}. In fact, \citet{ward2023} conjecture that careless responding is likely present in all survey data. However, to the best of our knowledge, the effects of careless responding on estimates of the polychoric model have not yet been studied. 

Existing model-based approaches to account for careless responding in various models typically explicitly model carelessness trough mixture models \citep[e.g.,][]{steinmann2022,vanlaar2022,ulitzsch2022time,ulitzsch2022notime,arias2020}. In contrast, our method does not model carelessness since we refrain from making assumptions on how the polychoric model might be misspecified. Another way to address careless responding is to directly detect them through person-fit indices and subsequently remove them from the sample \citep[e.g.,][]{patton2019}. As primary difference, our method simultaneously downweights aberrant observations during estimation rather than removing them. We refer to \citet{alfons2024spc} for a detailed overview of methods addressing careless responding in various settings.

Conceptually related to our approach, \citet{itaya2025} propose a way to robustly estimate parameters in item response theory (IRT) models. Their approach is conceptually similar to ours in the sense that it is based on minimizing a notion of divergence between an empirical density (from observed data) and a theoretical density of the IRT model. Like our approach, they achieve robustness by implicitly downweighting responses that the postulated model cannot fit well. Methodologically, our approach is different from \citet{itaya2025} because our method is based on $C$-estimation \citep{welz2024robcat}, which is designed specifically for categorical data, while they use \emph{density power divergence estimation} (DPD) theory \citep{basu1998}, which is not restricted to categorical data.\footnote{Another slightly more subtle difference is that \citet{itaya2025} robustify \emph{marginal} maximum ML, whereas we robustify \emph{joint} ML estimation. Marginal ML is often used to estimate IRT models because joint ML is known to yield inconsistent parameter estimates in such models \citep[e.g.,][]{lindsay1991}. The polychoric correlation model does not suffer from such problems and can therefore be estimated via joint~ML. Furthermore, marginal ML in IRT is a \emph{full information estimation} concept, whereas multivariate models that are being fitted to a given polychoric correlation matrix require \emph{limited information estimation}. Hence, the work of \citet{itaya2025} demonstrates that model misspecification is also a relevant issue in full information estimation.} A relevant consequence is that our estimator is fully efficient, whereas DPD estimators lose efficiency as a price for gaining robustness. To the best of our knowledge, DPD estimators have not yet been studied for estimating polychoric correlation.

Another related branch of literature is that of outlier detection in contingency tables \citep[see][for a recent overview]{sripriya2020}. In this literature, an outlier is a contingency table cell whose observed frequency is \emph{``markedly deviant''} from those of the remaining cells \citep{sripriya2020}. This literature is agnostic with respect to the observed contingency table and therefore does not impose a parameterization on each cell's probability. In contrast, the polychoric correlation model imposes such a parametrization through the assumption of latent bivariate normality. Another difference is that we are not primarily interested in outlier detection, but robust estimation of model parameters.

An alternative way to gain robustness against violations of latent nonnormality is to assume a different latent distribution, for instance one with heavier tails. Examples from the SEM literature  use elliptical distributions \citep{yuan2004} or skew-elliptical distributions \citep{asparouhov2016}, while \citet{lyhagen2023}, \citet{jin2017}, and \citet{roscino2006} consider nonnormal distributions specifically in the context of polychoric correlation. Furthermore, it is worth pointing out the term ``robustness'' is used in different ways in the methodological literature. Here, it refers to robustness against model misspecification. A popular but different meaning is robustness against heteroskedastic standard errors and corrected goodness-of-fit test statistics \citep[e.g.,][and references therein]{li2016,satorra2001,satorra1994}, which is, for instance, how the popular software package \pkg{lavaan} \citep{lavaan} uses the term. We refer to \citet{alfons2024robust} for an overview of the different meanings of ``robustness'' and a more detailed discussion.

\section{Polychoric correlation}\label{sec:polycor}
The polychoric correlation model \citep{pearson1922} models the association between two discrete ordinal variables by assuming that an observed pair of responses to two polytomous items is governed by an unobserved discretization process of latent variables that jointly follow a bivariate standard normal distribution. If both items are dichotomous, the polychoric correlation model reduces to the tetrachoric correlation model of \citet{pearson1901}. In the following, we first define the model and review maximum likelihood (ML) estimation thereof and then introduce a robust estimator in the next section.

\subsection{The polychoric model}
For %the 
ease of exposition, we restrict our presentation to the bivariate polychoric model. The model naturally generalizes to higher dimensions, see, e.g., \citet{muthen1984}. 

Let there be two ordinal random variables,~$X$ and~$Y$, that take values in the sets~$\X = \{1,2,\dots,K_X\}$ and~$\Y=\{1,2,\dots,K_Y\}$, respectively. The assumption that the sets contain adjacent integers is without loss of generality. 
Suppose there exist two continuous latent random variables,~$\xi$ and~$\eta$, that govern the ordinal variables through the discretization model
\begin{equation}\label{eq:polycormodel}
X=
\begin{cases}
    1 &\text{if }  \xi < a_1,\\
	2 &\text{if }  a_1 \leq \xi < a_2,\\
	3 &\text{if }  a_2 \leq \xi < a_3,\\
	\vdots\\
	K_X &\text{if }  a_{K_X-1} \leq \xi, \\
\end{cases}
\qquad 
\textnormal{ and }
\qquad
Y=
\begin{cases}
    1 &\text{if }  \eta < b_1,\\
	2 &\text{if }  b_1 \leq \eta < b_2,\\
	3 &\text{if }  b_2 \leq \eta < b_3,\\
	\vdots\\
	K_Y &\text{if }  b_{K_Y-1} \leq \eta, \\
\end{cases}
\end{equation}
where the fixed but unobserved parameters~$a_1 < a_2 < \dots < a_{K_X-1}$ and~$b_1 < b_2 < \dots < b_{K_Y-1}$ are called \textit{thresholds}. 

The primary object of interest is the population correlation between the two latent variables. To identify this quantity from the ordinal variables~$(X,Y)$, one assumes that the continuous latent variables follow a standard bivariate normal distribution with unobserved pairwise correlation coefficient~$\rho\in(-1,1)$, that is,
\begin{equation}\label{eq:latentnormality}
	\begin{pmatrix}
	\xi \\ \eta
	\end{pmatrix}
	\sim
	\gauss_2
	\left(
	\begin{pmatrix}
	0 \\ 0
	\end{pmatrix}
	,
	\begin{pmatrix}
	1 & \rho \\ \rho & 1
	\end{pmatrix}
	\right).
\end{equation}
Combining the discretization model~\eqref{eq:polycormodel} with the latent normality model~\eqref{eq:latentnormality} yields the \emph{polychoric model}. In this model, one refers to the correlation parameter~$\rho = \cor{\xi}{\eta}$ as the \emph{polychoric correlation coefficient} of the ordinal~$X$ and~$Y$. The polychoric model is subject to~$d=K_X+K_Y-1$ parameters, namely the polychoric correlation coefficient from the latent normality model~\eqref{eq:latentnormality} and the two sets of thresholds from the discretization model~\eqref{eq:polycormodel}. These parameters are jointly collected in a $d$-dimensional vector
\[
	\Btheta 
	=
	\left(\rho, a_1, a_2,\dots, a_{K_X-1}, b_1, b_2, \dots, b_{K_Y-1}\right)^\top.
\] 

Under the polychoric model, the probability of observing an ordinal response~$(x,y)\in\X\times\Y$ at a parameter vector~$\Btheta$ is given by
\begin{equation}\label{eq:pxy}
	\pxy{\Btheta} =
	\Pr{\Btheta}{X = x, Y=y} 
	= \int_{a_{x-1}}^{a_x}\int_{b_{y-1}}^{b_y} \PDF{t,s}{\rho}\d s\ \d t,
\end{equation}
where we use the conventions $a_0=b_0=-\infty, a_{K_X}=b_{K_Y}=+\infty$, and  
\[
	\PDF{u,v}{\rho} = \frac{1}{2\pi \sqrt{1-\rho^2}} \exp\left( -\frac{u^2-2\rho uv + v^2}{2(1-\rho^2)} \right)
\]
denotes the density of the standard bivariate normal distribution function with correlation parameter~$\rho\in (-1,1)$ at some~$u,v\in\R$, with corresponding distribution function
\[
 \CDF{u,v}{\rho} = \int_{-\infty}^u\int_{-\infty}^v \PDF{t,s}{\rho}\d s\ \d t.
\]

Regarding identification, it is worth mentioning that in the case where both~$X$ and~$Y$ are dichotomous, the polychoric model is exactly identified by the standard bivariate normal distribution. If at least one of the ordinal variables has more than two response categories, the polychoric model is over-identified, so it could identify more parameters than those in~$\Btheta$.\footnote{An ordinal sample provides $K_XK_Y$ statistics, namely the observed frequencies for each response category $(x,y)\in\X\times\Y$. Since the sum of the individual response frequencies must equal the sample size, one loses one degree of freedom so that only $K_X K_Y - 1$ of the frequency statistics are independent. Subsequently, the ordinal sample can identify at most $K_X K_Y - 1$ parameters. In particular, when~$X$ and~$Y$ are both dichotomous, only three parameters can be identified. In this case, the polychoric model (which reduces to the tetrachoric model here) depends on three parameters, namely a correlation parameter and two threshold parameters.} We refer to \citet[][Section~2]{olsson1979poly} for a related discussion.

To distinguish arbitrary parameter values~$\Btheta$ from a specific value under which the polychoric model generates ordinal data, denote the latter by~$\Btheta_* = \left(\rho_*, a_{*,1},\dots, a_{*,K_X-1}, b_{*,1}, \dots, b_{*,K_Y-1}\right)^\top$. Given a random sample of ordinal data generated by a polychoric model under parameter value~$\Btheta_*$, the statistical problem is to estimate the true~$\Btheta_*$, which is traditionally achieved by the maximum likelihood estimator of \citet{olsson1979poly}.\footnote{Alternatives to the commonly used maximum likelihood estimator of \citet{olsson1979poly} have been proposed in the literature, see, for instance, \citet{zhang2024}, \citet{joreskog1994}, and \citet{lau1985}.}

\subsection{Maximum likelihood estimation}
Suppose we observe a sample $\{(X_i, Y_i)\}_{i=1}^N$ of~$N$ independent copies of~$(X,Y)$ generated by the polychoric model under the true parameter~$\Btheta_*$. The sample may be observed directly or as a~$K_X\times K_Y$ contingency table that cross-tabulates the observed frequencies.
Denote by
\[
	N_{xy} = \sum_{i=1}^N\I{X_i = x, Y_i = y}
\]
the observed empirical frequency of a response $(x,y)\in\X\times\Y$, where the indicator function~$\I{E}$ takes value~1 if an event~$E$ is true, and~0 otherwise. 
The maximum likelihood estimator (MLE) of~$\Btheta_*$ can be expressed as
\begin{equation}\label{eq:MLE}
	\thetahatMLE
	=
	\arg\max_{\Btheta\in\BTheta}
	\left\{
		\sum_{x\in\X}\sum_{y\in\Y} N_{xy} \log\left(\pxy{\Btheta}\right)
	\right\},
\end{equation}
where the $\pxy{\Btheta}$ are the response probabilities in~\eqref{eq:pxy}, and
\begin{equation}\label{eq:BTheta}
	\BTheta 
	= \bigg(
	\Big(\rho, \big(a_i\big)_{i=1}^{K_X-1}, \big(b_j\big)_{j=1}^{K_Y-1}\Big)^\top
	\ \Big|\ 
	\rho\in(-1,1),\ a_1 < \dots < a_{K_X-1},\ b_1 < \dots < b_{K_Y-1}
	\bigg)
\end{equation}
is a set of parameters $\Btheta$ that rules out degenerate cases such as $\rho = \pm 1$ or thresholds that are not strictly monotonically increasing. This estimator, its computational details, as well as its statistical properties are derived in \citet{olsson1979poly}. In essence, if the polychoric model~\eqref{eq:polycormodel} is correctly specified---that is, the underlying latent variables~$(\xi, \eta)$ are indeed standard bivariate normal---then the estimator~$\thetahatMLE$ is consistent for the true~$\Btheta_*$. In addition,~$\thetahatMLE$ is asymptotically normally distributed with mean zero and covariance matrix being equal to the model's inverse Fisher information matrix, which makes it fully efficient.

As a computationally attractive alternative to estimating all parameters in~$\Btheta_*$ simultaneously in problem~\eqref{eq:MLE}, one may consider a ``2-step-approach'' where only the correlation coefficient~$\rho_*$ is estimated via ML, but not the thresholds. In this approach, one estimates in a first step the thresholds as quantiles of the univariate standard normal distribution, evaluated at the observed cumulative marginal proportion of each contingency table cell. Formally, in the 2-step-approach, thresholds $a_{*,x}$ and $b_{*,y}$ are respectively estimated via
\begin{equation}\label{eq:twostep}
	\hat{a}_{x} = \Phi_1^{-1}\left( \frac{1}{N}\sum_{k=1}^x\sum_{y\in\Y} N_{ky} \right)\qquad\text{and}\qquad
	\hat{b}_{y} = \Phi_1^{-1}\left( \frac{1}{N}\sum_{x\in\X}\sum_{l=1}^y N_{xl} \right),
\end{equation}
for $x=1,\dots,K_X-1$ and $y=1,\dots,K_Y-1$, where~$\Phi_1^{-1}(\cdot)$ denotes the quantile function of the univariate standard normal distribution. Then, taking these threshold estimates as fixed in the polychoric model, one estimates in a second step the only remaining parameter, correlation coefficient~$\rho_*$, via ML. The main advantage of the 2-step approach is reduced computational time, while it comes at the cost of being theoretically non-optimal because ML standard errors do not apply to the threshold estimators in~\eqref{eq:twostep} \citep{olsson1979poly}. Using simulation studies, \citet{olsson1979poly} finds that the two approaches tend to yield similar results in practice---both in terms of correlation and variance estimation---for small to moderate true correlations, while there can be small differences for larger true correlations.

Software implementations of polychoric correlation vary with respect to their estimation strategy. For instance, the popular \proglang{R} packages \pkg{lavaan} \citep{lavaan} and \pkg{psych} \citep{psych} only support the 2-step approach, while the package \pkg{polycor} \citep{polycor} supports both the 2-step-approach and simultaneous estimation of all model parameters, with the former being the default. Our implementation of ML estimation in package \pkg{robcat} also supports both strategies.

\section{Conceptualizing model misspecification}\label{sec:model-misspecification}
To study the effects of partial model misspecification from a theoretical perspective, we first rigorously define this concept and explain how it differs from distributional misspecification. 

\subsection{Partial misspecification of the polychoric model}\label{sec:hubermodel}
The polychoric model is partially misspecified when not all unobserved realizations of the latent variables~$(\xi,\eta)$ come from a standard bivariate normal distribution. Specifically, we consider a situation where only a fraction~$(1-\varepsilon)$ of those realizations are generated by a standard bivariate normal distribution with true correlation parameter~$\rho_*$, whereas a fixed but unknown fraction~$\varepsilon$ are generated by some different but unspecified distribution~$H$. 
Note that~$H$ being unspecified allows its correlation coefficient to differ from~$\rho_*$ so that realizations generated by~$H$ may be uninformative for the true polychoric correlation coefficient~$\rho_*$, such as, after discretization, careless responses, misresponses or responses stemming from item misunderstanding. 

Formally, we say that the polychoric model is partially misspecified if the latent variables~$(\xi,\eta)$ are jointly distributed according to  
\begin{equation}\label{eq:contaminationmodel}
	(u,v) \mapsto G_\varepsilon(u,v) = (1-\varepsilon) \CDF{u,v}{\rho_*} +\varepsilon H(u,v),
\end{equation} 
for $u,v\in\R$. 
Conceptualizing model misspecification in such a manner is standard in the robust statistics literature, going back to the seminal work of \citet{huber1964}.\footnote{In the robust statistics literature, a distribution like~\eqref{eq:contaminationmodel} is known as \emph{Huber contamination model}. For continuous random variables, this model is primarily used to model outliers and study the properties of outlier-robust estimators.} We therefore adopt terminology from robust statistics and call $\varepsilon$ the \emph{contamination fraction}, the uninformative~$H$ the \emph{contamination distribution} (or simply \emph{contamination}), and $G_\varepsilon$ the \emph{contaminated distribution}.  Observe that when the contamination fraction is zero, that is, $\varepsilon = 0$, there is no misspecification so that the polychoric model is correctly specified for all observations. 
However, neither the contamination fraction~$\varepsilon$ nor the contamination distribution~$H$ are assumed to be known. Thus, both quantities are left completely unspecified in practice and $\CDF{u,v}{\rho_*}$ remains the distribution of interest. That is, we only aim to estimate the model parameters $\Btheta$ of the polychoric model, while reducing the adverse effects of potential contamination in the observed ordinal data. The contaminated distribution~$G_\varepsilon$, on the other hand, is never estimated. It serves as purely theoretical construct that we use to study the theoretical properties of estimators of the polychoric model when that model is partially misspecified due to contamination.

Leaving the contamination distribution~$H$ and contamination fraction~$\varepsilon$ unspecified in the partial misspecification model~\eqref{eq:contaminationmodel} 
means that we are not making any assumptions on the degree, magnitude, or type of contamination (which is possibly absent altogether). Hence, in our context of responses to rating items, the polychoric model can be misspecfied due to an unlimited variety of reasons, for instance but not limited to careless/inattentive responding (e.g., straightlining, pattern responding, random-like responding), misresponses, or item misunderstanding.

Although we make no assumption on the specific value of the contamination fraction~$\varepsilon$~in the partial misspecification model~\eqref{eq:contaminationmodel}, we require the identification restriction $\varepsilon \in [0, 0.5)$. That is, we require that the polychoric model is correctly specified for the majority of observations, which is standard in the robust statistics literature \citep[e.g.,][p.~67]{hampel1986}.
While it is in principle possible to also consider a %misspecification
contamination fraction between~$0.5$ and~$1$, one would need to impose certain additional assumptions on the correct model to distinguish it from incorrect ones when the majority of observations are not generated by the correct model. Since we prefer refraining from imposing additional assumptions, we only consider $\varepsilon\in[0, 0.5)$. 
We discuss the link between identification and contamination fractions beyond~0.5 in more detail in Appendix~\ref{app:overlap-discussion}.

Furthermore, as another, more practical reason for considering $\varepsilon\in [0,0.5)$, having more than half of all observations in a sample being not informative for the quantity of interest would be indicative of serious data quality issues. When data quality is unreasonably low, it is doubtful whether the data are suitable for modeling analyses in the first place.

\subsection{Response probabilities under partial misspecification}\label{sec:feps}
Under contaminated distribution~$G_\varepsilon$ with contamination fraction~$\varepsilon\in [0, 0.5)$, the probability of observing an ordinal response $(x,y)\in\X\times\Y$ is given by  
\begin{equation}\label{eq:feps}
	\fepsxy =
	\Pr{G_\varepsilon}{X = x, Y=y} 
	=
	(1-\varepsilon) \pxy{\Btheta_*} + \varepsilon \int_{a_{\varepsilon,x-1}}^{a_{\varepsilon,x}} \int_{b_{\varepsilon,y-1}}^{b_{\varepsilon,y}} \d H,
\end{equation}
where the unobserved thresholds~$a_{\varepsilon,x},b_{\varepsilon,y}$ discretize the fraction~$\varepsilon$ of latent variables for which the polychoric model is misspecified. The thresholds $a_{\varepsilon,x},b_{\varepsilon,y}$ may be different from the true $a_{*,x},b_{*,y}$ and/or depend on contamination fraction~$\varepsilon$. However, it turns out that from a theoretical perspective, studying the case where the $a_{\varepsilon,x},b_{\varepsilon,y}$ are different from the $a_{*,x},b_{*,y}$ is equivalent to a case where they are equal.\footnote{Let $H'$ be an arbitrary contamination distribution of the latent variables. Let the thresholds that discretize these latent variables be given by arbitrary values $a'_{\varepsilon,x},b'_{\varepsilon,y}$, for $x\in\X,y\in\Y$. Since one makes no assumption on the contamination distribution in~\eqref{eq:contaminationmodel}, we can find another contamination distribution $H\neq H'$ that, when discretized with the true thresholds $a_{*,x},b_{*,y}$, yields the same discretization as~$H'$ with thresholds $a'_{\varepsilon,x},b'_{\varepsilon,y}$. Formally, $\forall a'_{\varepsilon,x},b'_{\varepsilon,y}, H'\ \exists H \text{ s.t. } \int_{a'_{\varepsilon,x-1}}^{a'_{\varepsilon,x}} \int_{b'_{\varepsilon,y-1}}^{b'_{\varepsilon,y}} \d H' = \int_{a_{*,x-1}}^{a_{*,x}} \int_{b_{*,y-1}}^{b_{*,y}} \d H$.} 

The population response probabilities $\fepsxy$ in~\eqref{eq:feps} are unknown in practice because they depend on unspecified and unmodeled quantities, namely the contamination fraction~$\varepsilon$, the contamination distribution~$H$, and the discretization thresholds of the latter. Consequently, we do not attempt to estimate the population response probabilities~$\fepsxy$. We instead focus on estimating the true polychoric model probabilities~$\pxy{\Btheta_*}$ while reducing bias stemming from potential contamination in the observed data.

\begin{figure}[t]
	\centering
	\includegraphics[width = \textwidth]{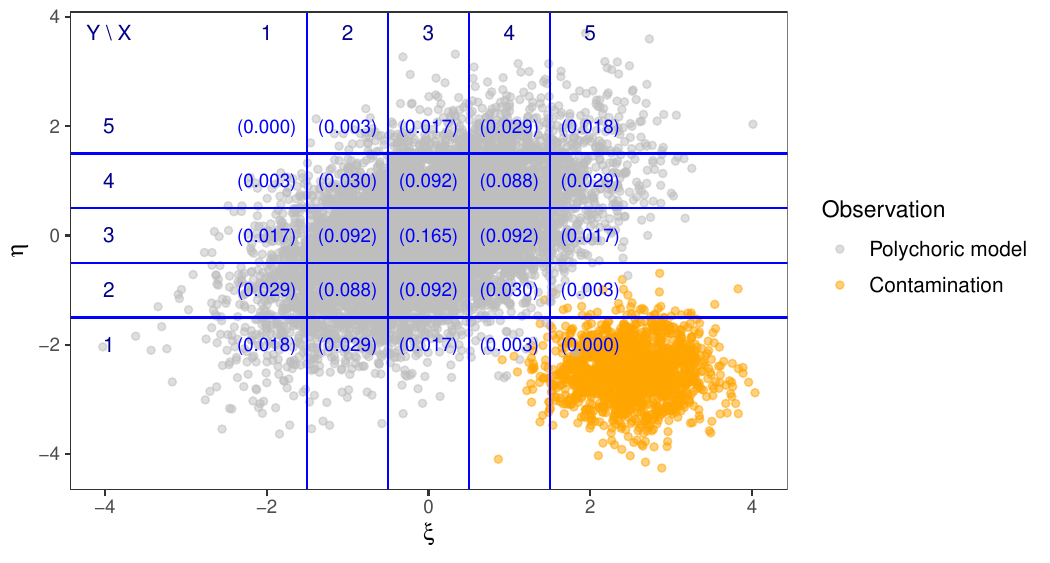}
\caption{Simulated data with $K_X=K_Y=5$ response options where the polychoric model is misspecified with contamination fraction $\varepsilon=0.15$. The gray dots represent random draws of $(\xi,\eta)$ from the polychoric model with~$\rho_*=0.5$, whereas the orange dots represent draws from a contamination distribution that primarily inflates the cell $(x,y)=(5,1)$. The contamination distribution is bivariate normal with a mean $(2.5,-2.5)^\top$, variances $(0.25, 0.25)^\top$, and zero correlation. The blue lines indicate the locations of the thresholds. In each cell, the numbers in parentheses denote the population probability of that cell under the true polychoric model.}
\label{fig:contamexample}
\end{figure}

Figure~\ref{fig:contamexample} visualizes a simulated example of bivariate data drawn from contaminated distribution~$G_\varepsilon$, where a fraction of $\varepsilon=0.15$ of the data follow a bivariate normal contamination distribution~$H$ (orange dots) with mean  $(2.5, -2.5)^\top$, variance $(0.25, 0.25)^\top$, and zero correlation, whereas the remaining data are generated by a  standard bivariate normal distribution with correlation $\rho_*=0.5$ (gray dots). In this example, the data from the contamination distribution~$H$ primarily inflate the cell~$(x,y) = (5,1)$ after discretization. That is, this cell will have a larger empirical frequency than the polychoric model allows for, since the probability of this cell is nearly zero at the polychoric model, yet many realized responses will populate it. Consequently, due to (partial) misspecification of the polychoric model, a maximum likelihood estimate of~$\rho_*$ on these data might be substantially biased for~$\rho_*$. Indeed, calculating the MLE using the data plotted in Figure~\ref{fig:contamexample} yields an estimate of $\hatNMLE{\rho}=-0.10$, which is far off from the true $\rho_*=0.5$. In contrast, our proposed robust estimator, which is calculated from the exact same information as the MLE and is defined in Section~\ref{sec:robust-polycor}, yields a fairly accurate estimate of~$0.47$.

It is worth addressing that there exist nonnormal distributions of the latent variables~$(\xi,\eta)$ that, after discretization with the same thresholds, result in the same response probabilities as under latent normality \citep{foldnes2019identification}. This implies that there may exist contamination distributions~$H$ and contamination fractions~$\varepsilon  > 0$ under which the population probabilities~$\fepsxy$ in~\eqref{eq:feps} are equal to the true population probabilities of the polychoric model,~$\pxy{\Btheta_*}$, that is, $\fepsxy = \pxy{\Btheta_*}$ for all $(x,y)\in\X\times\Y$. In this situation, the polychoric model is misspecified, but the misspecification does not have consequences because the response probabilities remain unaffected. To avoid cumbersome notation in the theoretical analysis of our robust estimator, we assume consequential misspecification throughout this paper, that is, $\fepsxy \neq \pxy{\Btheta_*}$ for some $(x,y)\in\X\times\Y$ whenever $\varepsilon > 0$. However, it is silently understood that misspecification need not be consequential, in which case there is no issue and both the MLE and our robust estimator are consistent for the true~$\Btheta_*$.

\subsection{Distributional misspecification}\label{sec:distmiss-theory}

A model is distributionally misspecified when all observations in a given sample are generated by a distribution that is different from the model distribution. 
In the context of the polychoric model, this means that all ordinal observations  are generated by a latent distribution that is nonnormal. 
Let~$G$ denote the unknown  nonnormal distribution that the latent variables~$(\xi, \eta)$ jointly follow under distributional misspecification. The object of interest is the population correlation between latent~$\xi$ and~$\eta$ under distribution~$G$, for which the normality-based MLE of \citet{olsson1979poly} turns out to be substantially biased in many cases \citep[e.g.,][]{lyhagen2023,foldnes2020polycor,foldnes2022,jin2017}.
As such, distributional misspecification is fundamentally different from partial misspecification: In the former, one attempts to estimate the population correlation of the nonnormal and unknown distribution that generated a sample, instead of estimating the polychoric correlation coefficient (which is the correlation under standard bivariate normality). In the latter, one attempts to estimate the polychoric correlation coefficient with a contaminated sample that has only been partly generated by latent normality (that is, the polychoric model). The assumption that the polychoric model is only partially misspecified for some uninformative observations enables one to still estimate the polychoric correlation coefficient of that model, which would not be feasible under distributional misspecification (at least not without additional assumptions).

Despite not being designed for distributional misspecification, the robust estimator introduced in the next section can offer a robustness gain in some situations where the polychoric model is distributionally misspecified. We discuss this in more detail in Section~\ref{sec:distributional-misspecification}.

\section{Robust estimation of polychoric correlation}\label{sec:robust-polycor}
The behavior of ML estimates of any model crucially depends on correct specification of that model. Indeed, ML estimation can be severely biased even when the assumed model is only slightly misspecified  \citep[e.g.,][]{huber2009,hampel1986,huber1964}. For instance, in many models of continuous variables like regression models, one single observation from a different distribution can be enough to make the ML estimator converge to an arbitrary value (\citealp{huber2009}; see also \citealp{alfons2022}, for the special case of mediation analysis). The non-robustness of ML estimation of the polychoric model has been demonstrated empirically by \citet{foldnes2020polycor,foldnes2022,gronneberg2022} for the case of distributional misspecification. In this section, we introduce an estimator that is designed to be robust to partial misspecification when present, but remains (asymptotically) equivalent to the ML estimator of \citet{olsson1979poly} when misspecification is absent. We furthermore derive the statistical properties of the proposed estimator. 

Throughout this section, let $\{(X_i, Y_i)\}_{i=1}^N$ be an observed %$N$-sized 
ordinal sample of size~$N$ generated by discretizing latent variables~$(\xi,\eta)$ that follow the unknown and unspecified %mixture
contaminated distribution~$G_\varepsilon$ in~\eqref{eq:contaminationmodel}. Hence, the polychoric model is possibly misspecified for an unknown fraction~$\varepsilon$ of the sample.

\subsection{The estimator}\label{sec:estimator}
The proposed estimator is a special case of a class of robust estimators for general models of categorical data called~$C$-estimators \citep{welz2024robcat}, and is based on the following idea. The empirical relative frequency of a response $(x,y)\in\X\times\Y$, denoted
\[
	\fhatxy = N_{xy} / N = \frac{1}{N}\sum_{i=1}^N \I{X_i = x, Y_i= y},
\]
is a consistent nonparametric estimator of the population response probability in~\eqref{eq:feps},
\[
\fepsxy =
	\Pr{G_\varepsilon}{X = x, Y=y} 
	=
	(1-\varepsilon) \pxy{\Btheta_*} + \varepsilon \int_{a_{\varepsilon,x-1}}^{a_{\varepsilon,x}} \int_{b_{\varepsilon,y-1}}^{b_{\varepsilon,y}} \d H,
\]
as $N\to\infty$ \citep[see, e.g., Chapter~19.2 in][]{vandervaart1998}. If the polychoric model is correctly specified $(\varepsilon = 0)$, then~$\fhatxy$ will converge (in probability) to the true model probability~$\pxy{\Btheta_*}$ because
\[
	f_0(x,y) = \pxy{\Btheta_*},
\] 
for all $(x,y)\in\X\times\Y$. 
Conversely, if the polychoric model is misspecified $(\varepsilon > 0)$, then~$\fhatxy$ may \emph{not} converge to the true $\pxy{\Btheta_*}$ because
\[
	\fepsxy \neq \pxy{\Btheta_*}
\]
for some $(x,y)\in\X\times\Y$, since we assume consequential misspecification.

It follows that if the polychoric model is misspecified, there exists no parameter value \mbox{$\Btheta\in\BTheta$} for which the nonparametric estimates~$\fhatxy$ converge pointwise to the associated model probabilities~$\pxy{\Btheta}$ for all $(x,y)\in\X\times\Y$.
Hence, it is indicative of model misspecification if there exists at least one response $(x,y)\in\X\times\Y$ for which~$\fhatxy$ does not converge to any polychoric model probability~$\pxy{\Btheta}$, resulting in a discrepancy between~$\fhatxy$ and~$\pxy{\Btheta}$.\footnote{Actually, if said convergence fails, it does so for at least two responses. Due to the relative frequencies summing up to~1, if the convergence fails for one response, it must also fail for at least one more.} This observation can be exploited in model fitting by minimizing the discrepancy between the empirical relative frequencies,~$\fhatxy$, and theoretical model probabilities,~$\pxy{\Btheta}$, to find the most accurate fit that can be achieved with the polychoric model for the observed data. Specifically, our estimator minimizes with respect to~$\Btheta$ the loss function
\begin{equation}\label{eq:loss}
	L\left( \Btheta,\ \fhat \right) = \sum_{x\in\X}\sum_{y\in\Y} \varphi\left(\residual{\fhatxy}{\Btheta}-1\right)\pxy{\Btheta},
\end{equation}
where $\varphi : [-1,\infty) \to \R$ is a prespecified \emph{discrepancy function} that will be defined momentarily. The proposed estimator~$\thetahat$ is given by the value minimizing the objective loss over parameter space~$\BTheta$,
\begin{equation}\label{eq:estimator}
	\thetahat = \arg\min_{\Btheta\in\BTheta} L\Big(\Btheta, \fhat\Big).
\end{equation}
For the choice of discrepancy function %choice 
$\varphi(z) = \varphi^{\mathrm{MLE}}(z) = (z+1) \log (z+1)$, it can be easily verified that~$\thetahat$ coincides with the MLE~$\thetahatMLE$ in~\eqref{eq:MLE}. In the following, we motivate a specific choice of discrepancy function~$\varphi(\cdot)$ that makes the estimator~$\thetahat$ less susceptible to misspecification of the polychoric model while preserving equivalence with ML estimation in the absence of misspecification.

\begin{figure}
 \centering
    \includegraphics[width = \textwidth]{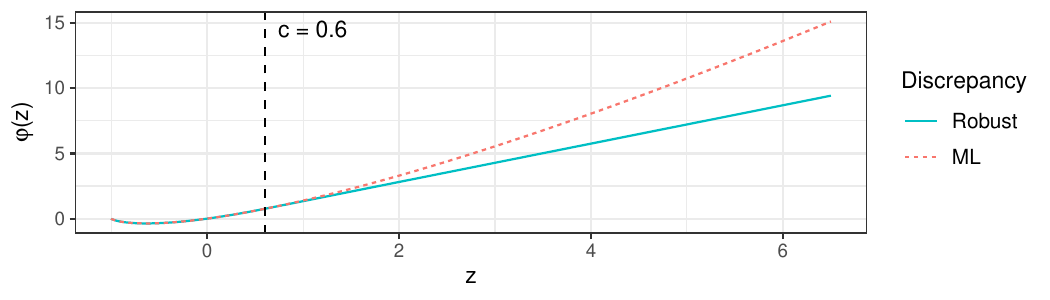}
\caption{Visualization of the robust discrepancy function $\varphi(z)$ in \eqref{eq:phifun} for~$c = 0.6$ (solid line) and the ML discrepancy function $\varphi^{\mathrm{MLE}}(z) = (z+1) \log (z+1)$ (dotted line).}
\label{fig:phifun}
\end{figure}

The fraction between empirical relative frequencies and model probabilities with value~1 deducted, 
\[
	\residual{\fhatxy}{\Btheta} -1,
\]
is referred to as \emph{Pearson residual} (PR) \citep{lindsay1994}. It takes values in $[-1,+\infty)$ and can be interpreted as a goodness-of-fit measure. PR values close to~0 indicate a good fit between data and polychoric model at~$\Btheta$, whereas values toward~$-1$ or~$+\infty$ indicate a poor fit because empirical response probabilities disagree with their model counterparts. To achieve robustness to misspecification of the polychoric model, responses that cannot be modeled well by the polychoric model, as indicated by their PR being away from~0, should receive less weight in the estimation procedure such that they do not over-proportionally affect the fit. Downweighting when necessary is achieved through a specific choice of discrepancy function~$\varphi(\cdot)$ proposed by \citet{welz2024robcat}, which is a special case of a function suggested by \citet{ruckstuhl2001}. The discrepancy function reads
\begin{equation}\label{eq:phifun}
	\varphi(z) = 
	\begin{cases}
		(z+1) \log(z+1)         &\textnormal{ if } z \in [-1, c],\\
		(z+1)(\log(c+1) + 1) - c-1&\textnormal{ if } z > c,
	\end{cases}
\end{equation}
where $c\in [0,\infty]$ is a prespecified tuning constant that governs the estimator's behavior at the PR of each possible response. Figure~\ref{fig:phifun} visualizes this function for the example choice~$c=0.6$ as well as the ML discrepancy function $\varphi^{\mathrm{MLE}}(z) = (z+1) \log (z+1)$. 
Note that deducting~1 in~\eqref{eq:loss} and adding it again in~\eqref{eq:phifun} is purely for keeping the interpretation that a PR close to~0 indicates a good fit.
We further stress that although the discrepancy function~\eqref{eq:phifun} can be negative, the loss function~\eqref{eq:loss} is always nonnegative \citep{welz2024robcat}.

For the choice~$c=+\infty$, minimizing the loss~\eqref{eq:loss} is equivalent to maximizing the log-likelihood objective in~\eqref{eq:MLE}, meaning that the estimator~$\thetahat$ is equal to~$\thetahatMLE$ for this choice of~$c$. More specifically, if a Pearson residual $z = \residual{\fhatxy}{\Btheta}-1$ of a response $(x,y)\in\X\times\Y$ is such that $z\in [-1, c]$ for fixed $c\geq 0$, then the estimation procedure behaves at this response like in classic ML estimation. As argued before, in the absence of misspecification,~$\fhatxy$ converges to~$\pxy{\Btheta_*}$ for all responses $(x,y)\in\X\times\Y$, therefore all PRs are asymptotically equal to~0. Hence, if the polychoric model is correctly specified, then estimator~$\thetahat$ is asymptotically equivalent to the MLE~$\thetahatMLE$ for any tuning constant value $c \geq 0$. On the other hand, if a response's PR~$z$ is larger than~$c$, that is, $z > c \geq 0$, then the estimation procedure does not behave like in ML, but the response's contribution to loss~\eqref{eq:loss} is linear rather than super-linear like in ML (Figure~\ref{fig:phifun}). It follows that the influence of responses that cannot be fitted well by the polychoric model is downweighted to prevent them from dominating the fit. The tuning constant~$c\geq 0$ is the threshold beyond which a PR will be downweighted, so the choice thereof determines what is considered an insufficient fit. The closer to~0 the tuning constant~$c$ is chosen, the more robust the estimator is in theory. In Appendix~\ref{app:simulation-c}, we explore different values of~$c$ in simulations and motivate a specific choice that we use for all numerical results in this paper, namely $c=0.6$.

Note that the discrepancy function~$\varphi(\cdot)$ in~\eqref{eq:phifun} may only downweight \emph{overcounts}, that is, the empirical probability~$\fhatxy$ exceeding the theoretical probability~$\pxy{\Btheta}$ for some cell $(x,y)\in\X\times\Y$. One might wonder why \emph{undercounts}---$\fhatxy$ being smaller than~$\pxy{\Btheta}$, resulting in negative Pearson residuals---are not downweighted as well. Indeed, the discrepancy function in~\eqref{eq:phifun} does not change its behavior compared to the MLE for Pearson residuals below~0. The empirical frequency~$\fhatxy$ is a \emph{relative} measure, so if a contingency table cell~$(x,y)$ has inflated counts, the other cells will have reduced values of~$\fhat$. If the discrepancy function would downweight undercounts, there is a risk of downweighting non-contaminated cells simply because these cells have reduced~$\fhat$ values if at least one cell is inflated due to contamination. Such behavior could result in bias since non-contaminated cells are not supposed to be downweighted. We refer to \citet[][p.~1128]{ruckstuhl2001} for a related discussion.

With the proposed choice of~$\varphi(\cdot)$, we stress that our estimator~$\thetahat$ in~\eqref{eq:estimator} has the same time complexity as ML,  namely~$O\big(K_X\cdot K_Y\big)$, since one needs to calculate the Pearson residual of all~$K_X\cdot K_Y$ possible responses for every candidate parameter value. Consequently, our proposed estimator does not incur any additional computational cost compared to ML.

\subsection{Statistical properties} \label{sec:properties}

We first address what quantity is estimated by the proposed estimator before discussing its asymptotic behavior.

\subsubsection{Estimand} 

The estimand of the estimator~$\thetahat$ in~\eqref{eq:estimator} is given by
\[
	\Btheta_0 = \arg\min_{\Btheta\in\BTheta} L\Big(\Btheta, f_\varepsilon\Big).
\]
This minimization problem is simply the population analogue to the minimization problem in~\eqref{eq:estimator} that the sample-based~$\thetahat$ solves because the probabilities~$\fepsxy$ are the population analogues to the empirical probabilities~$\fhatxy$. 

If the polychoric model is correctly specified, the estimand~$\Btheta_0$ equals the true parameter~$\Btheta_*$. Indeed, if $\varepsilon = 0$, then $f_0(x,y) = \pxy{\Btheta_*}$ for all $(x,y)\in\X\times\Y$, so it follows that the loss 
\begin{equation*}
\begin{split}
	L\Big(\Btheta, f_0\Big) 
	&=
	\sum_{x\in\X}\sum_{y\in\Y} \varphi\left(\residual{\pxy{\Btheta_*}}{\Btheta}-1\right)\pxy{\Btheta}
	\\
	&=
	\sum_{x\in\X}\sum_{y\in\Y} 
	\pxy{\Btheta_*} 
	\funresidual{\log}{\pxy{\Btheta_*}}{\Btheta}
	\I{\residual{\pxy{\Btheta_*}}{\Btheta}-1 \in [-1,c]}
	\\
	&\quad+ \Big(
		\pxy{\Btheta_*} \big(\log(c+1)+1\big)
		-\pxy{\Btheta}(c+1)	
	\Big)
	\I{\residual{\pxy{\Btheta_*}}{\Btheta}-1 >c}
\end{split}
\end{equation*}
attains its global minimum of zero if and only if $\Btheta = \Btheta_*$, for any choice of $c\geq 0$. Thus, in the absence of contamination, our estimator estimates the same quantity as the MLE, namely the true~$\Btheta_*$. In other words, it obtains the true~$\Btheta_*$ in population when the model is correctly specified, 
a property known as \emph{Fisher consistency}. We refer to \citet{welz2024robcat} for details.

\begin{figure}[t]
 \centering
    \includegraphics[width = \textwidth]{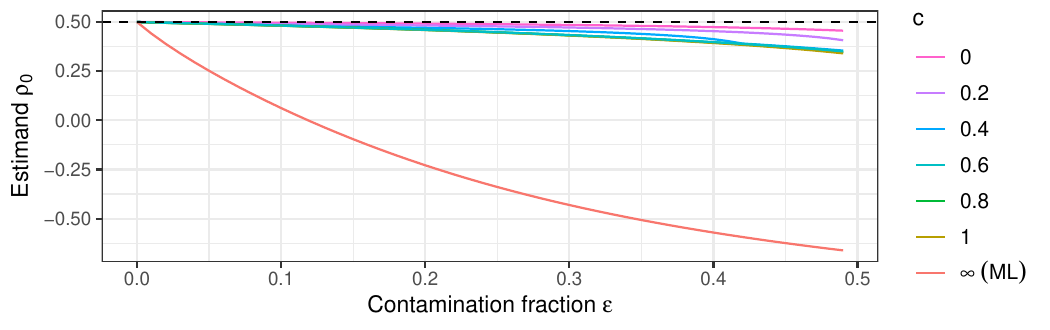}
\caption{The population estimand $\rho_0$ of the polychoric correlation coefficient for various degrees of contamination fractions~$\varepsilon$  ($x$-axis) and tuning constants~$c$ (line colors), for the same contamination distribution as in Figure~\ref{fig:contamexample}. The ML estimand corresponds to $c=+\infty$. There are $K_X=K_Y=5$ response options and the true value corresponds to $\rho_* = 0.5$ (dashed line).}
\label{fig:estimands}
\end{figure}

However, in the presence of misspecification ($\varepsilon > 0$), the sampling distribution differs from the model distribution such that the estimand~$\Btheta_0$---the parameter that minimizes the loss evaluated at the sampling distribution---is generally different from the true~$\Btheta_*$ \citep[cf.][]{white1982}.\footnote{\citet{white1982} is concerned with (quasi-)maximum likelihood estimation, but the same reasoning applies to our estimator. In the context of \citet{white1982}, the ML estimand of a misspecified model corresponds to the parameter value minimizing the Kullback-Leibler divergence between the sampling density and the model density. For $c=\infty$ (the ML case), the population analogue of our loss function~\eqref{eq:loss} corresponds to the Kullback-Leibler divergence between the sampling density~$\fepsxy$ and the model density~$\pxy{\cdot}$, that is, $\Btheta\mapsto L\left(\Btheta,\feps\right)$. For finite choices of~$c$, our loss function is designed to yield an estimand that is closer to the true~$\Btheta_*$ than that of the ML loss by downweighting responses that cannot be sufficiently well modeled.} The population estimand being different from the true value translates to biased estimates of the latter as a consequence of the misspecification.

How much the estimand~$\Btheta_0$ differs from the true~$\Btheta_*$ depends on the unknown fraction of contamination~$\varepsilon$, the unknown type of contamination~$H$, as well as the choice of tuning constant~$c$ in~$\varphi(\cdot)$. Mainly, the larger~$\varepsilon$ (more severe misspecification) and~$c$ (less downweighting of hard-to-fit responses), the further~$\Btheta_0$ is away from~$\Btheta_*$. Figure~\ref{fig:estimands} illustrates this behavior for the polychoric correlation coefficient at an example misspecified distribution that is described in Section~\ref{sec:feps} and in which the true polychoric correlation under the correct model amounts to ~$\rho_* = 0.5$. For increasing contamination fractions, the MLE ($c = +\infty)$ estimates a parameter value that is increasingly farther away from the true~$\Btheta_*$, where already a contamination fraction of less than~$\varepsilon = 0.15$ suffices for a sign flip in the correlation coefficient. Conversely, choosing tuning constant~$c$ to be near~0 results in a much less severe bias. For instance, even at contamination fraction $\varepsilon = 0.4$, the difference between estimand and true value is approximately~0.1 or less.

Overall, finite choices of~$c$ lead to an estimator that is at least as accurate as the MLE, and more accurate under misspecification of the polychoric model, thereby gaining robustness to misspecification. 

A relevant question is whether the true parameter~$\Btheta_*$ can be recovered when~$\varepsilon > 0$ such that it can be estimated using~$\thetahat$ combined with a  bias correction term. To derive such a bias correction term, one would need to impose assumptions on the %misspecification
contamination fraction and type of %misspecification
contamination. However, if one has strong prior beliefs about how the polychoric model is misspecified, modeling them explicitly rather than relying on the polychoric model seems more appropriate. Yet, one's beliefs about misspecification may not be accurate, so attempts to explicitly model the misspecification may themselves result in a misspecified model. Consequently, robust estimation traditionally refrains from making assumptions on how a model may potentially be misspecified by leaving~$\varepsilon$ and~$H$ unspecified in the contaminated %mixture
distribution~\eqref{eq:contaminationmodel}. Instead, one may use a robust estimator to identify data points that cannot be modeled with the model at hand, like the one presented in this paper. 

\subsubsection{Asymptotic analysis}

It can be shown that under certain standard regularity conditions that do not restrict the degree or type of possible partial misspecification beyond $\varepsilon \in [0,0.5)$, the robust estimator~$\thetahat$ is consistent for estimand~$\Btheta_0$ as well as asymptotically normally distributed. Specifically, under said regularity conditions and fixed tuning constant $c>0$, Theorem~\ref{thm:main} in Appendix~\ref{sec:asymptotics} establishes that 
\[
	\thetahat \convP \Btheta_0, 
\]
as well as
\[
	\sqrt{N}\left(\thetahat - \Btheta_0\right) \convweak \gauss_d \Big(\bm{0}, \BSigma{\Btheta_0} \Big), 
\]
as $N\to\infty$, where ``$\convP$'' and ``$\convweak$'' denote convergence in probability and distribution, respectively. The asymptotic 
covariance matrix has a sandwich-type construction
\[
\BSigma{\Btheta} = 
	\matinv{M}{\Btheta} \matfun{U}{\Btheta} \matinv{M}{\Btheta}, \qquad\Btheta\in\BTheta,
\]
where the $d\times d$ matrices
\begin{align*}
\matfun{M}{\Btheta} = 
	\partialderivativetwice{L\big(\Btheta, \feps \big)}{\Btheta}
	\quad \ \textnormal{and} \ \quad
\matfun{U}{\Btheta} = 
  \var{\feps}{  
  \partialderivative{\log\left( p_{XY}(\Btheta) \right)}{\Btheta}
  \I{\frac{\fepsfun{X,Y}}{p_{XY}(\Btheta)} -1\in [-1,c]}},
\end{align*}
respectively, are the Hessian matrix of the population loss and the covariance matrix (evaluated at~$\feps$) of the likelihood score function---that is, the gradient of $\log(p_{XY}(\Btheta))$---weighted by %certain 
stochastic binary weights %~$\psi_{XY}(\Btheta)$
whether the Pearson residual is smaller than or equal to the tuning constant~$c$.\footnote{It is worth mentioning that for the MLE ($c=+\infty$), our sandwich-type covariance matrix~$\BSigma{\Btheta_0}$ reduces to that of \citet{white1982} and \citet{huber1967}, who studied the limit distribution of ML in misspecified models.} 
We derive closed-form expressions of the matrices~$\matfun{M}{\Btheta}$ and~$\matfun{U}{\Btheta}$ as well as their properties in Appendix~\ref{sec:asymptotics}.

The asymptotic covariance matrix~$\BSigma{\Btheta_0}$ of our estimator is unobserved in practice because it depends on the unknown quantities~$\Btheta_0$ and~$\feps$. Yet,~$\BSigma{\Btheta_0}$ can be consistently estimated by replacing~$\Btheta_0$ and~$\feps$ by their corresponding consistent estimators~$\thetahat$ and~$\fhat$, respectively. Details are provided in Appendix~\ref{sec:asymptotics}.

With this limit theory, one can construct standard errors and confidence intervals for every element in~$\Btheta_0$. Importantly, in the absence of contamination (such that $\Btheta_0 = \Btheta_*$), the asymptotic covariance matrix~$\BSigma{\Btheta_0}$ of the robust estimator %is equal to
reduces to that of the fully efficient MLE (i.e., inverse Fisher information matrix) as long as $c>0$.  It follows that there is no loss of statistical efficiency when there is no contamination.\footnote{This property does not contradict standard theory on the Cramér-Rao lower bound (CRLB) for the variance of unbiased estimators \citep[e.g., Theorem~7.3.9 in][]{casella2002}. In brief, if a statistical model is correctly specified, the variance of the MLE asymptotically attains this lower bound such that no unbiased estimator can be asymptotically more efficient than the MLE. Nevertheless, this does not imply that no other estimators can (asymptotically) attain the CRLB. When the polychoric model is correctly specified (i.e., no contamination), our estimator has the same asymptotic variance as the MLE, which satisfies the Cramér-Rao lower bound with equality.} Hence, if the polychoric model is correctly specified, the robust estimator and the MLE are asymptotically first and second order equivalent. We refer to Appendix~\ref{sec:asymptotics} for a rigorous exposition and discussion of the robust estimator's asymptotic properties.

\subsection{Implementation}\label{sec:implementation}
We provide a free and open source implementation of our proposed methodology in a package for the statistical programming environment~\proglang{R} \citep{R}. The package is called \pkg{robcat} \citep[for ``ROBust CATegorical data analysis'';][]{robcat}, and it is available from CRAN (the Comprehensive \proglang{R} Archive Network) at \url{https://CRAN.R-project.org/package=robcat}. To maximize speed and performance, the package is predominantly developed in~\proglang{C++} and integrated to~\proglang{R} via \pkg{Rcpp} \citep{eddelbuettel2013}. All numerical results in this paper were obtained with this package.

The estimator's minimization problem in~\eqref{eq:estimator} can be solved with standard algorithms for numerical optimization. In our experience, using an unconstrained version of the quasi-Newton algorithm L-BFGS-B of \citet{byrd1995} works fine. However, additional stability might be gained from imposing the boundary constraint on the correlation coefficient and the monotonicity constraints on the thresholds, see~\eqref{eq:BTheta}, for which the the simplex algorithm of \citet{nelder1965} for constrained optimization can be used. In our implementation in package \pkg{robcat}, the default behavior is to first try unconstrained  optimization via \mbox{L-BFGS-B}. If numerical instability is encountered or a monotonicity constraint is violated, the constrained optimization algorithm of \citet{nelder1965} is used instead. While this is the default behavior, the package allows users to freely specify any supported optimization routine.

An important user choice is that of the tuning constant~$c$ in discrepancy function~\eqref{eq:phifun}. The closer~$c$ is to~0, the more robust the estimator will be to possible misspecification of the polychoric model (see, e.g., Figure~\ref{fig:estimands}). On the other hand, in the presence of model misspecification, the more robust the estimator is made, the larger its estimation variance becomes. Moreover, if the model is correctly specified, then \citet{welz2024robcat} shows that the most robust choice, $c=0$, is associated with two drawbacks, namely asymptotic nonnormality as well as certain finite sample issues. We therefore suggest choosing a value slightly larger than~0. In simulation experiments (see Appendix~\ref{app:simulation-c}), we find that the estimator is relatively insensitive to the specific choice of~$c > 0$, as long as it is reasonably small (for robustness) yet sufficiently far away from~0 (to avoid the aforementioned issues). The choice~$c = 0.6$ thereby yields a good compromise so that we use this value for all applications in this paper. 
However, we acknowledge that a detailed study, preferably founded in statistical theory, is necessary to provide guidelines on the choice of~$c$. We will explore this in future work.

Furthermore, a two-step estimation procedure like in~\eqref{eq:twostep} is not recommended for robust estimation. The possible presence of responses that have not been generated by the polychoric model can inflate the empirical cumulative marginal proportion of some responses, which may result in a sizable bias of threshold estimates~\eqref{eq:twostep}, possibly translating into biased estimates of polychoric correlation coefficients in the second stage. Our robust estimator therefore estimates all model parameters (thresholds and polychoric correlation) simultaneously.

\section{Simulation studies on partial misspecification}\label{sec:simulation}

In this section, we employ two simulation studies to demonstrate the robustness gain of our proposed estimator under partial misspecification of the polychoric model. The first simulation design (Section~\ref{sec:sim-individual}) is a simplified setting with respect to the partial misspecification, chosen specifically to illustrate the effects of a particular type of contamination with high leverage affecting only a small number of contingency table cells. The second design (Section~\ref{sec:polycormatrix}) is more involved and considers estimation of a polychoric correlation matrix with a contamination type that scatters in many directions so that nonnormal data points can occur in every contingency table cell. Section~\ref{sec:moresimulations} summarizes findings from additional simulations in the appendix. For all simulation designs, we perform $5,000$ repetitions.

\subsection{Individual polychoric correlation coefficient}
\label{sec:sim-individual}

Let there be~$K_X=K_Y = 5$ response categories for each of the two rating variables and define the true thresholds in the discretization process~\eqref{eq:polycormodel} as
\[
	a_{*,1}=b_{*,1} = -1.5,\quad a_{*,2}=b_{*,2} = -0.5,\quad a_{*,3}=b_{*,3}=0.5,\quad a_{*,4}=b_{*,4}=1.5,
\]
and let the true polychoric correlation coefficient in latent normality model~\eqref{eq:latentnormality} be~$\rho_*=0.5$. To simulate partial misspecification of the polychoric model according to~\eqref{eq:contaminationmodel}, we let a fraction~$\varepsilon$ of the data be generated by a particular contamination distribution $H$---which is left unspecified and therefore not explicitly modeled by our estimator---namely a bivariate normal distribution with mean $(2.5,-2.5)^\top$, variances $(0.25, 0.25)^\top$, and zero covariance (and therefore zero correlation). We discretize the realizations of the contamination distribution according to the same thresholds $a_{*,1}, \dots, a_{*,4}, b_{*,1}, \dots, b_{*,4}$ as the uncontaminated realizations. This contamination distribution will inflate the empirical frequency of contingency table cells $(x,y) \in \{(5,1), (4,1), (5,2)\}$, in the sense that they have a higher realization probability than under the true polychoric model.\footnote{Conceptually, contamination in the cell~$(5,1)$ would correspond to a straightlining careless respondent in~$(5,5)$ or~$(1,1)$ if one of the items is negatively keyed. Thus, after recoding, cell~$(5,1)$ will be inflated.} In fact, the data plotted in Figure~\ref{fig:contamexample} were generated by this process for contamination fraction~$\varepsilon = 0.15$, and one can see in this figure that particularly cell~$(x,y) = (5,1)$ is sampled frequently although it only has a near-zero probability at the true polychoric model. The data points causing these three cells to be inflated  are instances of \emph{negative leverage points}. Here, such leverage points drag correlational estimates away from a positive value towards zero or, if there are sufficiently many of them, even a negative value. For intuition, one may think of such points as the responses of careless or inattentive participants whose responses are not based on item content. Although careless responding is only one special case of the unlimited and unrestricted variety of uninformative responses generated by~$H$, we use careless responding as an illustrative running example throughout our simulations.

For contamination fraction~$\varepsilon \in\{0, 0.01, 0.05, 0.1, 0.15, 0.2, 0.3, 0.4, 0.49\}$, we sample \mbox{$N=1,000$} ordinal responses from this data generating process. We estimate the true parameter~$\Btheta_*$ with our proposed estimator with tuning constant set to~$c=0.6$, the MLE \citep{olsson1979poly}, and, for comparison, the Pearson sample correlation calculated on the integer-valued responses. 

Let~$\rhohat$ be the estimate on a simulated dataset and $\SEhat{\rhohat}$ the estimated standard error of~$\rhohat$, which is constructed using the limit theory developed in Theorem~\ref{thm:main} in Appendix~\ref{sec:asymptotics}. As performance measures, we calculate the average bias of the correlation estimates, the average bias of the standard error estimates (using the standard deviation of the correlation estimates across repetitions as an approximation of the true standard error), as well as coverage and average length of confidence intervals at significance level~$\alpha=0.05$. The coverage is defined as the proportion (across repetitions) of confidence intervals $\left[\rhohat\mp q_{1-\alpha/2}\cdot\SEhat{\rhohat}\right]$ 
that contain the true~$\rho_*$, where~$q_{1-\alpha/2}$ is the~$(1-\alpha/2)$~quantile of the standard normal distribution. The length of a confidence interval is given by $2 \cdot q_{1-\alpha/2}\cdot\SEhat{\rhohat}$.

\begin{figure}[!t]
\centering
\includegraphics[width = \textwidth]{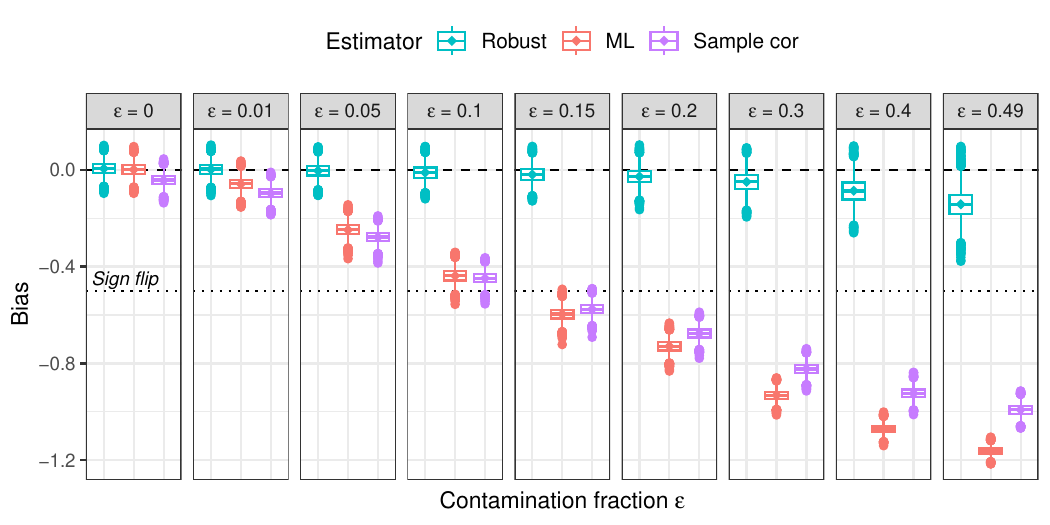}
\caption{Boxplot visualization of the bias of three estimators of the polychoric correlation coefficient, $\rhohat - \rho_*$, for various contamination fractions in the misspecified polychoric model across 5,000 repetitions. The estimators are the robust estimator with $c=0.6$ (left), the MLE (center), and the Pearson sample correlation (right). Diamonds represent the respective average bias. The dashed line denotes value~0 and the dotted line $-\rho_* = -0.5$, the latter of which indicating a sign flip in the correlation estimate.}
\label{fig:boxplot-sim}
\end{figure}

\begin{table}[!t]
\small
\centering
\setlength{\tabcolsep}{4.67pt}
\begin{tabular}{l l c r r c c c r c c c}
& & & \multicolumn{3}{c}{Point estimate} & & \multicolumn{2}{c}{Standard error} & & \multicolumn{2}{c}{Confidence interval} \\
\noalign{\smallskip}\cline{4-6}\cline{8-9}\cline{11-12}\noalign{\smallskip}
Contamination & Estimator & & \multicolumn{1}{c}{$\hatN{\rho}$} & \multicolumn{1}{c}{Bias} & SE & & $\widehat{\text{SE}}$ & Bias & & Coverage & Length \\
\noalign{\smallskip}\hline\noalign{\medskip}
\multirow{3}{*}{$\varepsilon = 0$} 
& Robust     & & 0.504 &   0.004  & 0.027 & & 0.027 & $-0.001$ & & 0.931 & 0.104 \\
& ML         & & 0.500 &   0.000  & 0.027 & & 0.026 & $-0.001$ & & 0.939 & 0.102 \\
& Sample cor & & 0.456 & $-0.044$ & 0.025 & & 0.025 & 0.003    & & 0.690 & 0.110 \\
\noalign{\medskip}
\multirow{3}{*}{$\varepsilon = 0.01$} 
& Robust     & & 0.501 &   0.001  & 0.028 & & 0.027 & $-0.001$ & & 0.938 & 0.107 \\
& ML         & & 0.441 & $-0.059$ & 0.027 & & 0.028 &   0.001  & & 0.446 & 0.110 \\
& Sample cor & & 0.402 & $-0.098$ & 0.025 & & 0.029 &   0.004  & & 0.043 & 0.114 \\ 
\noalign{\medskip}
\multirow{3}{*}{$\varepsilon = 0.05$} 
& Robust     & & 0.495 & $-0.005$ & 0.029 & & 0.029 & $-0.001$ & & 0.939 & 0.112 \\
& ML         & & 0.252 & $-0.248$ & 0.028 & & 0.033 &  0.005   & & 0.000 & 0.128 \\
& Sample cor & & 0.221 & $-0.279$ & 0.025 & & 0.031 &  0.006   & & 0.000 & 0.121 \\
\noalign{\medskip}
\multirow{3}{*}{$\varepsilon = 0.1$} 
& Robust     & & 0.488 & $-0.012$ & 0.031 & & 0.031 &  0.000 & & 0.933 & 0.120 \\
& ML         & & 0.062 & $-0.438$ & 0.028 & & 0.035 &  0.006 & & 0.000 & 0.135 \\
& Sample cor & & 0.052 & $-0.448$ & 0.025 & & 0.032 &  0.007 & & 0.000 & 0.124 \\
\noalign{\medskip}
\multirow{3}{*}{$\varepsilon = 0.15$} & 
Robust       & &   0.480  & $-0.020$ & 0.033 & & 0.033 &  0.000 & & 0.907 & 0.130 \\
& ML         & & $-0.097$ & $-0.597$ & 0.027 & & 0.035 &  0.007 & & 0.000 & 0.135 \\
& Sample cor & & $-0.076$ & $-0.576$ & 0.025 & & 0.032 &  0.007 & & 0.000 & 0.124 \\
\noalign{\medskip}
\multirow{3}{*}{$\varepsilon = 0.2$} & 
Robust       & &   0.472  & $-0.028$ & 0.036 & & 0.036 &  0.001 & & 0.882 & 0.142 \\
& ML         & & $-0.229$ & $-0.729$ & 0.026 & & 0.033 &  0.008 & & 0.000 & 0.131 \\
& Sample cor & & $-0.176$ & $-0.676$ & 0.025 & & 0.031 &  0.007 & & 0.000 & 0.122 \\
\noalign{\medskip}
\multirow{3}{*}{$\varepsilon = 0.3$} & 
Robust       & &   0.450  & $-0.050$ & 0.043 & & 0.045 &  0.002 & & 0.789 & 0.176 \\
& ML         & & $-0.431$ & $-0.931$ & 0.022 & & 0.030 &  0.008 & & 0.000 & 0.118 \\
& Sample cor & & $-0.322$ & $-0.822$ & 0.024 & & 0.030 &  0.007 & & 0.000 & 0.117 \\
\noalign{\medskip}
\multirow{3}{*}{$\varepsilon = 0.4$} & 
Robust       & &   0.413  & $-0.087$ & 0.053 & & 0.067 &  0.014 & & 0.595 & 0.261 \\
& ML         & & $-0.571$ & $-1.071$ & 0.019 & & 0.026 &  0.008 & & 0.000 & 0.103 \\
& Sample cor & & $-0.424$ & $-0.924$ & 0.024 & & 0.029 &  0.005 & & 0.000 & 0.112 \\
\noalign{\medskip}
\multirow{3}{*}{$\varepsilon = 0.49$} 
& Robust     & &   0.357  & $-0.143$ & 0.062 & & 0.071 &  0.009 & & 0.247 & 0.278 \\
& ML         & & $-0.662$ & $-1.162$ & 0.016 & & 0.024 &  0.008 & & 0.000 & 0.095 \\
& Sample cor & & $-0.492$ & $-0.992$ & 0.023 & & 0.028 &  0.004 & & 0.000 & 0.108
\end{tabular}
\caption{Results for the robust estimator with~$c=0.6$, the MLE, and the Pearson sample correlation, for various contamination fractions across~5,000 simulated datasets. The true polychoric correlation coefficient is $\rho_*=0.5$. We compute the average of point estimates~$\hatN{\rho}$ of the polychoric correlation coefficient, the average bias ($\hatN{\rho} - \rho_*$), the standard deviation of the~$\hatN{\rho}$ (SE; an approximation of the true standard error), the average standard error estimate $\widehat{\text{SE}}$, the corresponding average bias ($\widehat{\text{SE}} - \text{SE}$), confidence interval coverage with respect to the true~$\rho_*$ at nominal level 95\%, and the average length of the respective confidence intervals.} 
\label{tab:simresults}
\end{table}

Figure~\ref{fig:boxplot-sim} visualizes the bias of each estimator with respect to the true polychoric correlation~$\rho_*$ across the~5,000 simulated datasets. An analogous plot for the whole parameter~$\Btheta_*$ can be found in Appendix~\ref{app:sim-individual}; the results are similar to those of~$\rho_*$. Additional performance measures are shown in Table~\ref{tab:simresults}. For all considered contamination fractions, the estimates of the MLE and sample correlation are somewhat similar, which is expected because these two estimators are known to yield similar results when there are five or more rating options and the discretization thresholds are symmetric \citep[cf.,][]{rhemtulla2012}. In the absence of contamination, the MLE and the robust estimator yield accurate estimates. Both estimators are nearly equivalent to one another in the sense that their point estimates, standard deviation, and coverage at significance level~$\alpha=0.05$ are very similar. However, when contamination is introduced, MLE, sample correlation, and the robust estimator yield noticeably different results. Already at the small contamination fraction~$\varepsilon=0.01$ (corresponding to only~10 observations), MLE and sample correlation are noticeably biased, resulting in poor coverage of only about~0.45 and~0.04, respectively. Increasing the contamination fraction to the still relatively small value of~$\varepsilon = 0.05$, MLE and sample correlation start to be substantially biased with average biases of~$-0.25$ and~$-0.28$, respectively, leading to zero coverage. The biases of these two methods further deteriorate as the contamination fraction is gradually increased. At $\varepsilon \geq 0.15$, MLE and sample correlation produce estimates that are not only severely biased but also sign-flipped: while the true correlation is positive~(0.5), both estimates are negative. In stark contrast, the proposed robust estimator remains accurate throughout nearly all considered contamination fractions. At the small $\varepsilon = 0.01$, the robust estimator is nearly unaffected, while at $\varepsilon = 0.15$, it only exhibits a minor bias of about~$-0.02$. Even at extreme contamination $\varepsilon = 0.4$, its bias amounts to less than~0.1.  In addition, coverage of the robust method remains above or close to~0.9 for contamination fractions $\varepsilon \leq 0.2$.

It is worth noting that the confidence intervals of the robust estimator grow wider with increasing contamination fraction~$\varepsilon$. We also observe that the standard deviation of the robust estimates over the repetitions grow similarly. This indicates that the derived asymptotic distribution used to estimate standard errors matches well with the simulated distribution of the estimator across the repetitions. Indeed, the bias of standard error estimation remains at near-zero for the robust estimator, except for extremely large contamination fractions ($\varepsilon \geq 0.4$). We investigate this in more detail in Appendix~\ref{app:sim-individual}.

\subsection{Polychoric correlation matrix}
\label{sec:polycormatrix}

The goal of this simulation study is to robustly estimate a polychoric correlation matrix, that is, a matrix comprising of pairwise polychoric correlation coefficients. The simulation design is based on \citet{foldnes2020polycor}. 

Let there be~$r$ observed ordinal random variables and assume that a latent variable underlies each ordinal variable. In accordance with the multivariate polychoric model \citep[e.g.,][]{muthen1984}, the latent variables are assumed to jointly follow an~$r$-dimensional normal distribution with mean zero and a covariance matrix with unit diagonal elements so that the covariance matrix is a correlation matrix. Each individual latent variable is discretized to its corresponding observed ordinal variable akin to discretization process~\eqref{eq:polycormodel}. 

Following the five-dimensional simulation design in \citet{foldnes2020polycor}, there are $r=5$ ordinal variables with polychoric correlation matrix as in Table~\ref{tab:cormat-sim} such that the pairwise correlations vary from a low~0.2 to a moderate~0.56.\footnote{The correlation matrix in Table~\ref{tab:cormat-sim} has the additional interpretation of being the covariance matrix of a factor model for a single factor with loadings vector $(0.8, 0.7, 0.6, 0.5, 0.4)^\top$.} For all latent variables, the discretization thresholds are set to, in ascending order, $\Phi_1^{-1}(0.1) = -1.28,\  \Phi_1^{-1}(0.3) = -0.52,\ \Phi_1^{-1}(0.7) = 0.56$, and $\Phi_1^{-1}(0.9) = 1.28$, such that each ordinal variable can take five possible values. A visualization of the implied distribution of each ordinal variable can be found in Figure~5 in \citet{foldnes2020polycor}. 

As contamination distribution, we choose an $r$-dimensional Gumbel distribution comprising of mutually independent Gumbel marginal distributions, each with location and scale parameters equal to~0 and~3, respectively. To obtain ordinal observations, the unobserved realizations from this distribution are discretized via the same threshold values as realizations from the model (normal) distribution. 
As such, the uninformative ordinal observations generated by this contaminated distribution emulate the erratic behavior of a careless respondent. Unlike in the previous simulation design (Section~\ref{sec:sim-individual}), the uninformative responses are not concentrated around a few response options, but may occur in every response option. 

For contamination fraction~$\varepsilon \in\{0, 0.01, 0.05, 0.1, 0.15, 0.2, 0.3, 0.4, 0.49\}$, we sample \mbox{$N=1,000$} ordinal five-dimensional responses from this data generating process and use them to estimate the polychoric correlation matrix in Table~\ref{tab:cormat-sim} via %ML and
our robust estimator (again with tuning constant $c=0.6$) as well as the MLE. 

\begin{table}[!t]
\centering
\begin{tabular}{c | c c c c c}
Variable  & 1    & 2     & 3  & 4     & 5      \\\hline
1 	   & 1.00 &       &		&		&				\\
2	   & 0.56 & 1.00  &		&		&				\\
3	   & 0.48 & 0.42 & 1.00	&		&				\\
4 	   & 0.40 & 0.35 & 0.30	& 1.00	&				\\
5 	   & 0.32 & 0.28 & 0.24	& 0.20  & 1.00
\end{tabular}
\caption{Correlation matrix of $r=5$ latent variables as in \citet{foldnes2020polycor}. In line with the multivariate polychoric correlation model \citep[e.g.,][]{muthen1984}, the latent variables are jointly normally distributed with mean zero and this correlation matrix as covariance matrix.}
\label{tab:cormat-sim}
\end{table}

\begin{figure}[!t]
\centering
\includegraphics[width = \textwidth]{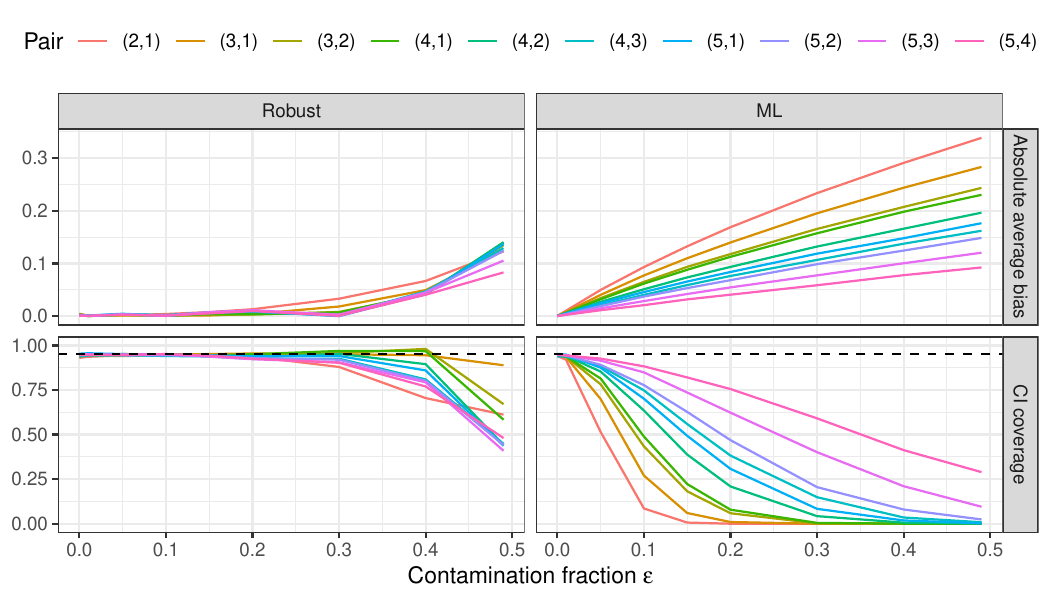}
\caption{Absolute average bias (top) and confidence interval coverage (bottom) %with 95\%
at nominal level 95\% (dashed horizontal lines) of the robust estimator with~$c=0.6$ (left) and the MLE (right) for each unique pairwise polychoric correlation coefficient in the true correlation matrix (Table~\ref{tab:cormat-sim}), expressed as a function of the contamination fraction~$\varepsilon$ ($x$-axis). %Averages are taken 
Results are aggregated over 5,000 repetitions.}
\label{fig:cormat-sim}
\end{figure}

Figure~\ref{fig:cormat-sim} visualizes the absolute average bias as well as confidence interval coverage at 95\% nominal level (calculated over the 5,000 repetitions) of the robust estimator and the MLE for each pairwise polychoric correlation coefficient.  
As expected, when the model is correctly specified ($\varepsilon = 0$), %then 
both estimators coincide with accurate estimates. However, in the presence of contamination ($\varepsilon > 0$), the two estimators deviate. The MLE exhibits a notable bias for all correlation coefficients, which increases gradually with increasing contamination fraction. The magnitude of the bias tends to be larger for pairs with larger true correlation, such as~$0.56$ for pair $(2,1)$, than for pairs with weaker true correlation, such as~$0.20$ for pair $(5,4)$. In addition, coverage of the MLE drops quickly for many pairs and gradually for the remaining ones. Conversely, the robust estimator remains nearly unaffected for a broad range of contamination fractions for each correlation coefficient ($\varepsilon \leq 0.2$ or $\varepsilon \leq 0.3$ depending on the variable pair), with bias only somewhat increasing afterwards. Furthermore, its coverage remains close to 0.9 or higher even at the high contamination fraction of $\varepsilon = 0.3$. This reflects excellent performance of the proposed estimator with respect to robustness to uninformative responses. Appendix~\ref{app:polycormatrix} contains additional evaluations. In essence, the robust estimator yields accurate standard errors, but its confidence intervals tend to be wider than those of the MLE in the presence of contamination.

We stress that polychoric correlation matrices need not be positive definite \citep[e.g.,][p.~22]{mair2018}, although all estimated polychoric correlation matrices in this simulation turned out positive definite. If an estimated polychoric correlation matrix is not positive definite, one may opt to apply a smoothing procedure like in \citet{yuan2011} or \citet{bock1988}.

\subsection{Discussion and additional simulation experiments} \label{sec:moresimulations}

The two simulation studies above demonstrate that already a small degree of partial misspecification due to uninformative responses, such as careless responses, can render the commonly employed MLE unreliable, while the proposed robust estimator retains good accuracy and coverage even in the presence of a considerable number of uninformative responses. On the other hand, when the polychoric model is correctly specified, the MLE and the robust estimator produce equivalent estimates.

To further evaluate our robust estimator and investigate its limitations, we conduct additional simulation experiments in Appendix~\ref{app:additional-sims}.

The first experiment, described in Appendix~\ref{app:meanshift}, is a generalization of the design in Section~\ref{sec:sim-individual} with different \emph{mean shifts} in the contamination distribution~$H$. For small mean shifts, the proposed estimator does not improve upon the MLE, but the bias of both estimators remains reasonable. The larger the mean shift, the larger the detrimental effect on the MLE and the higher the robustness gain of our proposed estimator.

The second experiment, described in Appendix~\ref{app:overlap}, focuses on \emph{correlation shifts} in the contamination distribution~$H$. Specifically, the contamination distribution is the same as the true model distribution except for a sign-flipped correlation coefficient $-\rho_*$. For moderate correlation $\rho_*$, the proposed estimator does not improve upon the MLE due to substantial overlap between the true model distribution and the contamination. However, the gain in robustness increases substantially for higher correlation coefficients~$\rho_*$. We expect the gain in robustness to increase for a higher number of response options and decrease for fewer response options. In the most extreme case of two dichotomous rating variables, no improvement can be expected.

\section{Empirical application}
\label{sec:application}

We now demonstrate our proposed method on empirical data by using a subset of the~100 unipolar markers of the Big Five personality traits \citep{goldberg1992}. 

\subsection{Background and study design}
Each marker is a an item comprising a single English adjective (such as ``bold'' or ``timid'') asking respondents to indicate how accurately the adjective describes their personality using a 5-point Likert-type rating scale (\emph{very inaccurate, moderately inaccurate, neither accurate nor inaccurate, moderately accurate}, and \emph{very accurate}). Here, each Big Five personality trait is measured with six pairs of adjectives that are polar opposites to one another (such as ``talkative'' vs. ``silent''), that is, twelve items in total for each trait. It seems implausible that an attentive respondent would choose to agree (or disagree) to \emph{both} items in a pair of polar opposite adjectives. Consequently, one would expect a strongly negative correlation between polar adjectives if all respondents respond attentively \citep{arias2020}.

\citet{arias2020} collect measurements of three Big Five traits in this way, namely \emph{extroversion, neuroticism}, and \emph{conscientiousness}.\footnote{\citet{arias2020} synonymously refer to \emph{neuroticism} as \emph{emotional stability}. Furthermore, in addition to the three listed traits, \citet{arias2020} collect measurements of the trait \emph{dispositional optimism} by using a different instrument, and another scale that is designed to not measure any construct. We do not consider these scales in this empirical demonstration. Furthermore, \citet{arias2020} have made their data publicly available at \url{https://osf.io/n6krb}.} The sample that we shall use, Sample~1 in \citet{arias2020}, consists of~$N=725$ online respondents who are all U.S. citizens, native English speakers, and tend to have relatively high levels of reported education (about~90\% report to hold an undergraduate or higher degree). Concerned about respondent inattention in their data, \citet{arias2020} construct a factor mixture model for detecting inattentive/careless participants. Their model crucially relies on response inconsistencies to polar opposite adjectives and is designed to primarily detect careless straightlining responding. They find that careless responding is a sizable problem in their data. 
Their model finds evidence of straightliners, and the authors conclude that if unaccounted for, they can substantially deteriorate the fit of theoretical models, produce spurious variance, and overall jeopardize the validity of research results.

Due to the suspected presence of careless respondents, we apply our proposed method to estimate the polychoric correlation coefficients between all~$\binom{12}{2}=66$ unique item pairs in the \emph{neuroticism} scale to obtain an estimate of the scale's (polychoric) correlation matrix. The results of the remaining two scales are qualitatively similar and are reported in Appendix~\ref{app:application-moreresults}. We estimate the polychoric correlation matrix via the MLE and via our proposed robust alternative with tuning parameter~$c=0.6$. As a robustness check, we further investigate the effect of the choice of~$c$.

\subsection{Results}
\label{sec:application-results}

\begin{figure}[t]
\centering
\includegraphics[width = 0.7\textwidth]{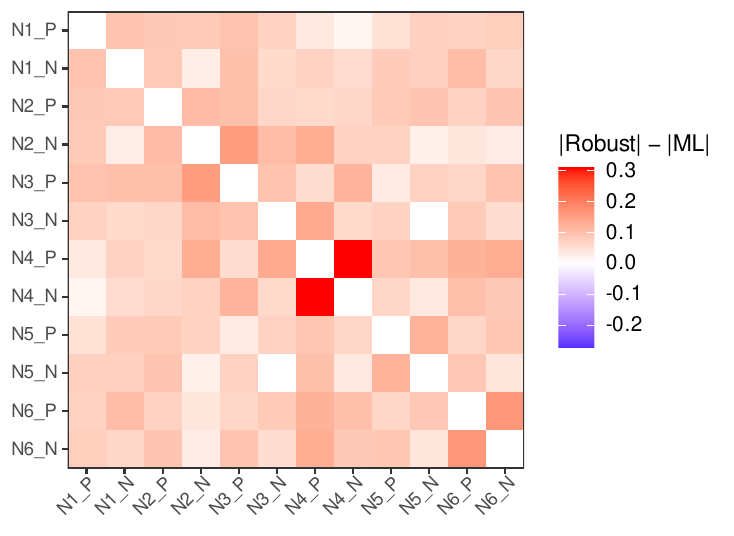}
\caption{Difference between absolute estimates for the polychoric correlation coefficient of %our
the robust estimator with $c=0.6$ and the MLE for each item pair in the \emph{neuroticism} scale, using the data of \citet{arias2020}. The items are ``calm'' (N1\_P), ``angry'' (N1\_N), ``relaxed'' (N2\_P), ``tense'' (N2\_N), ``at ease'' (N3\_P), ``nervous'' (N3\_N), ``not envious'' (N4\_P), ``envious'' (N4\_N), ``stable'' (N5\_P), ``unstable'' (N5\_N), ``contented'' (N6\_P), and ``discontented'' (N6\_N). For the item naming given in parentheses, items with identical identifier (the integer after the first ``N'') are polar opposites, where a last character ``P'' refers to the positive opposite and ``N'' to the negative opposite. The individual estimates of each method are provided in Table~\ref{tab:corrmatN} in Appendix~\ref{app:application-moreresults}.}
\label{fig:application-neuroticism-heatmap}
\end{figure}

Figure~\ref{fig:application-neuroticism-heatmap} visualizes the difference in absolute estimates for the polychoric correlation coefficient between all~66 unique item pairs in the \emph{neuroticism} scale. For all unique pairs, our method estimates a stronger correlation coefficient than the MLE. The differences in absolute estimates on average amount to~0.083, ranging from only marginally larger than zero to a substantive~0.314. For correlations between polar opposite adjectives, the average absolute difference between our robust method and the MLE is~0.151. The fact that a robust method consistently yields stronger correlation estimates than the MLE, particularly between polar opposite adjectives, is indicative of the presence of leverage points, which drag negative correlation estimates towards zero, that is, they attenuate the estimated strength of correlation. Here, such leverage points could be the responses of careless respondents who report agreement or disagreement to \emph{both} items in item pairs that are designed to be negatively correlated. For instance, recall that it is implausible that an attentive respondent would choose to agree (or disagree) to \emph{both} adjectives in the pair ``envious'' and ``not envious'' \citep[cf.,][]{arias2020}. If sufficiently many such respondents are present, then the presumably strongly negative correlation between these two opposite adjectives will be estimated to be weaker than it actually is.

\begin{table}
\centering
\small
\begin{tabular}{c c r c c r c c r c}
& & \multicolumn{2}{c}{Robust} & & \multicolumn{2}{c}{ML} & & \multicolumn{2}{c}{Sample cor}
\\
\noalign{\smallskip}\cline{3-4}\cline{6-7}\cline{9-10}\noalign{\smallskip}
Parameter & & Estimate & $\widehat{\text{SE}}$ & & Estimate & $\widehat{\text{SE}}$ & & Estimate & $\widehat{\text{SE}}$ 
\\
\noalign{\smallskip}\hline\noalign{\smallskip}
$\rho$ & & $-0.925$ & 0.062 & & $-0.618$ & 0.025 & & $-0.562$ & 0.031
\\
\noalign{\smallskip}
$a_1$  & & $-1.567$ & 0.276 & & $-1.373$ & 0.061 & & &
\\
$a_2$  & & $-0.560$ & 0.203 & & $-0.476$ & 0.043 & & &
\\
$a_3$  & &  $0.110$ & 0.187 & &  $0.121$ & 0.042 & & &
\\
$a_4$  & &  $1.076$ & 0.105 & &  $1.059$ & 0.054 & & &
\\
\noalign{\smallskip}
$b_1$  & & $-0.905$ & 0.073 & & $-0.857$ & 0.049 & & &
\\
$b_2$  & & $-0.040$ & 0.091 & & $-0.004$ & 0.041 & & &
\\
$b_3$  & &  $0.640$ & 0.364 & &  $0.608$ & 0.044 & & &
\\
$b_4$  & &  $1.171$ & 0.811 & &  $1.583$ & 0.071 & & &
\end{tabular}
\caption{Parameter estimates %with 
and standard error estimates ($\widehat{\text{SE}}$) for the correlation between the \emph{neuroticism} adjective pair %``envious'' and ``not envious''
``not envious'' and ``envious'' in the data of \citet{arias2020}, using the robust estimator with $c=0.6$, the MLE, and the Pearson sample correlation. Each adjective item has five answer categories. Note that the Pearson sample correlation does not model thresholds.}
\label{tab:application-neuroticism-results}
\end{table}

To further investigate the presence of careless respondents who attenuate correlational estimates, we study in detail the adjective pair ``not envious'' and ``envious'', which featured the largest discrepancy between the ML estimate and the robust estimate in Figure~\ref{fig:application-neuroticism-heatmap}, with an absolute difference of~0.314. The results of the two estimators and, for completeness, the sample correlation, are summarized in Table~\ref{tab:application-neuroticism-results}. The ML estimate of~$-0.618$ and sample correlation estimate of~$-0.562$ for the (polychoric) correlation coefficient seem remarkably weak considering that the two adjectives in question are polar opposites. In contrast, the  robust correlation estimate is given by~$-0.925$, which seems much more in line with what one would expect if all participants responded accurately and attentively \citep[cf.,][]{arias2020}.

\begin{table}[!t]
\centering
\begin{subtable}{0.99\textwidth}
\centering
\begin{tabular}{c|*{5}{c}}
$X$\textbackslash $Y$   & 1     & 2     & 3     & 4     & 5     \\
\hline
1  & 0.019 & 0.007 & 0.003 & 0.028 & 0.022 \\
2  & 0.007 & 0.040 & 0.050 & 0.138 & 0.014 \\
3  & 0.006 & 0.047 & 0.143 & 0.030 & 0.003 \\
4  & 0.054 & 0.189 & 0.029 & 0.019 & 0.007 \\
5  & 0.108 & 0.018 & 0.006 & 0.008 & 0.007 \\
\end{tabular}
\caption{Empirical relative frequencies $\fhatxy$\medskip}
\end{subtable}
\begin{subtable}{0.99\textwidth}
\centering
\begin{tabular}{c|*{5}{r}}
$X$\textbackslash $Y$   & 1     & 2     & 3     & 4     & 5     \\
\hline
1  & $<0.001$ & $<0.001$ & $<0.001$ & 0.024 & 0.034 \\
2  & $<0.001$ & 0.004 & 0.062 & 0.153 & 0.010 \\
3  & 0.001 & 0.072 & 0.145 & 0.038 & $<0.001$ \\
4  & 0.061 & 0.205 & 0.047 & 0.002 & $<0.001$ \\
5  & 0.120 & 0.020 & $<0.001$ & $<0.001$ & $<0.001$ \\
\end{tabular}
\caption{Estimated response probabilities $\pxy{\thetahat}$}
\end{subtable}
\begin{subtable}{0.99\textwidth}
\centering
\begin{tabular}{c|*{5}{r}}
$X$\textbackslash $Y$     & 1        & 2        & 3        & 4        & 5        \\
\hline
1     & $>1,000$  & $>1,000$ & 10.81   & 0.14     & $-0.35$  \\
2     & $>1,000$  & 9.06     & $-0.20$ & $-0.10$  &  0.42    \\
3     & 4.48      & $-0.35$  & $-0.01$ & $-0.20$  & 76.11    \\
4     & $-0.12$   & $-0.08$  & $-0.39$ & 11.66    & $>1,000$ \\
5     & $-0.11$   & $-0.12$  & 34.98   & $>1,000$ & $>1,000$ \\
\end{tabular}
\caption{Pearson residuals $\fhatxy \big/ \pxy{\thetahat}-1$\medskip}
\end{subtable}
\caption{Empirical relative frequency (top), estimated response probability (center), and Pearson residual (PR) (bottom) of each response $(x,y)$ for the item pair ``not envious'' ($X$) and ``envious''~($Y$) in the measurements of \citet{arias2020} of the \emph{neuroticism} scale. Estimate~$\thetahat$ was computed via the robust estimator with tuning constant $c=0.6$. The complete PR values are provided in Table~\ref{tab:arias2020-DPR-complete} in the appendix.}
\label{tab:application-neuroticism-summary}
\end{table}

To study the potential presence of careless responses in each contingency table cell~$(x,y)$ for item pair %``envious'' and ``not envious'',
``not envious'' and ``envious'', Table~\ref{tab:application-neuroticism-summary} lists the PRs at the robust estimate, alongside the associated model probabilities and empirical relative frequencies.\footnote{A visualization of Table~\ref{tab:application-neuroticism-summary} is provided in Figure~\ref{fig:application-neuroticism-dotplot} in the appendix.} A total of six cells have extremely large PR values of higher than~1,000, and, in addition, five  cells have a PR of higher than~9, and one cell has a PR of higher than~4. Such PR values are far away from ideal value~0 at which the model would fit perfectly, thereby indicating a poor fit of the polychoric model for such responses.  It stands out that all such poorly fitted cells are those whose responses might be viewed as inconsistent. Indeed, response cells %$(x,y) = (1,1), (1,2), (2,1), (2,2)$
$(x,y) \in \{(1,1), (1,2), (2,1), (2,2)\}$ indicate that a participant reports that \emph{neither} %``envious'' nor ``not envious''
``not envious'' nor ``envious'' characterizes them accurately, which are mutually contradicting responses, while for response cells %$(x,y) = (4,4), (4,5), (5,4), (5,5)$
$(x,y) \in \{(4,4), (4,5), (5,4), (5,5)\}$ \emph{both} adjectives characterize them accurately, which is again contradicting. As discussed previously, such responses might be due to careless responding. The robust estimator suggests that such responses cannot be fitted well by the polychoric model and subsequently downweighs their influence in the estimation procedure by mapping their Pearson residual with the linear part of the discrepancy function~$\varphi(\cdot)$ in~\eqref{eq:phifun}.  
Notably, also cells %$(x,y) = (1,3), (3,1), (3,5), (5,3)$
$(x,y) \in \{(1,3), (3,1), (3,5), (5,3)\}$ are poorly fitted. These responses report (dis)agreement to one opposite adjective, while being neutral about the other opposite. It is beyond the scope of this paper to assess whether such response patterns are also indicative of careless responding, but the robust estimator suggests that such responses at least cannot be fitted well by the polychoric model with the data of \citet{arias2020}. 

\begin{figure}[!t]
\centering
\includegraphics[width = 0.95\textwidth]{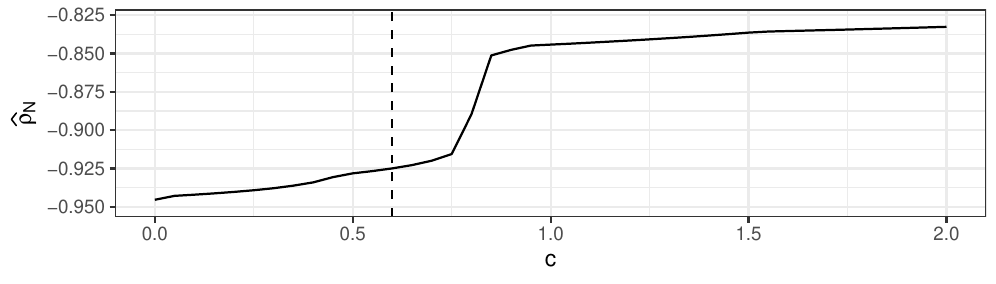}
\caption{Estimates of the polychoric correlation between the items  ``not envious'' and ``envious'' in the data of \citet{arias2020} for various choices of the tuning constant~$c$ ($x$-axis). The dashed vertical line marks the default value of $c=0.6$.}
\label{fig:application-neuroticism-vary}
\end{figure}

As a robustness check on the role of tuning constant~$c$, Figure~\ref{fig:application-neuroticism-vary} visualizes the point estimate~$\hatN{\rho}$ of the polychoric correlation between the item pair ``not envious'' and ``envious'' for various values of~$c$. The point estimate stays relatively constant for~$c$ between~0 and~0.75, with~$\hatN{\rho}$ ranging between $-0.95$ and $-0.92$. Just after $c=0.75$, $\hatN{\rho}$ abruptly jumps to about~$-0.85$, before it stabilizes again and slowly transitions to the ML estimate of $-0.62$ (see Table~\ref{tab:application-neuroticism-results}) for $c \rightarrow \infty$. Since~$\hatN{\rho}$ very slowly approaches the value of the ML estimate, we only visualize choices of~$c$ up to~2 in Figure~\ref{fig:application-neuroticism-vary}. The instability of the estimate around $c=0.75$ suggests that $c$ should be chosen below this value to avoid disproportionate influence of poorly fitting cells. For the broad range of $c \leq 0.75$, we obtain a robust finding of ``not envious'' and ``envious'' having a very strong negative correlation after accounting for likely careless responding.

Overall, leveraging our robust estimator, we find evidence for the presence of careless respondents in the data of \citet{arias2020}. While they substantially affect the correlation estimate of the MLE, amounting to about~$-0.62$, which is much weaker than one would expect for polar opposite items, our robust estimator can withstand their influence with an estimate of about~$-0.93$ and also help identify them through unreasonably large PR values. 
On a final note, our findings from the empirical application align remarkably well with those of the simulation from Appendix~\ref{app:overlap}, therefore strenghening their validity. We provide a detailed discussion of this similarity in Appendix~\ref{app:sim-application}.

\section{Estimation under distributional misspecification}
\label{sec:distributional-misspecification}

The simulation studies in Section~\ref{sec:simulation} have been concerned with a situation in which the polychoric model is misspecified for a subset of a sample (partial misspecification). However, as outlined in Section~\ref{sec:distmiss-theory}, another misspecification framework of interest is that of \emph{distributional misspecification} where the model is misspecified for the entire sample. Suppose that instead of a bivariate standard normal distribution, the latent variables~$(\xi, \eta)$ jointly follow an unknown and unspecified distribution~$G$.  In this framework, the object of interest is the population correlation between~$\xi$ and~$\eta$ under distribution~$G$, that is, $\rho_G = \corF{G}{\xi}{\eta}$, rather than a polychoric correlation coefficient. Estimators for such situations where the population distribution~$G$ is nonnormal have been proposed by \citet{lyhagen2023}, \citet{jin2017}, and \citet{roscino2006}.

\subsection{Distributional vs. partial misspecification}
\label{sec:distributional-approximation}

\citet[][p.~4]{huber2009} note that, although conceptually distinct,  robustness to distributional misspecification and partial misspecification are \textit{``practically synonymous notions''}. Hence, despite distributional misspecification not being covered by the 
%misspecified distribution~\eqref{eq:contaminationmodel}
partial misspecification framework under which we study the theoretical properties of our proposed estimator, our estimator could still offer a gain in robustness compared to %ML estimation
the MLE of the polychoric model under distributional misspecification. Specifically, the robust estimator may still be useful if the central part of the nonnormal distribution~$G$ is not too different from a standard bivariate normal distribution. Intuitively, if the difference between~$G$ and a standard normal distribution is mainly in the tails,~$G$ can be approximated by a %mixture of a standard normal distribution (to cover the central part) and some other distribution~$H$ (to cover the tails). 
contaminated distribution as in~\eqref{eq:contaminationmodel}, with the standard normal distribution covering the central part and some contamination distribution~$H$ covering the tails.
The %ML estimator
polychoric MLE tries to treat influential observations from the tails---which cannot be fitted well by the polychoric model---as if they were normally distributed, resulting in a possibly large estimation bias. In contrast, the robust estimator uses the normal distribution only for observations from the central part---which may fit the polychoric model well enough---and downweights observations from the tails. Thus, as long as such a %mixture
contaminated normal distribution is a decent approximation of the nonnormal distribution~$G$, the robust estimator should perform reasonably well. However, if~$G$ cannot be approximated by such a %mixture,
contaminated normal distribution, neither the %ML estimator
polychoric MLE nor our estimator can be expected to perform well. Overall, though, our estimator could offer an improvement in terms of robustness to distributional misspecification. 

In the following, we perform a simulation study to investigate the performance of our estimator when the polychoric model is distributionally misspecified. 

\subsection{Simulation study}
To simulate ordinal variables that were generated by a nonnormal latent distribution~$G$, we employ the VITA simulation method of \citet{gronneberg2017vita}. 
For a pre-specified value of the population correlation~$\rho_G$, the VITA method models the latent random vector~$(\xi,\eta)$ such that the individual variables~$\xi$ and~$\eta$ both possess standard normal marginal distributions with population correlation set equal to $\rho_G = \corF{G}{\xi}{\eta}$, but are \emph{not} jointly normally distributed. Instead, their joint distribution~$G$ is equal to a pre-specified nonnormal copula distribution, such as the Clayton or Gumbel copula. \citet{gronneberg2017vita} show that discretizing such VITA latent variables yields ordinal observations that could  not have been generated by a standard bivariate normal distribution, thereby ensuring proper violation of the latent normality assumption.

\begin{figure}[!t]
\centering
\includegraphics[width = \textwidth]{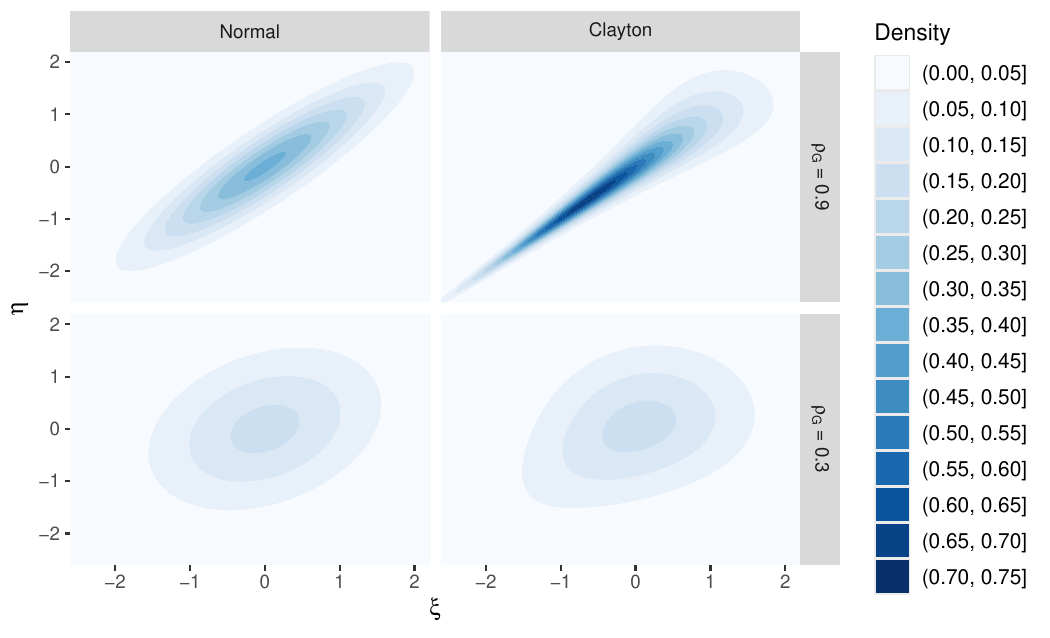}
\caption{Bivariate density plots of the standard normal distribution (left) and Clayton copula with standard normal marginals (right), for population correlations~0.9 (top) and~0.3 (bottom).}
\label{fig:copula}
\end{figure}

To investigate the robustness of our estimator to distributional misspecification, we use the VITA  %method
implementation in package \pkg{covsim} \citep{gronneberg2022covsim} to generate draws of the latent variables $(\xi,\eta)$ such that the latent variables are jointly distributed according to a Clayton copula~$G$ with population correlation~$\rho_G \in \{0.9, 0.3\}$ (see Figure~\ref{fig:copula} for visualizations).  
Following the discretization process~\eqref{eq:polycormodel}, we discretize both latent variables via discretization thresholds
\begin{align*}
\begin{tabular}{r r r r r}
\centering
$a_1 = {-1.5},$ & $a_2 = {0},$& $a_3 = {0.5},$ &$a_4 = {1},$ & \quad and\\ 
$b_1 = {-1},$    & $b_2 ={1},$& $b_3 ={1.5},$     &$b_4 ={2},$ &
\end{tabular}
\end{align*}
such that both resulting ordinal variables have five response options each. We generate \mbox{$N=1,000$} ordinal responses according to this data generation process and compute across~5,000 repetitions the same estimators and performance measures as in the simulations in Section~\ref{sec:simulation}. 

\begin{figure}[!t]
\centering
\includegraphics[width = 0.8\textwidth]{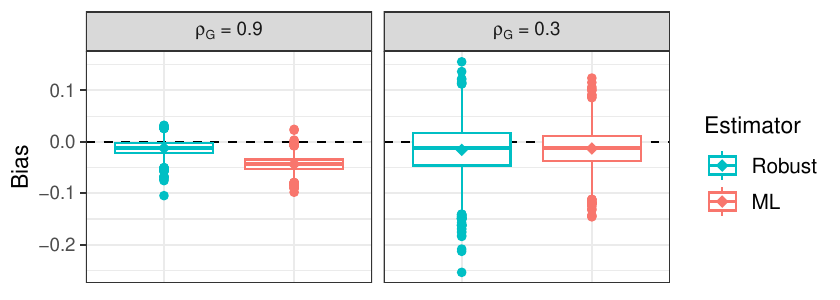}
\caption{Boxplot visualization of the bias of the robust %and ML estimator for the polychoric correlation coefficient
estimator and the polychoric MLE, $\hatN{\rho} - \rho_G$, under distributional misspecification via a Clayton copula with correlation $\rho_G = 0.9$ (left) and $\rho_G = 0.3$ (right), across 5,000 repetitions. Diamonds represent the respective average bias. The tuning constant of the robust estimator is set to $c=0.6$.}
\label{fig:simresults-distributional}
\end{figure}

Figure~\ref{fig:simresults-distributional} visualizes the bias of the robust estimator and the polychoric MLE under both Clayton copulas across the repetitions.\footnote{In 32 of the 5,000 repetitions for the Clayton copula, the robust estimator experienced numerical instability and subsequently did not converge to a solution. The polychoric MLE failed to converge~56 times. Consequently, these unconverged estimates are omitted from Figure~\ref{fig:simresults-distributional}. We explain this numerical issue in detail in Appendix~\ref{sec:instability}. Note that due to the shape of the Clayton copula with high correlation, stability issues with the robust estimator are in our experience amplified if the true thresholds are not well distributed over the domain of the distribution.} For correlation~0.9, the polychoric MLE exhibits a noteworthy bias, whereas the robust estimator remains accurate, albeit with a larger estimation variance as compared to most simulations configurations of partial misspecification (cf.~Section~\ref{sec:simulation}). Conversely, for the weaker correlation~0.3, both estimators are fairly accurate with average  biases of about~$-0.015$. 
Table~\ref{tab:simresults-distributional} contains additional performance measures regarding inference. For $\rho_G = 0.3$, the average standard error estimate of the robust method is accurate, but for $\rho_G = 0.9$, it notably overestimates. We therefore also computed the median of the standard error estimates in the latter case: at 0.017, it  
is fairly close to the true standard error of 0.014. 
It turns out that there is a small number of simulated datasets with a large majority of empty cells in the contingency table, resulting in numerical instability of the standard error estimates and inflating their average.
As discussed in Section~\ref{sec:distmiss-theory}, distributional misspecification is not covered by our partial misspecification framework, so it is not surprising that standard errors derived under partial misspecification are not always valid. Bootstrap inference may therefore be an attractive alternative. Nevertheless, unlike the polychoric MLE with coverage of only about~13\% for $\rho_G = 0.9$, the robust estimator maintains high coverage of over~90\%.

\begin{table}[!t]
\small
\centering
\setlength{\tabcolsep}{4.67pt}
 \begin{tabular}{l l c r r c c c r c c c}
 & & & \multicolumn{3}{c}{Point estimate} & & \multicolumn{2}{c}{Standard error} & & \multicolumn{2}{c}{Confidence interval} \\
 \noalign{\smallskip}\cline{4-6}\cline{8-9}\cline{11-12}\noalign{\smallskip}
 Correlation & Estimator & & \multicolumn{1}{c}{$\hatN{\rho}$} & \multicolumn{1}{c}{Bias} & SE & & $\widehat{\text{SE}}$ & Bias & & Coverage & Length \\
 \noalign{\smallskip}\hline\noalign{\medskip}
 \multirow{2}{*}{$\rho_G = 0.9$} 
 & Robust     & & 0.888 &   $-0.012$  & 0.014 & & 0.037 & 0.023 & & 0.915 & 0.146 \\
 & ML         & & 0.857 &   $-0.043$  & 0.013 & & 0.016 & 0.002 & & 0.132 & 0.061 \\
 \noalign{\medskip}
 \multirow{2}{*}{$\rho_G = 0.3$} 
 & Robust     & & 0.284 & $-0.016$ & 0.047 & & 0.052 & 0.006    & & 0.938 & 0.205 \\
 & ML         & & 0.287 & $-0.013$ & 0.036 & & 0.036 & $-0.001$ & & 0.932 & 0.141 \\
 \end{tabular}
\caption{Results for the robust estimator with~$c=0.6$ and the polychoric MLE across~5,000 simulated datasets under distributional misspecification via a Clayton copula with true population correlation~$\rho_G\in\{0.9, 0.3\}$. See Table~\ref{tab:simresults} for explanatory notes on the performance measures.}
\label{tab:simresults-distributional}
\end{table}

These results suggest that at least for point estimation, the Clayton copula with correlation~0.9 might be reasonably approximable by a 
contaminated normal distribution where the normal distribution covers the center of the probability mass and some contamination distribution covers the tails (cf.~Section~\ref{sec:distributional-approximation}). Indeed, Figure~\ref{fig:copula} indicates that the normal distribution and Clayton copula at correlation~0.9 seem to behave similarly in the center, but deviate from one another towards the tails. On the other hand, at correlation~0.3, the two densities do not appear to be drastically different from one another, which may explain why \emph{both} %ML
the polychoric MLE and the robust estimator work reasonably well for such distributional misspecification. If the latent distribution is not too different from a normal distribution, then the polychoric model may offer a satisfactory fit despite being technically misspecified.

Overall, this simulation study demonstrates that in some cases of distributional misspecification, robustness can be gained with our estimator, compared to %ML estimation
the polychoric MLE. However, it also demonstrates that there are cases of distributional misspecification for which the %ML estimator
polychoric MLE still works quite well such that the robust estimator offers little gain. 
Nevertheless, the fact that in some situations robustness can be gained under distributional misspecification represents an overall gain in robustness.

\section{Discussion and conclusion}\label{sec:conclusion}
We consider a situation where the polychoric correlation model is potentially misspecified for a subset of responses in a sample, that is, a set of uninformative observations not generated by a latent standard normal distribution. This model misspecification framework, called \emph{partial misspecification} here, stems from the robust statistics literature, and a relevant special case is that of careless respondents in questionnaire studies. We demonstrate that maximum likelihood estimation is highly susceptible to the presence of such uninformative responses, resulting in possibly large estimation biases and low coverage of confidence intervals.

As a remedy, we propose an estimator based on the $C$-estimation framework of \citet{welz2024robcat} that is designed to be robust to partial model misspecification. Our estimator generalizes maximum likelihood estimation, does not make any assumption on the magnitude or type of potential misspecification, comes at no additional computational cost, and possesses attractive statistical guarantees such as asymptotic normality.  It furthermore allows to pinpoint the sources of potential model misspecification through the notion of Pearson residuals (PRs). Each possible response option is assigned a PR, where values substantially larger than the ideal value~0 imply that the response in question cannot be fitted well by the polychoric correlation model. 
In addition, the methodology proposed in this paper is implemented in the free open source package \pkg{robcat} \citep{robcat} for the statistical programming environment~\proglang{R} and is publicly available at \url{https://CRAN.R-project.org/package=robcat}.

Although not covered by our partial misspecification framework, we also discuss how and when our estimator can offer a robustness gain (compared to %ordinary 
the polychoric maximum likelihood estimator) when the polychoric model is misspecified for \emph{all} observations, which has been a subject of interest in recent literature. In essence, there can be a robustness gain if the latent nonnormal distribution that generated the data can be reasonably well approximated by a contaminated normal distribution where the normal distribution reflects the central part and some unspecified contamination distribution reflects the tails.

We verify the enhanced robustness and theoretical properties of our robust estimator in simulation studies. Furthermore, we demonstrate the estimator's practical usefulness in an empirical application on a Big Five administration, where we find compelling evidence for the presence of careless respondents as a source of partial model misspecification. 

However, our estimator depends on a user-specified choice of a tuning constant~$c$, which governs a tradeoff between robustness and efficiency in case of misspecification. While simulation experiments suggest that the choice~$c=0.6$ provides a good tradeoff and estimates do not change considerably for %other
a broad range of finite choices of~$c$, a detailed investigation on this tuning constant needs to be carried out in future work. As an alternative, one could consider other discrepancy functions that do not depend on tuning constants, like the ones discussed in \citet{welz2024robcat}. As practical guidelines, we recommend always comparing robust estimates to that of ML and running the robust estimator for various choices of~$c$, like in Figure~\ref{fig:application-neuroticism-vary}. Doing so not only helps assess the estimates' stability, but also the severity of (partial) model misspecification. If the ML and robust estimates strongly differ, one may want to opt for choices of $c>0$ %that are closer to~0
not too far from~0 to achieve larger robustness gains.

A practical consideration of polychoric correlation is computing time. Our estimator simultaneously estimates all model parameters (i.e., correlation coefficient and thresholds) for robustness reasons, hence it is computationally more intensive than the non-robust two-stage approach of \citet{olsson1979poly}. To alleviate the computational burden, our implementation in package \pkg{robcat} is written in fast and efficient \proglang{C++} code. Furthermore, its default behavior first tries a fast unconstrained numerical optimization routine, which in our experience almost always suffices and executes in about half a second for five-point rating variables on a regular laptop. For estimating polychoric correlation matrices, package \pkg{robcat} supports parallel computing to keep computation time low. Since our method generalizes maximum likelihood and has the same time complexity, these functionalities also provide a fast implementation of the maximum likelihood estimator.

The methodology proposed in this paper suggests a number of extensions. For instance, one could use a robustly estimated polychoric correlation matrix in structural equation models to robustify such models and their fit indices against misspecification. Similar robustification could by achieved in, for instance but not limited to, principal component analyses, multidimensional scaling, or clustering. In addition, the theory of \citet{welz2024robcat} may allow to pinpoint possible sources of model misspecification on the individual response level. That is, it may enable the derivation of statistically sound cutoff values for Pearson residuals in order to detect whether a given response can be fitted well by the polychoric correlation model.
We leave these avenues to further research.

Overall, our novel robust estimator could open the door for a new line of research that is concerned with making the correlation-based analysis of rating data more reliable by reducing dependence on modeling assumptions.

\bibliography{references_polycor}

%%%%%%%%%%%%%%%%%%%%%%%%%%%%%%%%%%%%%%%%%%%%%%%%%%%%%%%%
%%%%%%%%%%%%%%%%%%% APPENDIX %%%%%%%%%%%%%%%%%%%%%%%%%%%
%%%%%%%%%%%%%%%%%%%%%%%%%%%%%%%%%%%%%%%%%%%%%%%%%%%%%%%%
\newpage
\appendix

\renewcommand\thefigure{\thesection.\arabic{figure}}   
\renewcommand\thetable{\thesection.\arabic{table}}   
\renewcommand{\theequation}{\thesection.\arabic{equation}}
\renewcommand{\thefootnote}{\roman{footnote}}
\setcounter{footnote}{0}   

\section{Asymptotic properties of the robust estimator} \label{sec:asymptotics}

\setcounter{figure}{0} 
\setcounter{table}{0}   
\setcounter{equation}{0} 

This appendix section presents the limit theory of the proposed robust estimator~$\thetahat$ in~\eqref{eq:estimator}.
Throughout this section, we assume that the number of response categories,~$K_X$ and~$K_Y$, are fixed and known, and that the sample $\{(X_i,Y_i)\}_{i=1}^N$ used to compute an estimate~$\thetahat$ has been generated by the process in~\eqref{eq:polycormodel} where the latent $\{(\xi_i,\eta_i)\}_{i=1}^N$ are draws from distribution~$G_\varepsilon$ in~\eqref{eq:contaminationmodel} with unobserved contamination fraction~$\varepsilon\in[0,0.5)$ and unknown contamination distribution~$H$. 

The following three subsections first state the main theorem, then discuss its assumptions, and then provide closed-form expressions of quantities used in the asymptotic analysis. 

\subsection{Main theorem}

We start by introducing additional notation.
%Define
Denote the likelihood score function, i.e., the $d$-dimensional gradient vector of $\log\left(\pxy{\Btheta}\right)$ for response $(x,y)\in\X\times\Y$ at parameter~$\Btheta\in\BTheta$, by %as
\[
	\sxy{\Btheta} = 
	\partialderivative{}{\Btheta} \log\left( \pxy{\Btheta} \right) = 
	\frac{1}{\pxy{\Btheta}} \left(\partialderivative{}{\Btheta}\ \pxy{\Btheta}\right).
\]
Further, given the tuning constant $c\in [0,\infty]$, define the $d\times d$ matrix 
\[
	\mat{Q}_{xy}(\Btheta) = 
	\Bigg(
		\residual{\fepsxy}{\Btheta} \I{\residual{\fepsxy}{\Btheta}-1\in [-1,c]}
		+ (c+1)\I{\residual{\fepsxy}{\Btheta}-1\ > c}
	\Bigg)
	\partialderivativetwice{}{\Btheta}\pxy{\Btheta}
\]
for $(x,y)\in\X\times\Y$ and $\Btheta\in\BTheta$.
We derive closed-form expressions of the gradient and Hessian matrix of~$\pxy{\Btheta}$ in Subsection~\ref{app:derivatives}. 
%
% \edit{$\psi_{XY}(\Btheta) \partialderivative{\log\left( p_{XY}(\Btheta) \right)}{\Btheta}$}
%
In addition, for $K=K_X\cdot K_Y$ the total number of contingency table cells and~$d$-dimensional vectors
\[
	\wxy{xy} = \sxy{\Btheta}\I{\residual{\fepsxy}{\Btheta}-1\in[-1,c]},
\] 
define the $d\times K$ matrix 
\begin{align*}
	&\BW{\Btheta}
	=\\
	&\quad\bigg(
		\wxy{11},\cdots, \wxy{1,K_Y}, \wxy{21},\cdots, \wxy{2,K_Y},\cdots, \wxy{K_X,1}, \wxy{K_X,2},\cdots,\wxy{K_X,K_Y}
	\bigg)
\end{align*}
that row-binds all~$K$ vectors~$\sxy{\Btheta}$ multiplied by an indicator that takes value~1 when associated population Pearson residual is in the MLE-part of the function~$\varphi(\cdot)$ in~\eqref{eq:phifun} and~0 otherwise.
In similar fashion, define the~$K$-dimensional vector 
\[
	\Bfeps 
	=	
	\Big(
	\fepsfun{1,1}, \dots, \fepsfun{1,K_Y}, \fepsfun{2,1},\dots, \fepsfun{2,K_Y},\dots, \fepsfun{K_X, 1}, \fepsfun{K_X,2},\dots,\fepsfun{K_X,K_Y}
	\Big)^\top
\]
that holds all~$K$ evaluations of the function $\feps$, and put
\[
	\BOmega = \mathrm{diag}(\Bfeps) - \Bfeps\Bfeps^\top,
\]
where $\mathrm{diag}(\Bfeps)$ is a $K\times K$ diagonal matrix that holds
the coordinates of~$\Bfeps$ on its principal diagonal.

The following theorem establishes consistency and asymptotic normality of the estimator. This theorem follows immediately from Theorems~1 and~2 in \citet{welz2024robcat}.

\begin{theorem}\label{thm:main}
For $c\in [0,\infty]$ the tuning constant in the discrepancy function, assume that $\residual{\fepsxy}{\Btheta_0}-1\neq c$ for %any
all $(x,y)\in\X\times\Y$. Then, under certain regularity conditions that do not restrict the degree or type of possible misspecification of the polychoric model beyond $\varepsilon \in [0,0.5)$, when $N\to\infty$ it holds true that
\[
	\thetahat \convP \Btheta_0, 
\]
as well as 
\[
	\sqrt{N}\left(\thetahat - \Btheta_0\right) \convweak \gauss_d \Big(\bm{0}, \BSigma{\Btheta_0} \Big), 
\]
where, as a function of $\Btheta\in\BTheta$, the estimator's invertible asymptotic covariance matrix is given by
\[
	\BSigma{\Btheta}
	=
	\matinv{M}{\Btheta} \matfun{U}{\Btheta} \matinv{M}{\Btheta},
\]
where
\begin{align*}
	\matfun{U}{\Btheta} 
	&=
	\var{\feps}{\vec{w}_{XY}(\Btheta)} =  \matfun{W}{\Btheta} \BOmega \matfun{W}{\Btheta}^\top \qquad\text{and}
	\\
	\matfun{M}{\Btheta} 
	&= \partialderivativetwice{}{\Btheta}L\big(\Btheta, \feps \big) \\
	&=
	\sum_{x\in\X}\sum_{y\in\Y} 
		\Bigg[
			\I{\residual{\fepsxy}{\Btheta}-1\in[-1,c]} \fepsxy\sxy{\Btheta}\sxy{\Btheta}^\top
			-
			\mat{Q}_{xy}(\Btheta)
		\Bigg]
\end{align*}
are $d\times d$ symmetric and invertible matrices. 
\end{theorem}

The regularity  conditions are presented and discussed in \citet{welz2024robcat}. We stress again that no assumption is made that would restrict the degree or type of potential misspecification of the polychoric model.

It can be shown that in the absence of misspecification (such that $\Btheta_0 = \Btheta_*$), the asymptotic covariance matrix~$\BSigma{\Btheta_*}$ is equal to the inverted Fisher information matrix of the polychoric model \citep[Lemma~B.5 in][]{welz2024robcat}, which is well-known to be the asymptotic covariance matrix of the MLE. It follows that under correct specification of the polychoric model, the %generalized ML
proposed robust estimator is indeed asymptotically first and second order equivalent to the consistent and efficient MLE of \citet{olsson1979poly}.

It can furthermore be shown that Theorem~\ref{thm:main} generalizes well-known results on the behavior of ML estimation in misspecified models. Indeed, for the MLE ($c=+\infty$ in the discrepancy function), the sandwich-type asymptotic covariance matrix~$\BSigma{\Btheta_0}$ reduces to that of \citet[][Theorem~3.2]{white1982} and \citet[][Corollary on p.~231]{huber1967}. However, our asymptotic results are more general because they encompass a broad class of estimators, including and beyond ML.

A consistent estimator of the unobserved asymptotic covariance matrix~$\BSigma{\Btheta_0}$ can be constructed as follows. Replace all population class probabilities~$\fepsxy$ by their corresponding empirical counterparts~$\fhatxy$ in matrices~$\matfun{W}{\Btheta}, \matfun{M}{\Btheta}$, and~$\BOmega$, and denote the resulting observable matrices by $\hatmatfun{W}{\Btheta}, \hatmatfun{M}{\Btheta}$, and~$\hat{\BOmega}_N$, respectively. Each of these observable matrices is a (pointwise) consistent estimator of its corresponding population counterpart by the  continuous mapping theorem. Then apply the formulas in Theorem~\ref{thm:main} to construct
\[
	\hatmatfun{\Sigma}{\Btheta} = \hatmatinv{M}{\Btheta} \hatmatfun{U}{\Btheta} \hatmatinv{M}{\Btheta},
\]
where $\hatmatfun{U}{\Btheta} = \hatmatfun{W}{\Btheta} \hat{\BOmega}_N \hatmatfun{W}{\Btheta}^\top$. The observable matrix $\hatmatfun{\Sigma}{\Btheta}$ is a pointwise consistent estimator of~$\BSigma{\Btheta}$ for a given parameter value~$\Btheta\in\BTheta$. It now follows from the fact that $\thetahat\convP\Btheta_0$ (Theorem~\ref{thm:main}) and the continuous mapping theorem that~$\hatmatfun{\Sigma}{\thetahat}$ is consistent for the population asymptotic covariance matrix~$\BSigma{\Btheta_0}$.

For the choice $c=0$, note that the condition $\residual{\fepsxy}{\Btheta_0}-1\neq c$ in Theorem~\ref{thm:main} rules out the zero-misspecification case ($\varepsilon = 0$). In fact, in this case, the condition is not met for any $(x,y)\in\X\times\Y$. Indeed, \citet{welz2024robcat} shows that for $c=0$, the estimator is \emph{not} asymptotically Gaussian if the model is correctly specified. We therefore recommend to choose $c>0$ (see Section~\ref{sec:implementation} for a discussion). The reason why we require $\residual{\fepsxy}{\Btheta_0}-1\neq c$ for all $(x,y)\in\X\times\Y$ is because the value~$c$ is the threshold at which the discrepancy function~\eqref{eq:phifun} transitions from superlinear growth to linear growth. The discrepancy function is not twice differentiable at this point, which causes a particular Taylor expansion in the theorem's proof to fail to exist. This Taylor expansion plays a crucial role in establishing asymptotic normality. If it cannot be performed for at least one response $(x,y)\in\X\times\Y$, the estimator is not asymptotically normal. We refer to \citet{welz2024robcat} for a detailed discussion.

\subsection{Expressions of first and second order derivatives}\label{app:derivatives}
This section presents closed-form expressions of all components of the gradient and Hessian matrix of the probability mass function of the polychoric model,~$\pxy{\Btheta}$.

\subsubsection{First order terms}
For response~$(x,y)\in\X\times\Y$ and $\Btheta\in\BTheta$, the gradient of $\pxy{\Btheta}$ can be expressed as
\begin{equation}\label{eq:pxygradientapp}
	\partialderivative{\pxy{\Btheta}}{\Btheta}
	=
	\partialderivative{}{\Btheta}\CDF{a_x,b_y}{\rho} 
	- \partialderivative{}{\Btheta}\CDF{a_{x-1},b_{y}}{\rho}
	- \partialderivative{}{\Btheta}\CDF{a_{x},b_{y-1}}{\rho}
	+ \partialderivative{}{\Btheta}\CDF{a_{x-1},b_{y-1}}{\rho},
\end{equation}
see, e.g., \citet[][Equation~4]{olsson1979poly}. To characterize this gradient, we provide expressions for individual partial derivatives of~$\pxy{\Btheta}$, that is,
\[
	\partialderivative{\pxy{\Btheta}}{\Btheta}
	=
	\left(
		\partialderivative{\pxy{\Btheta}}{\rho},
		\partialderivative{\pxy{\Btheta}}{a_1},
		\dots,
		\partialderivative{\pxy{\Btheta}}{a_{K_X-1}},
		\partialderivative{\pxy{\Btheta}}{b_{1}},
		\dots,
		\partialderivative{\pxy{\Btheta}}{b_{K_Y-1}}
	\right)^\top.
\]
First, for any $u,v\in\R$, it can be shown \citep[e.g.,][]{drezner1990} that
\[
	\partialderivative{}{\rho}\CDF{u,v}{\rho} = \PDF{u,v}{\rho},
\]
as well as \citep[e.g.,][]{tallis1962}
\[
	\partialderivative{}{u}\CDF{u,v}{\rho} = \PDFuni{u}\CDFuni{\frac{v-\rho u}{\sqrt{1-\rho^2}}},
\]
where $\PDFuni{\cdot}$ and $\CDFuni{\cdot}$ denote the density and distribution function, respectively, of the \emph{univariate} standard normal distribution. The complementary partial derivative with respect to~$v$ follows analogously by symmetry.

It now follows immediately from~\eqref{eq:pxygradientapp} that the partial derivative of~$\pxy{\Btheta}$ with respect to~$\rho$ is given by
\[
	\partialderivative{\pxy{\Btheta}}{\rho} 
	=
	\PDF{a_x,b_y}{\rho}
	-\PDF{a_{x-1},b_{y}}{\rho}
	-\PDF{a_{x},b_{y-1}}{\rho}
	+\PDF{a_{x-1},b_{y-1}}{\rho},
\]
whereas the partial derivatives with respect to the individual thresholds are characterized by
\[
	\partialderivative{\pxy{\Btheta}}{a_k}
	=
	\begin{dcases}
	\partialderivative{}{a_x}\CDF{a_x,b_y}{\rho} - \partialderivative{}{a_x}\CDF{a_x,b_{y-1}}{\rho} &\text{if } k=x,
	\\
	-\partialderivative{}{a_{x-1}}\CDF{a_{x-1},b_y}{\rho} + \partialderivative{}{a_{x-1}}\CDF{a_{x-1},b_{y-1}}{\rho} &\text{if } k=x-1,
	\\
	0
	&\text{otherwise},
	\end{dcases} 
\]
for $k=1,\dots,K_X-1$. An expression for $\partialderivative{\pxy{\Btheta}}{b_k}$ can be derived analogously.

\subsubsection{Second order terms}
Here we provide expressions for the individual coordinates of the~$d\times d$ symmetric Hessian matrix of~$\pxy{\Btheta}$, that is,
\[
\partialderivativetwice{\pxy{\Btheta}}{\Btheta}
=
\begin{pmatrix}	
\partialderivativetwicesame{\pxy{\Btheta}}{\rho} & \partialderivativetwiced{\pxy{\Btheta}}{\rho}{a_1} & \cdots & \partialderivativetwiced{\pxy{\Btheta}}{\rho}{a_{K_X-1}} & \partialderivativetwiced{\pxy{\Btheta}}{\rho}{b_{1}} & \cdots & \partialderivativetwiced{\pxy{\Btheta}}{\rho}{b_{K_Y-1}}
\\
\partialderivativetwiced{\pxy{\Btheta}}{a_1}{\rho} & \partialderivativetwicesame{\pxy{\Btheta}}{a_1} & \cdots & \partialderivativetwiced{\pxy{\Btheta}}{a_1}{a_{K_X-1}} & \partialderivativetwiced{\pxy{\Btheta}}{a_1}{b_{1}} & \cdots & \partialderivativetwiced{\pxy{\Btheta}}{a_1}{b_{K_Y-1}}
\\
\vdots & \vdots & \ddots & \vdots & \vdots & \ddots & \vdots
\\
\partialderivativetwiced{\pxy{\Btheta}}{a_{K_X-1}}{\rho} & \partialderivativetwiced{\pxy{\Btheta}}{a_{K_X-1}}{a_1} & \cdots & \partialderivativetwicesame{\pxy{\Btheta}}{a_{K_X-1}} &  \partialderivativetwiced{\pxy{\Btheta}}{a_{K_X-1}}{b_1} & \cdots & \partialderivativetwiced{\pxy{\Btheta}}{a_{K_X-1}}{b_{K_Y-1}}
\\
\partialderivativetwiced{\pxy{\Btheta}}{b_{1}}{\rho} & \partialderivativetwiced{\pxy{\Btheta}}{b_{1}}{a_1} & \cdots & \partialderivativetwiced{\pxy{\Btheta}}{b_1}{a_{K_X-1}} &  \partialderivativetwicesame{\pxy{\Btheta}}{b_1} & \cdots & \partialderivativetwiced{\pxy{\Btheta}}{b_1}{b_{K_Y-1}}
\\
\vdots & \vdots & \ddots & \vdots & \vdots & \ddots & \vdots
\\
\partialderivativetwiced{\pxy{\Btheta}}{b_{K_Y-1}}{\rho} & \partialderivativetwiced{\pxy{\Btheta}}{b_{K_Y-1}}{a_1} & \cdots & \partialderivativetwiced{\pxy{\Btheta}}{b_{K_Y-1}}{a_{K_X-1}} &  \partialderivativetwiced{\pxy{\Btheta}}{b_{K_Y-1}}{b_1} & \cdots & \partialderivativetwicesame{\pxy{\Btheta}}{b_{K_Y-1}}
\end{pmatrix}.
\]
This Hessian matrix can alternatively be expressed as follows, which follows by~\eqref{eq:pxygradientapp}:
\begin{equation}\label{eq:pxyhessianapp}
\begin{split}
	&\partialderivativetwice{\pxy{\Btheta}}{\Btheta}
	=
	\\
	&\quad\partialderivativetwice{}{\Btheta}\CDF{a_x,b_y}{\rho} 
	- \partialderivativetwice{}{\Btheta}\CDF{a_{x-1},b_{y}}{\rho}
	- \partialderivativetwice{}{\Btheta}\CDF{a_{x},b_{y-1}}{\rho}
	+ \partialderivativetwice{}{\Btheta}\CDF{a_{x-1},b_{y-1}}{\rho}.
\end{split}
\end{equation}

First, by means of repeated applications of the product rule and chain rule it can be shown that for any $u,v\in\R$,
\[
	\partialderivativetwicesame{}{\rho}\CDF{u,v}{\rho} = \partialderivative{}{\rho}\PDF{u,v}{\rho}
	=
	\frac{\PDF{u,v}{\rho}}{(1-\rho^2)^2} \Big( (1-\rho^2)(\rho + uv) - \rho\big(u^2-2\rho u v + v^2 \big)  \Big),
\]
as well as
\[
	\partialderivativetwicesame{}{u}\CDF{u,v}{\rho}
	=
	\phi_1'(u) \CDFuni{\frac{v-\rho u}{\sqrt{1-\rho^2}}} - \frac{\rho}{\sqrt{1-\rho^2}} \PDFuni{u}\PDFuni{\frac{v-\rho u}{\sqrt{1-\rho^2}}},
\]
where
\[
	\phi_1'(u) = - \frac{u}{\sqrt{2\pi}} \exp\left(-u^2/2 \right),
\]
which follows immediately by the chain rule.

Next, for the second order cross-derivatives, it can be shown that
\[
	\partialderivativetwiced{}{u}{\rho}\CDF{u,v}{\rho} = \PDFuni{u}\PDFuni{\frac{v-\rho u}{\sqrt{1-\rho^2}}} \frac{\rho v - u}{(1-\rho^2)^{3/2}}
\]
and
\[
	\partialderivativetwiced{}{u}{v}\CDF{u,v}{\rho}
	=
	\PDFuni{u}\PDFuni{\frac{v-\rho u}{\sqrt{1-\rho^2}}}
	\frac{1}{\sqrt{1-\rho^2}},
\]
both by applications of the chain rule and product rule.

It now follows by \eqref{eq:pxyhessianapp} combined with these second order cross-derivatives that
\[
	\partialderivativetwiced{\pxy{\Btheta}}{a_k}{\rho}
	=
	\begin{dcases}
	\partialderivativetwiced{}{a_x}{\rho}\CDF{a_x,b_y}{\rho} - \partialderivativetwiced{}{a_x}{\rho}\CDF{a_x,b_{y-1}}{\rho} &\text{if } k=x,
	\\
	-\partialderivativetwiced{}{a_{x-1}}{\rho}\CDF{a_{x-1},b_y}{\rho} + \partialderivativetwiced{}{a_{x-1}}{\rho}\CDF{a_{x-1},b_{y-1}}{\rho} &\text{if } k=x-1,
	\\
	0
	&\text{otherwise},
	\end{dcases}
\] 
and 
\[
	\partialderivativetwiced{\pxy{\Btheta}}{a_k}{b_l}
	=
	\begin{dcases}
	\partialderivativetwiced{}{a_k}{b_l}\CDF{a_x,b_y}{\rho} &\text{if } (k,l)\in\big\{ (x,y), (x-1,y-1) \big\},
	\\
	-\partialderivativetwiced{}{a_k}{b_l}\CDF{a_x,b_y}{\rho} &\text{if } (k,l)\in\big\{ (x-1,y), (x,y-1) \big\},
	\\
	0
	&\text{otherwise},
	\end{dcases}
\] 
and
\[
	\partialderivativetwiced{\pxy{\Btheta}}{a_k}{a_l}
	=
	\begin{dcases}
	\partialderivativetwicesame{\pxy{\Btheta}}{a_k} &\text{if } k=l,
	\\
	0 &\text{otherwise},
	\end{dcases}
\]
and
\[
	\partialderivativetwicesame{\pxy{\Btheta}}{a_k}
	=
	\begin{dcases}
	\partialderivativetwicesame{}{a_x}\CDF{a_x,b_y}{\rho} - \partialderivativetwicesame{}{a_x}\CDF{a_x,b_{y-1}}{\rho} &\text{if } k=x,
	\\
	-\partialderivativetwicesame{}{a_{x-1}}\CDF{a_{x-1},b_y}{\rho} + \partialderivativetwicesame{}{a_{x-1}}\CDF{a_{x-1},b_{y-1}}{\rho} &\text{if } k=x-1,
	\\
	0
	&\text{otherwise},
	\end{dcases}
\] 
and, finally,
\begin{align*}
	&\partialderivativetwicesame{\pxy{\Btheta}}{\rho}
	=
	\\
	&\quad\partialderivativetwicesame{}{\rho}\CDF{a_x,b_y}{\rho} 
	- \partialderivativetwicesame{}{\rho}\CDF{a_{x-1},b_{y}}{\rho}
	- \partialderivativetwicesame{}{\rho}\CDF{a_{x},b_{y-1}}{\rho}
	+ \partialderivativetwicesame{}{\rho}\CDF{a_{x-1},b_{y-1}}{\rho}.
\end{align*}

\newpage

\section{Remark on numerical optimization}
\label{sec:instability}

\setcounter{figure}{0} 
\setcounter{table}{0}   
\setcounter{equation}{0}  

In the simulation study on distributional misspecification from Section~\ref{sec:distributional-misspecification}, 
the robust estimator did not converge to a solution in~32 out of 5,000 repetitions. This issue occurred 27 times for the Clayton copula with correlation~0.9, and five times with correlation~0.3. The polychoric MLE failed to converge~56 times, all at the Clayton copula with correlation~0.9. 
For the robust estimator, this situation tends to occur in the rare event that single rows or columns in a contingency table contain only one non-zero cell. It turns out that for such data, the estimator may attempt to effectively eliminate a threshold by either pushing the outermost finite thresholds to $\pm\infty$ or pulling adjacent thresholds to be as close to each other as numerically feasible, thereby practically merging these thresholds into one. Such behavior often leads to degeneracy of the final Nelder-Mead simplex (if used). Thus, package \pkg{robcat} identifies numerical instability either by degeneracy of Nelder-Mead simplexes and/or adjacent thresholds being unreasonably far away from each other, and subsequently throws a warning. We decided to set the cutoff for two adjacent thresholds being unreasonably far away to a minimum distance of~3.92. Under the polychoric model, this distance can cover as much as~95\% of all probability mass of the corresponding standard normal marginal distribution. We believe that having adjacent threshold values separated by this much probability mass may be an indication of poor model fit and/or severe measurement issues. Hence, being able to detect such numerical instability may in fact be viewed as a useful feature of our robust estimator: it alerts the user to potential broader issues where the use of the model may not be recommended for the data at hand.

\newpage

\section{The tuning constant $c$}
\label{app:simulation-c}

\setcounter{figure}{0} 
\setcounter{table}{0}   
\setcounter{equation}{0}

\subsection{Background}

Our proposed discrepancy function~$\varphi(\cdot)$ in~\eqref{eq:phifun} depends on the choice of a tuning constant $c \geq 0$. By construction of the discrepancy function, the larger the choice of~$c$, the more similar the robust estimate will become to the ML estimate (which occurs for $c=+\infty$): as one gradually increases~$c$ from~0 to larger values, the corresponding estimate gradually approaches the ML estimate (see, e.g., Figures~\ref{fig:estimands} and~\ref{fig:application-neuroticism-vary}). This behavior begs the question which value of~$c$ one should choose. This section motivates and explains the reasoning behind the choice of $c=0.6$, which we use throughout this paper.

In theory, the closer~$c$ is chosen to~0, the more robust the estimator becomes against contamination because cells whose Pearson residuals exceed the ideal value~0 will be downweighted more stringently. On the other hand, if contamination is absent, Theorem~\ref{thm:main} reveals that any choice of~$c$ that is \emph{strictly} larger than~0 will result in an asymptotically fully efficient estimator. Therefore, from a purely theoretical perspective, one should choose a~$c$ that infinitesimally exceeds~0.

Needless to say, asymptotic theory and empirical practice are two very different animals: In practice, one only has access to a finite number of observations,~$N$, so a number of finite sample issues might arise that could affect practical recommendations for the choice of~$c$. Therefore, this section carries out simulation studies to explore the effects of different choices of~$c$.

\subsection{Simulation design}

To explore the effects of various choices of~$c$, we use the same data generating process as in the simulation design in Section~\ref{sec:sim-individual}, in which we are interested in the polychoric correlation between two ordinal variables~$(X, Y)$ with $K_X = K_Y = 5$ response categories. The true value of the polychoric correlation coefficient is set to $\rho_*\in\{0, 0.5\}$, which by construction corresponds to the population correlation between the latent $(\xi, \eta)$ under the polychoric model. The latent variables are then discretized according to the true thresholds
\[
	a_{*,1}=b_{*,1} = -1.5,\quad a_{*,2}=b_{*,2} = -0.5,\quad a_{*,3}=b_{*,3}=0.5,\quad a_{*,4}=b_{*,4}=1.5.
\]
To simulate contamination, we let a fraction~$\varepsilon\in\{0,0.1,0.2\}$ of the data be generated by a contamination distribution~$H$, which here is a bivariate normal distribution with population mean $(2.5, -2.5)^\top$, variances $(0.25, 0.25)^\top$, and zero correlation between the two latent variables. We then discretize the latent realizations from this contamination distribution according to the thresholds $(a_{*,j}, b_{*,j}, j=1,\dots,4)$. We sample $N=1,000$ ordinal observations $(X_i, Y_i)_{i=1}^N$ from this data generating process. We then use this sample to estimate the polychoric correlation between the two ordinal variables with our proposed robust estimator at various choices of tuning parameter~$c$. Specifically, the set of considered tuning parameters~$c$ is given by the granular grid $\{0, 0.1, 0.2, \dots, 14.8, 14.9, 15, +\infty\}$, where $c=+\infty$ is understood as the maximum likelihood estimator (MLE). We repeat this procedure for~$T=5,000$ simulation runs.

\subsection{Performance measures}
\label{app:performancemeasures}

Let $\rhohatt$ denote the point estimate of the $t$-th simulation repetition, $t=1,\dots,T$, where $T=5,000$. For performance evaluation, we compute the following three statistics.
\begin{enumerate}
	\item Sample mean of the 5,000 point estimates: 
	\[
		\rhohatmean = \frac{1}{T}\sum_{t=1}^T\hatN{\rho}^{(t)},
	\]
	being a performance measure for accuracy in the estimation of the true~$\rho_*$. When contamination is present ($\varepsilon > 0$), we expect that the further~$c$ is away from~0, the larger the estimation error becomes. Conversely, in the absence of contamination ($\varepsilon = 0$), the estimates should be accurate no matter the choice of~$c$.
	\item Sample mean of the 5,000 standard error estimates associated with each point estimate: 
	\[
		\SEbarhat{\rhohat} = \frac{1}{T}\sum_{t=1}^T \SEhat{\rhohatt},
	\]
	capturing dispersion estimation. Every individual standard error estimate~$\SEhat{\rhohatt}$ estimates the true standard error~$\SE{\rhohat}$, hence also~$\SEbarhat{\rhohat}$ estimates~$\SE{\rhohat}$. To approximate the estimand~$\SE{\rhohat}$ we calculate the standard deviation of the simulated distribution of point estimates as the third statistic. 
	\item Sample standard deviation of the 5,000 point estimates: 
	\[
		\SEapprox{\rhohat} = \sqrt{\frac{1}{T-1}\sum_{t=1}^T \left(\rhohatt - \rhohatmean \right)^2}.
	\]
	We call this statistic $\SEapprox{\rhohat}$ because it approximates the 
	finite-sample standard error~$\SE{\rhohat}$. Although an asymptotic expression for the standard error is known from Theorem~\ref{thm:main} (the square root of first diagonal element of $\BSigma{\Btheta_0}$), the true standard error~$\SE{\rhohat}$ in finite samples is unknown. We therefore use the standard deviation of the simulated sampling distribution as an approximation.
\end{enumerate}

Comparing~$\SEbarhat{\rhohat}$ with~$\SEapprox{\rhohat}$ can be seen as a sanity check for the correctness of the estimator's limit theory in Theorem~\ref{thm:main}. Hence, if the limit theory is correct, we expect for a sufficiently large sample size that~$\SEbarhat{\rhohat}$ will be close to~$\SEapprox{\rhohat}$. 
However, recall from Theorem~\ref{thm:main} that the estimator is \emph{not} asymptotically Gaussian for the choice $c=0$ when contamination is absent $(\varepsilon = 0)$. Thus, in this case ($c=\varepsilon = 0$), we expect~$\SEbarhat{\rhohat}$ to differ from~$\SEapprox{\rhohat}$ because the former is derived under the then-invalid asymptotic normality of estimator~$\thetahat$.

\subsection{Simulation results}

\begin{figure}[!t]
\centering
\includegraphics[width = \textwidth]{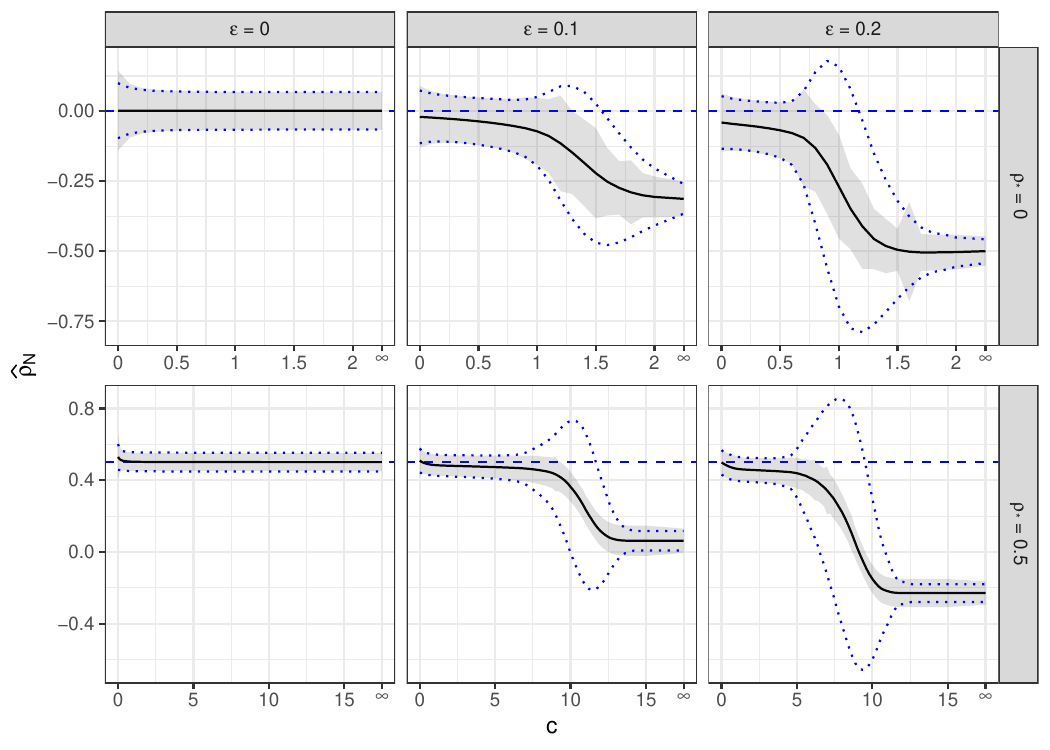}
\caption{Results across 5,000 simulated datasets for various choices of the tuning constant~$c$ ($x$-axis), contamination fractions~$\varepsilon$ (columns), and true polychoric correlation coefficients~$\rho_*$ (rows). Solid black lines indicate the average point estimate~$\rhohatmean$ and dashed blue lines the true~$\rho_*$. Shaded gray areas visualize confidence bands using the average standard error estimate $\SEbarhat{\rhohat}$, while dotted blue lines represent confidence bands using the approximate true standard error $\SEapprox{\rhohat}$.}
\label{fig:simresults-c-choice}
\end{figure}

Figure~\ref{fig:simresults-c-choice} visualizes the results as a function of the tuning constant~$c$ for the considered contamination fractions~$\varepsilon$ (columns) and true correlation coefficients~$\rho_*$ (rows). The average point estimate~$\rhohatmean$ is represented by a solid black line, while a dashed blue line indicates the true~$\rho_*$. In addition, the average standard error estimate $\SEbarhat{\rhohat}$ and the approximate true standard error $\SEapprox{\rhohat}$ are visualized by confidence bands: a shaded gray area covers $\rhohat \pm q_{1-\alpha/2} \cdot \SEbarhat{\rhohat}$ and dotted blue lines indicate $\rhohat \pm q_{1-\alpha/2} \cdot \SEapprox{\rhohat}$, where~$q_{1-\alpha/2}$ denotes the~$(1-\alpha/2)$~quantile of the standard normal distribution for $\alpha = 0.05$.

We first focus on the top row, which corresponds to the zero-correlation setting ($\rho_* = 0$). In the absence of contamination ($\varepsilon = 0$), the point estimates are accurate for all choices of~$c$. For strictly positive~$c$, the standard error estimates are also accurate in the sense that they closely follow the approximate true standard errors. On the other hand, if $c=0$, the standard error estimate deviates from the approximate true standard error. This is expected, because we estimate the standard errors based on Theorem~\ref{thm:main}, which explicitly excludes $c=0$ since the estimator~$\thetahat$ in that case is \emph{not} asymptotically normal in the absence of contamination (see Appendix~\ref{sec:asymptotics}).
Introducing contamination $(\varepsilon > 0)$, the point estimates stay relatively stable and accurate for~$c$ up to~1 (for $\varepsilon = 0.1$) or~0.7 (for $\varepsilon = 0.2$). For choices of~$c$ beyond these values, the estimates deteriorate fairly quickly and approach the ML estimates ($c = \infty$), which are strongly biased for either contamination fraction (absolute biases of~0.3 and~0.5, respectively). 
Furthermore, the standard error estimates are accurate for the same range of the tuning constant~$c$ where the point estimates are stable and close to the true value. Once~$c$ is large enough so that the point estimates abruptly move away from the true value, also the standard error estimates deteriorate and only stabilize again when the point estimates get close to the ML estimate with sufficiently large~$c$. It follows that if the choice of~$c$ is suitable for the point estimate, it also seems to be suitable for the standard error estimate. For such a suitable choice of~$c$, we further observe that the standard errors of robust estimates (that is, sufficiently small values of $c$) tend to be larger than those of the ML estimate when contamination is present.

We now shift our attention to the bottom row in Figure~\ref{fig:simresults-c-choice}, corresponding to a true polychoric correlation coefficient of $\rho_* = 0.5$. For nonzero contamination ($\varepsilon = 0.1$ or $\varepsilon = 0.2$), the point estimates and standard error estimates stay fairly stable and accurate for choices of~$c$ up to about~8 and~5, respectively. For values beyond that, the point estimates again abruptly drop---and standard error estimates deteriorate---before they stabilize again around the strongly biased ML estimates. 
For the zero-contamination case $(\varepsilon = 0)$, the standard error estimates are accurate for all choices of~$c$, and the same applies to the point estimates except for $c$ very close to 0. Indeed, for $c=0$, there seems to be a small upward bias. This might be surprising because the estimator~$\thetahat$ is Fisher consistent for~$\Btheta_*$ in the zero-contamination case for all $c\geq 0$ (see Section~\ref{sec:properties}). It turns out that this empirical bias is a finite-sample issue that vanishes in larger sample sizes. Specifically, it is due to a finite-sample approximation error of the nonparametric estimator~$\fhat$ of the population probability mass function~$f_0(x,y) = \pxy{\Btheta_*}$, as we explain in the following.

Recall from Section~\ref{sec:estimator} that the empirical Pearson residual at the true parameter value~$\Btheta_*$ of a cell $(x,y)\in\X\times\Y$ is defined as
\[
	\residual{\fhatxy}{\Btheta_*} -1.
\]
In the absence of contamination, $\fhatxy\convP f_0(x,y) = \pxy{\Btheta_*}$ as $N\to\infty$, so all Pearson residuals are asymptotically zero whenever $\varepsilon = 0$. Consequently, no cells are downweighted by the discrepancy function~$\varphi$ in~\eqref{eq:phifun}, no matter the choice of $c\geq 0$. However, note that if the true~$\rho_*$ is moderate to large in magnitude, some of the~$\pxy{\Btheta_*}$ may be extremely small. Then, in finite sample sizes~$N$, a nonzero~$\fhatxy$ due to natural variation may be relatively large in comparison such that
\[
	\residual{\fhatxy}{\Btheta_*} -1 > c = 0,
\]
meaning that the corresponding cell is downweighted by the discrepancy function in~\eqref{eq:phifun} despite contamination being absent. It follows from this incorrect downweighting that the ensuing estimate~$\thetahat$ may suffer from a finite-sample bias for the true~$\Btheta_*$ that is due to the finite-sample approximation error of~$\fhat$. Similar reasoning applies for positive choices of~$c$ that are close to~0. To avoid this issue, it may be preferable to choose a~$c$ that is large enough to give the Pearson residual sufficient leeway to avoid incorrect downweighting from finite-sample approximation errors when contamination is absent, while simultaneously being sufficiently small to retain robustness against contamination if present.

We stress that finite-sample approximation errors in the absence of contamination tend to be negligibly small when the true correlation is zero, i.e., $\rho_* = 0$ (see the top row of Figure~\ref{fig:simresults-c-choice}). Consequently, in sufficiently sized samples, we believe the risk of a false positive (significantly nonzero correlation estimate when the true correlation is zero) is small for~$c$ sufficiently away from~0.

\subsection{Discussion and implications}

The simulation results suggest that for a broad range of sufficiently small values of~$c$, point estimates are relatively constant and robust to contamination, and the respective standard error estimates are accurate. However, if the true correlation is nonzero and contamination is absent, values of~$c$ very close to 0 may result in bias due to finite-sample approximation error. Hence, in practice, one should choose~$c$ sufficiently small to retain robustness and stability, but sufficiently large to avoid finite-sample bias in certain situations. It turns out that the choice $c=0.6$ seems to be a reasonable compromise. We therefore use this choice in all other simulations in this paper. Nevertheless, we stress that clear practical recommendations for the choice of~$c$, preferably grounded in statistical theory, are an important avenue for future research.

\newpage

\section{Additional results of simulations from the main text}
\label{app:maintext-simulations-moreresults}

\setcounter{figure}{0} 
\setcounter{table}{0}   
\setcounter{equation}{0}  

Sections~\ref{app:sim-individual} and~\ref{app:polycormatrix} present additional results for the simulations from Sections~\ref{sec:sim-individual} and~\ref{sec:polycormatrix}, respectively.

\subsection{Individual polychoric correlation coefficient}
\label{app:sim-individual}

\begin{figure}[!b]
\centering
\includegraphics[width = \textwidth]{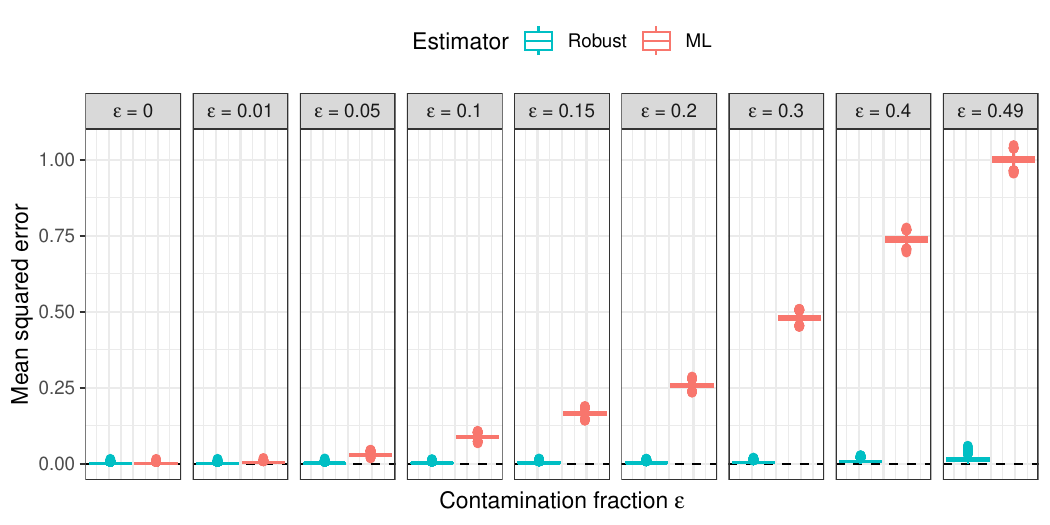}
\caption{Boxplot visualization of the mean squared error of the estimated parameter vector~$\thetahat$ with respect to the true~$\Btheta_*$, for various contamination fractions in the misspecified polychoric model across 5,000 repetitions. The estimators are the proposed robust estimator with $c=0.6$ (left) and the MLE (right).}
\label{fig:polycorsimtheta}
\end{figure}

Figure~\ref{fig:polycorsimtheta} visualizes the mean squared error of the estimated parameter vector in the polychoric model. We do not include the Pearson sample correlation coefficient because it does not estimate threshold parameters. Clearly, the robust estimator remains accurate across all contamination fractions, whereas the MLE becomes increasingly biased.

\begin{figure}
\centering
\includegraphics[width = \textwidth]{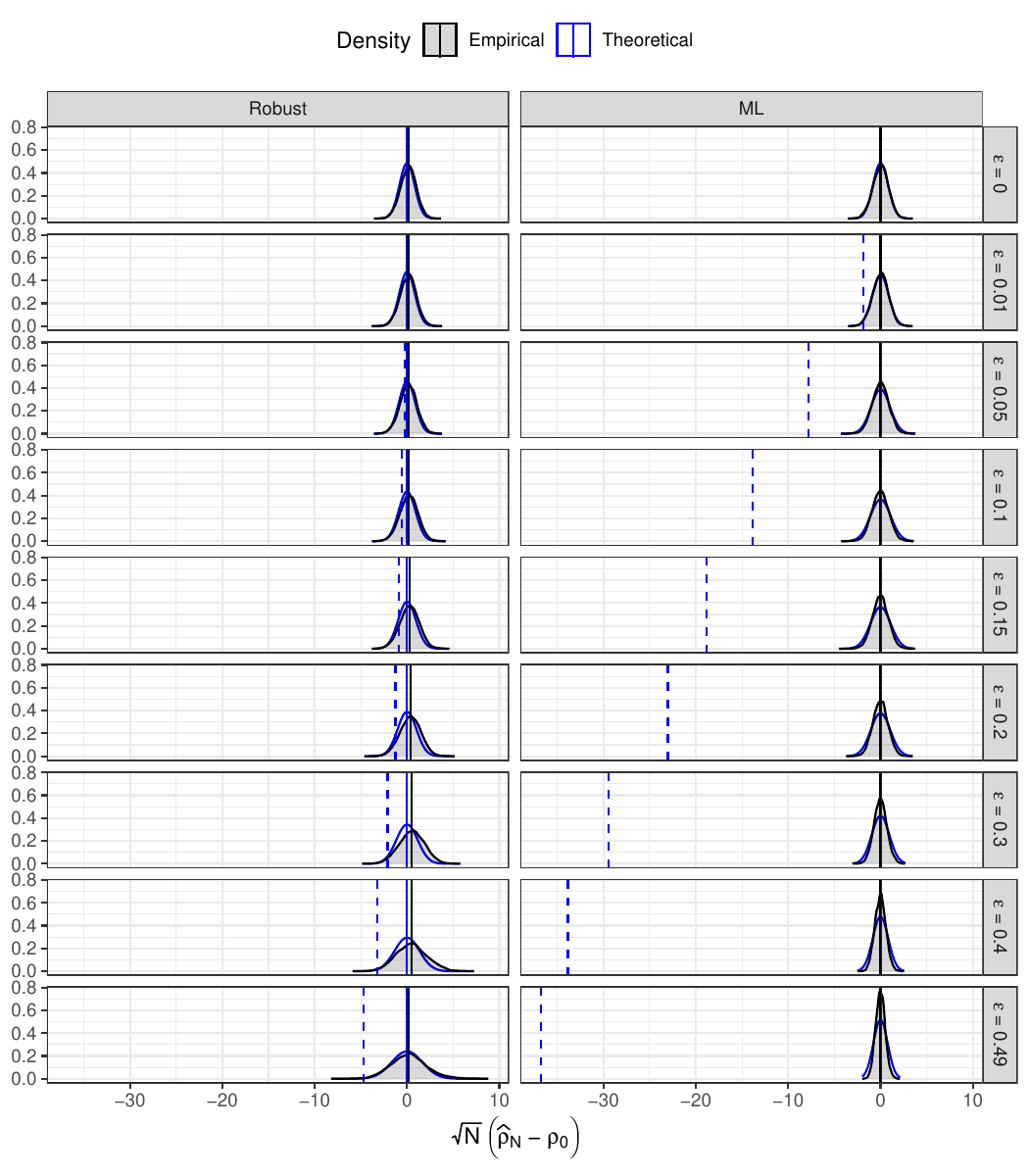}
\caption{Visualization of the empirical density across 5,000 simulated datasets and the theoretical density of $\sqrt{N}\left(\rhohat - \rho_0\right)$ for the robust estimator (left) and the MLE (right), for various contamination fractions (rows). The solid vertical lines correspond to the mean of a density. The blue dashed vertical lines correspond to the (scaled) population bias with respect to the true parameter, $\sqrt{N}\left(\rho_0 - \rho_*\right)$.}
\label{fig:polycorsimdensity}
\end{figure}

In order to verify the correctness of the robust estimator's asymptotic behavior established in Theorem~\ref{thm:main}, we further compare the correlation estimator's theoretical density with its empirical density in the simulation. Specifically, we compare these two densities for the bias with respect to the estimand, scaled by the square root of the sample size, that is, $\sqrt{N}\left(\rhohat - \rho_0\right)$. Theorem~\ref{thm:main} tells us that the asymptotic distribution of this term is given by $\gauss\big(0, \SE{\rho_0}^2)$, where $\SE{\rho_0}$ is square root of first diagonal element of the asymptotic covariance matrix~$\BSigma{\Btheta_0}$, i.e., the asymptotic standard error of the correlation estimator~$\rhohat$. The density of this distribution constitutes the theoretical density. For the empirical density, we apply a kernel density estimator to the estimates $\sqrt{N}\left(\hatN{\rho}^{(t)} - \rho_0\right)$, $t = 1,\dots,T$, from the $T=5,000$ simulation repetitions. If the asymptotic theory of Theorem~\ref{thm:main} is correct, the empirical and theoretical densities should be close to each other, save for some finite-sample approximation error. We repeat this procedure for the MLE. However, in the presence of contamination, the two densities are expected to be quite different from each other because the MLE's theoretical density (based on the inverse Fisher information matrix) is derived under the then-violated assumption of latent normality.

Figure~\ref{fig:polycorsimdensity} visualizes the empirical and theoretical densities of the robust estimator and the MLE (columns) for the considered contamination fractions (rows). To get a sense of bias with respect to the true correlation~$\rho_* = 0.5$, the black dotted vertical line is the (appropriately scaled) difference between estimand~$\rho_0$ and true value, that is, $\sqrt{N}(\rho_0 - \rho_*)$. The figure reveals that the for the robust estimator, the empirical and theoretical density are indeed (relatively) close throughout all contamination fractions. In contrast, the MLE's empirical density increasingly differs from the theoretical density as the contamination fraction increases, with the former having larger and larger kurtosis. In addition, the MLE's bias with respect to the true parameter is much larger than that of the robust estimator. We conclude that this exercise confirms the correctness of our asymptotic theory and further demonstrates the enhanced robustness of our proposed estimator.

\subsection{Polychoric correlation matrix}
\label{app:polycormatrix}

\begin{figure}
\centering
\includegraphics[width = \textwidth]{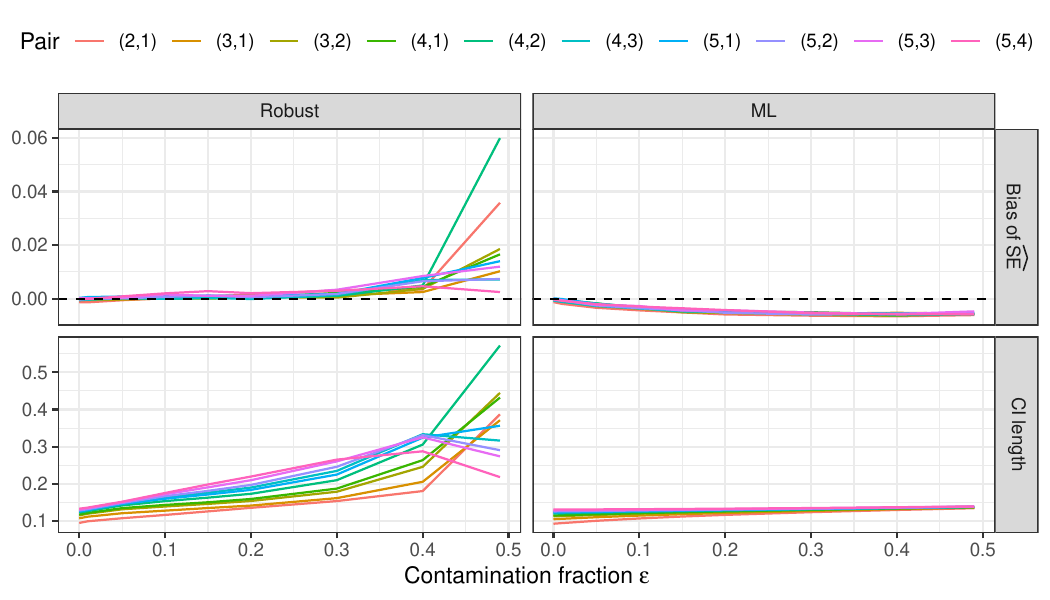}
\caption{Average bias of the standard error estimates (top) and average length of 95\% confidence intervals (bottom) of the robust estimator with~$c=0.6$ (left) and the MLE (right) for each unique pairwise polychoric correlation coefficient (see Table~\ref{tab:cormat-sim}), expressed as a function of the contamination fraction~$\varepsilon$ ($x$-axis). Averages are taken over 5,000 repetitions.}
\label{fig:cormatsim-additional}
\end{figure}

Figure~\ref{fig:cormatsim-additional} visualizes additional performance measures regarding inference, namely the (approximate) average bias of the standard error estimates, $\SEbarhat{\rhohat} - \SEapprox{\rhohat}$, defined in Appendix~\ref{app:performancemeasures}, and the average length of 95\% confidence intervals. We observe that the standard error estimates of the robust estimator are accurate except for extremely large contamination fractions ($\epsilon \geq 0.4$), whereas those of the MLE are underestimated in the presence of contamination. Furthermore, the confidence intervals of the robust estimator tend to be wider when contamination is present compared to those of the MLE.

\newpage

\section{Additional simulations with distributional overlap}
\label{app:additional-sims}

\setcounter{figure}{0} 
\setcounter{table}{0}   
\setcounter{equation}{0}  

This appendix contains two additional simulation designs. The first %one 
generalizes the design from Section~\ref{sec:sim-individual} by considering different mean shifts in the contamination distribution. The second uses contamination that manifests through correlation shifts while keeping the same mean as the model distribution. Before introducing specifics, we first discuss the role of distributional overlap, as it plays a prominent role in both designs.

\subsection{The role of distributional overlap}
\label{app:overlap-discussion}

Both simulation designs explore, among other things, the behavior of the robust estimator when the contamination distribution~$H$ substantially overlaps with the latent model distribution~$\CDF{\cdot,\cdot}{\rho_*}$ in the contaminated distribution~$G_\varepsilon$ from~\eqref{eq:contaminationmodel}. In the first design, we achieve such distributional overlap by moving the mean of~$H$ towards the origin, while in the second design, overlap is induced by letting~$H$ be a bivariate standard normal as well, but with a flipped correlation coefficient compared to the latent model distribution, i.e., $H = \CDF{\cdot,\cdot}{-\rho_*}$. It follows that contamination in the first design manifests as mean shifts in~$H$, while in the second design it manifests as a correlation shift in~$H$.

In general and beyond polychoric correlation, substantial overlap between the the model distribution and the contamination distribution in the contamination model of \citet{huber1964}, a particular identification issue may arise. For ease of exposition, even though similar arguments hold for estimating polychoric correlation, we explain this identification issue within the broader context of covariance matrix estimation. In this setting, said  identification issue affects any affine equivariant robust estimator of the covariance matrix. Consider first a contamination fraction $\varepsilon = 0.5$ and a contamination distribution~$H$ so that we obtain two perfectly separated point clouds of exactly the same size. An affine equivariant robust   estimator would have no way of knowing whether it should return the covariance parameters of the first group or the second group \citep[cf.~Section~6.17.1 in][]{maronna2018}, which is the reason why the contamination fraction~$\varepsilon$ is typically restricted to the interval $[0, 0.5)$ in the Huber contamination model. If the point clouds are no longer separated but overlap, fewer points from the groups are separated such that the identification issue already occurs for contamination fractions~$\varepsilon$ of less than~0.5. This is because an affine equivariant covariance matrix estimator may successfully identify some points as outliers and downweight them, but it would have no way of identifying whether observations from the overlapping part come from the model distribution of interest or the contamination distribution, causing bias in the estimates. \citet{riani2014} have formalized and studied this issue for various robust estimators in a regression setting. 

Importantly, this is a general identification issue for all robust estimators with the relevant equivariance/invariance properties. Although such properties may not be meaningful in case of ordinal data, which need not admit a numeric interpretation, we expect this issue to also arise for our robust estimator of the polychoric correlation model due to the use of a latent continuous space. This issue could only be overcome by placing additional assumptions on the contamination distribution~$H$ in~\eqref{eq:contaminationmodel}. However, since such assumptions may also be violated in practice, we believe that it is undesirable to introduce new assumptions beyond those needed for ML estimation of the polychoric model, and we refrain from doing so.

\subsection{Robustness against mean shifts}
\label{app:meanshift}

In the simulations from Section~\ref{sec:sim-individual}, the true polychoric correlation coefficient is given by \mbox{$\rho_*=0.5$} and the contamination distribution~$H$ in the contaminated distribution~$G_\varepsilon$ from~\eqref{eq:contaminationmodel} is set to
\begin{equation*}
	H = \gauss_2 \left(\vec{\mu}_H, 
		\begin{pmatrix}
			0.25 & 0 \\ 0 & 0.25
		\end{pmatrix}			
	 \right)
\end{equation*}
with mean vector $\vec{\mu}_H = (2.5, -2.5)^\top$. Compared to the polychoric model where the latent distribution is standard bivariate normal with correlation~$\rho_*$, this contamination distribution is mean-shifted and has a different covariance structure.

\subsubsection{Simulation design}

\begin{figure}
\centering
\includegraphics[width = 0.95\textwidth]{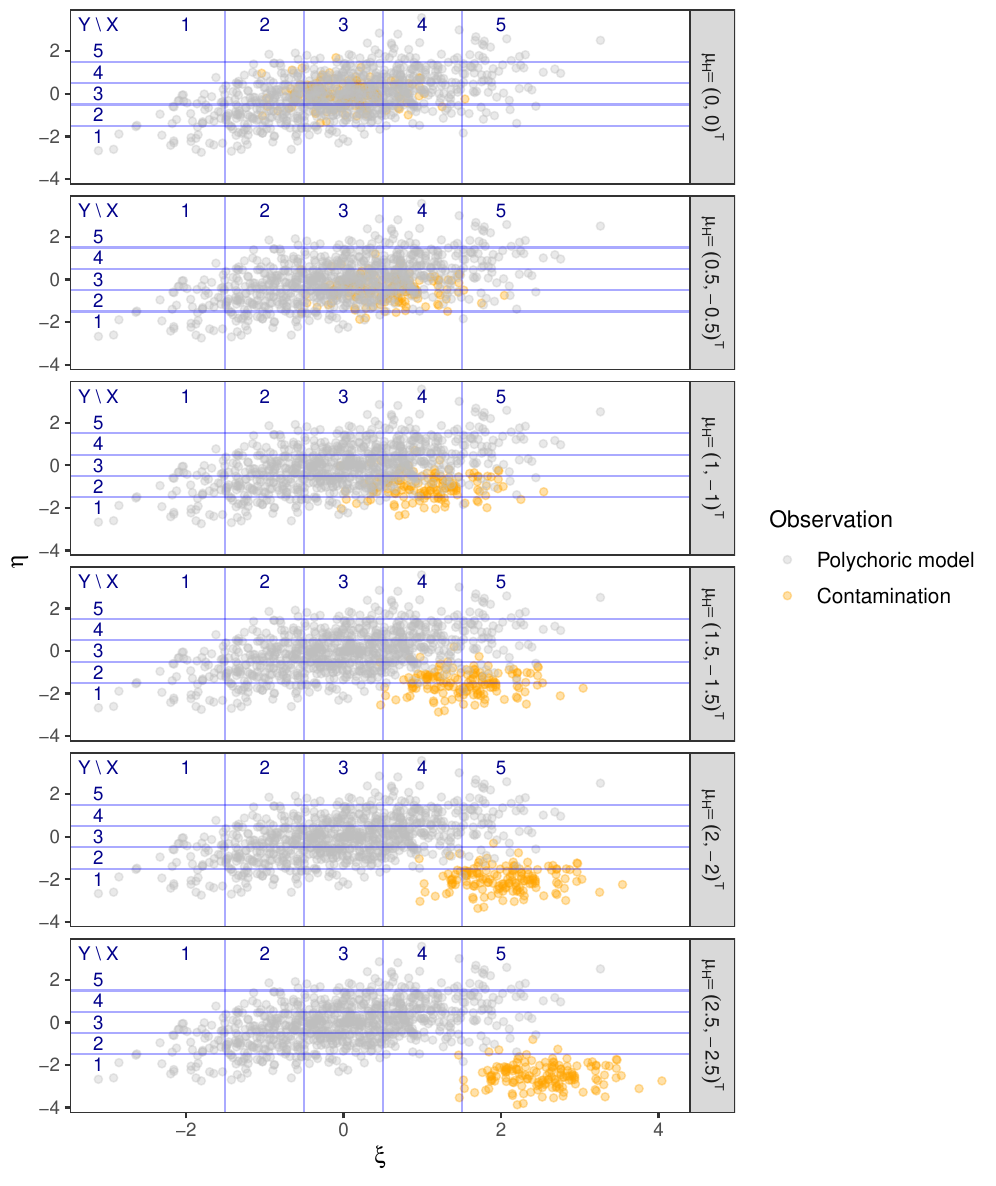}
\caption{Simulated example dataset with different mean vectors~$\vec{\mu}_H$ of the contamination distribution (rows). The gray dots represent random draws of~$(\xi, \eta)$ from the polychoric model with $\rho_* = 0.5$, whereas orange dots represent draws from the contamination distribution with mean~$\vec{\mu}_H$, variances $(0.25, 0.25)^\top$, and zero correlation. The contamination fraction is $\varepsilon = 0.15$ here, and we simulate $N=1,000$ data points. The blue lines indicate the locations of the thresholds.}
\label{fig:simdesign-careless-appendix}
\end{figure}

Here, we extend the simulation design from Section~\ref{sec:sim-individual} with different mean shifts. That is, we set all parameters as in Section~\ref{sec:sim-individual}, but
we consider the mean vectors 
\[
	\vec{\mu}_H
	\in 
	\Bigg\{
	\begin{pmatrix} 0 \\ 0\end{pmatrix},
	\begin{pmatrix} 0.5 \\ -0.5\end{pmatrix}, 
	\begin{pmatrix} 1 \\ -1\end{pmatrix}, 
	\begin{pmatrix} 1.5 \\ -1.5\end{pmatrix}, 
	\begin{pmatrix} 2 \\ -2\end{pmatrix}, 
	\begin{pmatrix} 2.5 \\ -2.5\end{pmatrix}
	\Bigg\}.
\]
Figure~\ref{fig:simdesign-careless-appendix} visualizes an example dataset, which illustrates that the considered mean shifts result in negative leverage points with increasing leverage. For instance, the choice $\vec{\mu}_H = (0, 0)^\top$ primarily inflates the center cell $(x,y) = (3,3)$ after discretization, $\vec{\mu}_H = (1, -1)^\top$ the cell $(x,y) = (4,2)$, and $\vec{\mu}_H = (2.5, -2.5)^\top$ the cell $(x,y) = (5,1)$. As such, these choices of mean vectors gradually move the contaminated data points (orange point clouds in Figure~\ref{fig:simdesign-careless-appendix} away from the origin. We expect that the stronger the leverage, the more bias the non-robust estimators will incur \citep[cf.][]{welz2024maxbias}. Hence, the added value of our proposed estimator should gradually increase as~$\vec{\mu}_H$ moves away from the origin.

Intuitively, one may think of the considered mean-shifted contamination as careless respondents who consider only certain response patterns irrespective of the item content. Some may be straighliners after recoding of negatively keyed items: an inflated cell $(5,1)$ corresponds to recoded responses of straightliners who exclusively chose the 1st or 5th response category,  $(4,2)$ to straightlining at the 2nd or 4th category, and $(3,3)$ to straightlining at the central response category.

For evaluation, we compute the same performance measures as in Section~\ref{sec:sim-individual}. In addition, we compute the average empirical loss function at the point estimate,~$L\left(\thetahat,\fhat \right)$. With increasing contamination fraction (i.e., the polychoric model becoming more and more misspecified), we expect the empirical loss to gradually increase (indicating deteriorating fit of the polychoric model).

\subsubsection{Results}

\begin{figure}
\centering
\includegraphics[width = \textwidth]{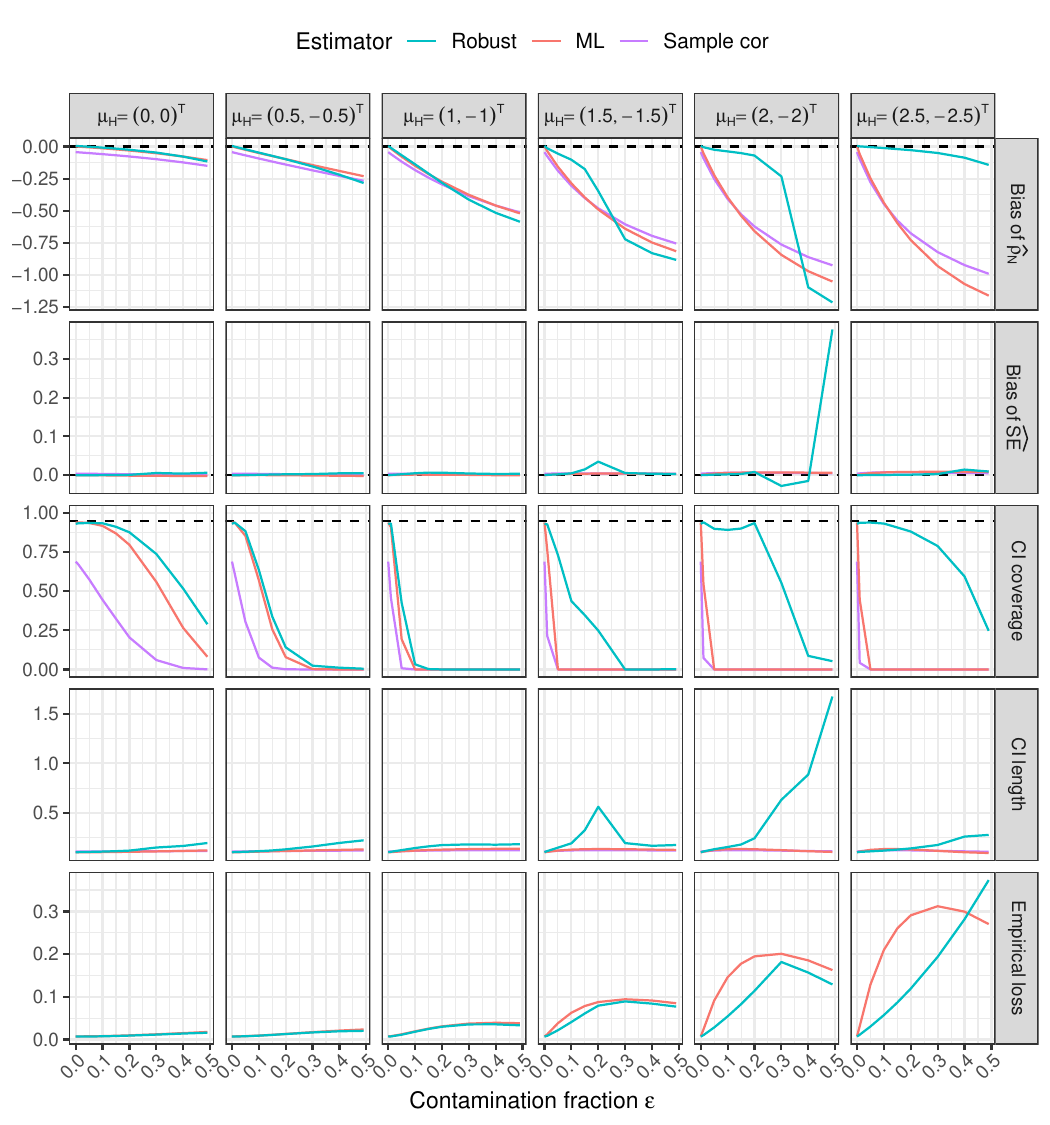}
\caption{Average bias of the point estimates and standard error estimates, coverage and average length of 95\% confidence intervals, and average empirical loss (rows), for different mean vectors~$\vec{\mu}_H$ of the contamination distribution (columns), expressed as a function of the contamination fraction~$\varepsilon$ ($x$-axis). Results are aggregated over 5,000 repetitions. The estimators are the robust estimator with~$c=0.6$, the MLE, and the Pearson sample correlation. Note that we include the empirical loss only for estimators of the polychoric model.}
\label{fig:simresults-meanshift}
\end{figure}

Figure~\ref{fig:simresults-meanshift} visualizes the simulation results with different performance measures in the rows and different contamination mean vectors~$\vec{\mu}_H$ in the columns.
The results for $\vec{\mu}_H = (2.5, -2,5)^\top$ in the rightmost column have already been described in Section~\ref{sec:sim-individual} and demonstrate a substantial improvement of the proposed estimator over the existing methods, even at high contamination fractions.
Generally speaking, the robustness benefit of our estimator increases as the contamination mean~$\vec{\mu}_H$ moves away from the origin.

For small mean shifts (up to $\vec{\mu}_H  = (1, -1)^\top$), the latent data points from the polychoric model and the contamination completely or mostly overlap to form one coherent point cloud (see Figure~\ref{fig:simdesign-careless-appendix}). In these settings, the three estimators perform similarly: while standard errors are estimated accurately, bias increases with higher contamination fractions and a larger mean shift (due to higher leverage). Yet, at least for $\vec{\mu}_H  = (0, 0)^\top$, the robust estimator yields better confidence interval coverage close to 90\% or higher for contamination fraction~$\varepsilon \leq 0.2$. Looking at the empirical loss, we find that the loss of the robust estimator is nearly identical to that of the MLE for these three mean shifts, meaning that the Pearson residuals are rarely large enough to to fall within the linear part of the discrepancy function of the robust estimator. Since the latent data form one coherent point cloud, the polychoric model still fits (reasonably) well but with different parameter values. 

Then at $\vec{\mu}_H  = (1.5, -1.5)^\top$, the latent point clouds start to separate to some extent, and even though the robust estimator is still biased, it clearly improves upon the nonrobust estimators for contamination fractions $\varepsilon \leq 0.15$, where its  standard errors are also estimated accurately. At \mbox{$\vec{\mu}_H  = (2, -2)^\top$}, the robust estimator exhibits very little bias, accurate standard error estimates, and confidence interval coverage close to the nominal level for contamination fractions $\varepsilon \leq 0.2$. The nonrobust estimators, on the other hand, are severely influenced by the contamination even for small $\varepsilon$. At the high contamination fraction $\varepsilon = 0.3$, the bias of the robust estimator is quite pronounced, but much less than that of the nonrobust estimators. However, the robust estimator then sharply deteriorates such that for extreme contamination fractions $\varepsilon \geq 0.4$, its bias is even worse than that of the nonrobust estimators. To understand this phenomenon, we again look at model fit as indicated by the empirical loss. For the robust estimator, the model fit at first deteriorates as the contamination fraction increases, but it \emph{improves} again for extreme contamination fractions. What we observe here is the identification issue discussed in Section~\ref{app:overlap-discussion}. Although the latent data points from the polychoric model and the contamination are nearly separated, there is still some overlap between the point clouds (see Figure~\ref{fig:simdesign-careless-appendix}). Hence, for extreme contamination fractions, the robust estimator can achieve a better fit by modeling observations from the contamination and the overlapping part of the model distribution---while downweighting other observations that are generated by the true polychoric model. The same phenomenon, albeit less pronounced, occurs for the shifted mean $\vec{\mu}_H = (1.5, -1.5)^\top$.\footnote{Another consequence of this identification issue is that the robust estimator's estimated asymptotic covariance matrix~$\hatmatfun{\Sigma}{\thetahat}$ may not be invertible. In our simulation results with 5,000 repetitions, for mean vector~$\vec{\mu}_H = (2, -2)^\top$, this occurred 9 times at $\varepsilon = 0.4$ and once at $\varepsilon = 0.49$. For~$\vec{\mu}_H = (1.5, -1.5)^\top$, it occurred 13 times at $\varepsilon = 0.4$ and once at $\varepsilon = 0.49$. Since standard errors cannot be computed in such cases, they were excluded from the aggregation of the inference performance measures in Figure~\ref{fig:simresults-meanshift}.}
Nevertheless, we stress that such an issue is unlikely in practice, unless there are severe problems with data quality. Contamination fractions beyond~30\% are extreme, and it is questionable whether modeling such severely compromised data is meaningful to begin with.

In summary, the robust estimator requires at least partial separation between the model distribution and the contamination to offer an improvement over the nonrobust estimators. Yet, the larger the overlap between the distributions, the smaller the bias for all estimators due to the lower leverage of the contamination. Moreover, the robust estimator yields \mbox{substantial} gains in robustness where it matters most: in settings with high leverage, where the nonrobust estimators are severely influenced even by small amounts of contamination, whereas the robust estimator remains stable across realistic contamination fractions.

\subsection{Robustness against correlation shifts}
\label{app:overlap}

Another intuitive manifestation of contamination lies in the correlation structure of the latent variables $(\xi, \eta)$. Consider, for instance, a (perhaps poorly designed) negatively-worded item. Some participants may be attentive enough to catch the general context of the item, but they may miss crucial details such as a negation. When correlating this item to another, responses of such participants would therefore invert the correlation pattern of fully attentive respondents.
In general, contamination due to correlation shifts is difficult to detect and robustify against when the mean remains equal, as it usually induces substantial overlap with the model distribution (cf.~Section~\ref{app:overlap-discussion}).

\subsubsection{Simulation design}

We again follow the same basic simulation design as in Section~\ref{sec:sim-individual}, with the model distribution of interest being the bivariate standard normal distribution $\CDF{\cdot,\cdot}{\rho_*}$. But now the contamination distribution~$H$ is also a bivariate standard normal distribution with sign-flipped correlation, i.e., $H = \CDF{\cdot,\cdot}{-\rho_*}$. Note that the contamination distribution $H$ has the same zero mean as the model distribution. All other parameters are kept as in Section~\ref{sec:sim-individual}, except that we extend the values of the true correlation to $\rho_* \in\{0.5, 0.7, 0.9\}$.

\begin{figure}
\centering
\includegraphics[width = 0.95\textwidth]{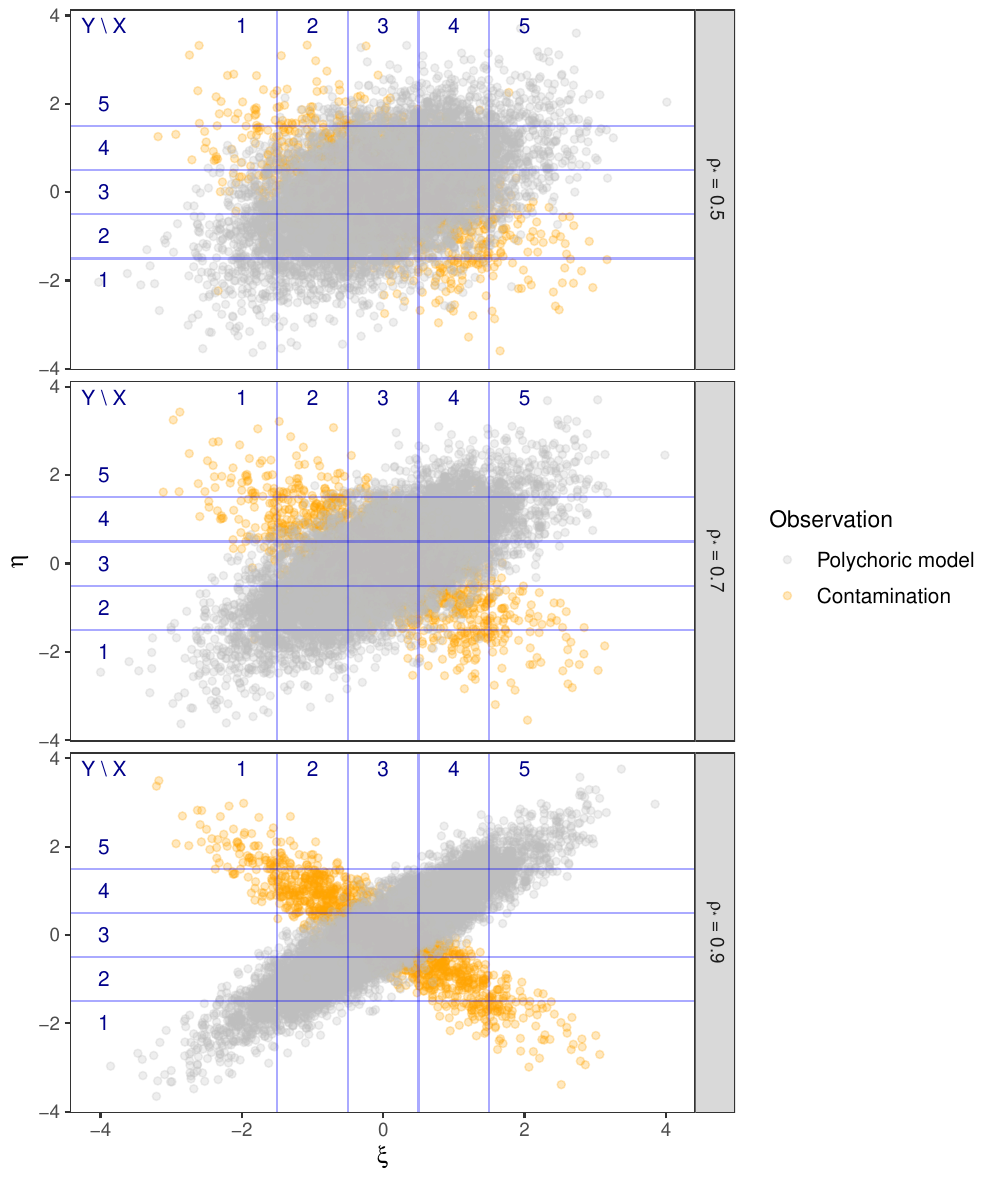}
\caption{Illustrative examples of the data generating process with different correlation coefficients~$\rho_*$ (rows). The gray dots represent random draws of~$(\xi, \eta)$ from the polychoric model with correlation $\rho_*$, whereas orange dots represent draws from the contamination distribution with correlation~$-\rho_*$. The contamination fraction is $\varepsilon = 0.15$ here, and we generate $N=10,000$ data points for a clearer visualization. The blue lines indicate the locations of the thresholds.}
\label{fig:simdesign-overlap}
\end{figure}

Figure~\ref{fig:simdesign-overlap} provides an illustrative visualization of the data generating process. While we generate $N=1,000$ observations in the simulations, the plot shows 10,000 latent data points for a clearer visualization. Clearly, the model distribution (gray points) and the contamination (orange points) largely overlap for $\rho_* = 0.5$, but they become more dissimilar as $\rho_* = 0.5$ increases except for some remaining overlap around the shared mean. We therefore expect the robust estimator to struggle to improve over the nonrobust estimators for $\rho_* = 0.5$, but it should yield robustness gains as $\rho_*$ increases.

\subsubsection{Results}

\begin{figure}
\centering
\includegraphics[width = \textwidth]{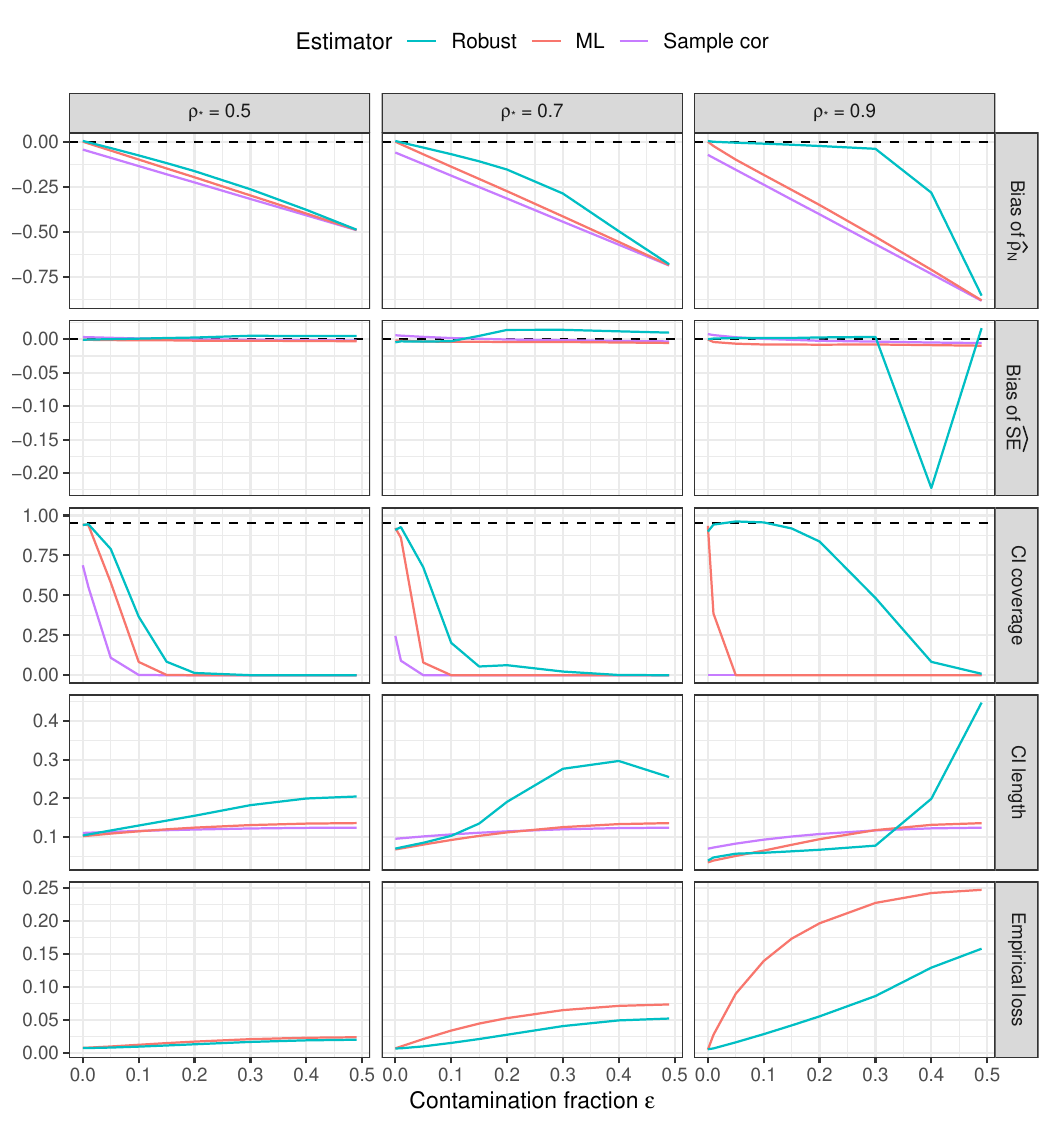}
\caption{Average bias of the point estimates and standard error estimates, coverage and average length of 95\% confidence intervals, and average empirical loss (rows), for different correlation coefficients~$\rho_*$ (columns), expressed as a function of the contamination fraction~$\varepsilon$ ($x$-axis). Results are aggregated over 5,000 repetitions. The estimators are the robust estimator with~$c=0.6$, the MLE, and the Pearson sample correlation. Note that we include the empirical loss only for estimators of the polychoric model.}
\label{fig:simresults-overlap-all}
\end{figure}

Figure~\ref{fig:simresults-overlap-all} visualizes the performance measures and the empirical loss in the rows and the different true correlation coefficients $\rho_*$ in the columns. At correlation $\rho_* = 0.5$, the robust estimator yields perhaps a minor improvement over the nonrobust estimators, at least in terms of confidence interval coverage, but all three estimators are very similar and become gradually more biased with increasing contamination fraction~$\varepsilon$. Looking at the empirical loss confirms that the loss of the robust estimator is only marginally smaller than that of the MLE, indicating only a minor downweighting of Pearson residuals.

Increasing the true correlation to $\rho_* = 0.7$, the robust estimator performs notably better than the nonrobust estimators. Although it still exhibits considerable and gradually increasing bias, this bias is roughly half that of the nonrobust estimators for contamination fractions $\varepsilon \leq 0.2$, and its standard error estimates are accurate for $\varepsilon \leq 0.15$.

For true correlation $\rho_* = 0.9$, the model distribution and the contamination distribution are sufficiently different (see Figure~\ref{fig:simdesign-overlap}) such that our robust estimator yields substantial robustness gains. While the nonrobust estimators gradually become more biased, the robust estimator displays minimal bias and accurate standard error estimates up to high contamination fractions of~\mbox{$\varepsilon \leq 0.3$}. Furthermore, confidence interval coverage remains near the nominal level of~95\% for~$\varepsilon \leq 0.15$ and still close to~85\% for $\varepsilon = 0.2$.

Finally, it is no coincidence that for all three choices of the correlation coefficient~$\rho_*$, the three estimators arrive at a bias of about $-\rho_*$ at contamination level $\varepsilon = 0.49$, which indicates that a near-zero correlation is estimated. 
Indeed, we have two groups of observations of about equal size that are symmetric to each other about the origin, one with correlation~$\rho_*$ and one with correlation~$-\rho_*$. It is not surprising------due to the overlap between the groups even for the robust estimator---that the best fit can be attained at a correlation estimate around~0.

To summarize, we observe again that if the model distribution and the contamination distribution have large overlap, it is difficult for the robust estimator to identify the contamination (cf.~the discussion in Appendix~\ref{app:overlap-discussion}) and consequently, to improve upon the nonrobust estimators. Nevertheless, we observe a clear benefit of the robust estimator as correlation increases and overlap decreases, even for (relatively) high contamination fractions.

\subsubsection{Implications for the empirical application}
\label{app:sim-application}

It turns out that our findings for $\rho_* = 0.9$ are remarkably similar to that of the empirical application from Section~\ref{sec:application} with the item pair corresponding to the polar opposite adjectives ``not envious'' and ``envious''. The main difference is that in the simulations, the true correlation coefficient is strongly positive (with the correlation coefficient of the contamination being sign-flipped), whereas we expect a strongly negative correlation between the aforementioned items in the empirical application.

In our simulation with true correlation $\rho_* = 0.9$ and contamination fraction~$\varepsilon = 0.15$, we obtain average correlation estimates of just below 0.9 for the robust estimator, just above 0.6 for the MLE, and just below 0.6 for the sample correlation (see Figure~\ref{fig:simresults-overlap-all}). In the empirical application, on the other hand, we obtain estimates of $-0.925$ with the robust estimator, $-0.618$ with the MLE, and $-0.562$ with the Pearson sample correlation (see Table~\ref{tab:application-neuroticism-results}). The robust estimator thereby assigns high Pearson residuals to the response cells $(x,y) \in \{(1,1), (1,2), (2,1), (2,2), (4,4), (4,5), (5,4), (5,5) \}$, and the empirical frequencies of these cells sum up to 11.4\% (see Table~\ref{tab:application-neuroticism-summary}). As discussed in Section~\ref{sec:application-results}, these cells likely correspond to careless respondents, as it is contradictory that both adjectives or neither adjective describe their personality accurately \citep[cf.][]{arias2020}. It seems plausible that these respondents simply overlooked the negation and responded to the item ``envious'' twice, in which case the correlation within this group is sign-flipped compared to the correlation within attentive respondents. Assuming that similar careless respondents also choose the middle response categories on both items, an overall prevalence of careless respondents around 15\% seems plausible, too.

In short, the findings from the empirical application seem to match both the simulation design and the corresponding results. Not only does this strengthen the validity of our findings in the empirical application, it also demonstrates that our simulation design with such a strong correlation is of practical relevance.

\newpage
\section{Additional results for the empirical application}
\label{app:application-moreresults}

\setcounter{figure}{0} 
\setcounter{table}{0}   
\setcounter{equation}{0}  

\begin{figure}
\centering
\includegraphics[width = 0.7\textwidth]{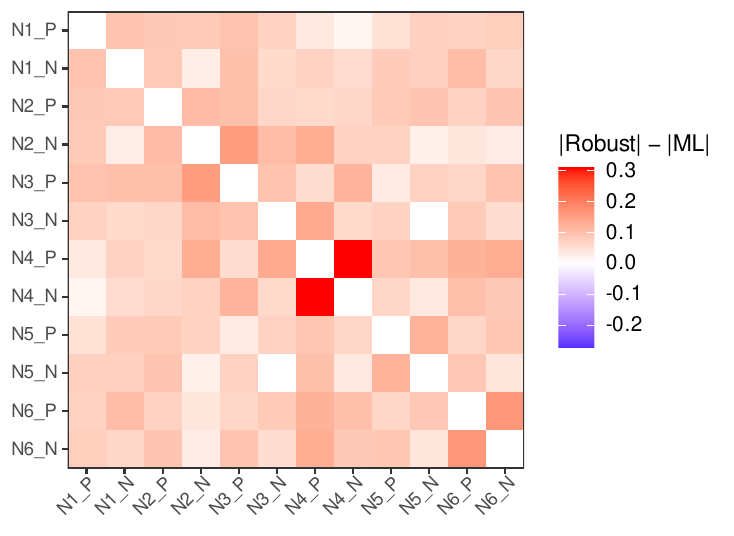}
\caption{Difference between absolute estimates for the polychoric correlation coefficient of %our
the robust estimator with $c=0.6$ and the MLE for each item pair in the \emph{neuroticism} scale, using the data of \citet{arias2020}. The items are ``calm'' (N1\_P), ``angry'' (N1\_N), ``relaxed'' (N2\_P), ``tense'' (N2\_N), ``at ease'' (N3\_P), ``nervous'' (N3\_N), ``not envious'' (N4\_P), ``envious'' (N4\_N), ``stable'' (N5\_P), ``unstable'' (N5\_N), ``contented'' (N6\_P), and ``discontented'' (N6\_N). For the item naming given in parentheses, items with identical identifier (the integer after the first~``N'') are polar opposites, where a last character~``P'' refers to the positive opposite and~``N'' to the negative opposite.}
\label{fig:application-neuroticism-heatmapAPP}
\end{figure}

\begin{figure}
\centering
\includegraphics[width = 0.7\textwidth]{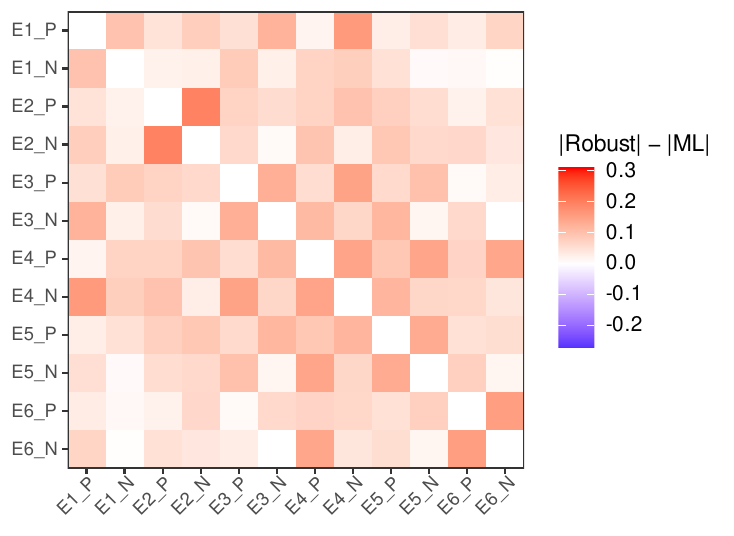}
\caption{Difference between absolute estimates for the polychoric correlation coefficient of %our 
the robust estimator with $c=0.6$ and the MLE for each item pair in the \emph{extroversion} scale, using the data of \citet{arias2020}. The items are ``extraverted'' (E1\_P), ``introverted'' (E1\_N), ``energetic'' (E2\_P), ``unenergetic'' (E2\_N), ``talkative'' (E3\_P), ``silent'' (E3\_N), ``bold'' (E4\_P), ``timid'' (E4\_N), ``assertive'' (E5\_P), ``unassertive'' (E5\_N), ``adventurous'' (E6\_P), and ``unadventurous'' (E6\_N). For the item naming given in parentheses, items with identical identifier (the integer after the first~``N'') are polar opposites, where a last character~``P'' refers to the positive opposite and~``N'' to the negative opposite.}
\label{fig:application-extroversion-heatmap}
\end{figure}

\begin{figure}
\centering
\includegraphics[width = 0.7\textwidth]{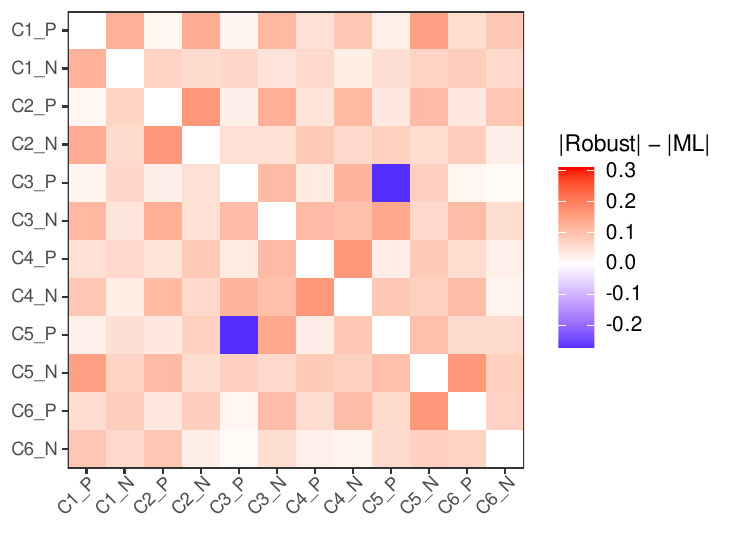}
\caption{Difference between absolute estimates for the polychoric correlation coefficient of %our
the robust estimator with $c=0.6$ and the MLE for each item pair in the \emph{conscientiousness} scale, using the data of \citet{arias2020}. The items are ``calm'' (C1\_P), ``angry'' (C1\_N), ``relaxed'' (C2\_P), ``tense'' (C2\_N), ``at ease'' (C3\_P), ``nervous'' (C3\_N), ``not envious'' (C4\_P), ``envious'' (C4\_N), ``stable'' (C5\_P), ``unstable'' (C5\_N), ``contented'' (C6\_P), and ``discontented'' (C6\_N). For the item naming given in parentheses, items with identical identifier (the integer after the first~``N'') are polar opposites, where a last character~``P'' refers to the positive opposite and~``N'' to the negative opposite.}
\label{fig:application-conscientiousness-heatmap}
\end{figure}

\begin{figure}[h!]
	\centering
	\includegraphics[width = 0.7\textwidth]{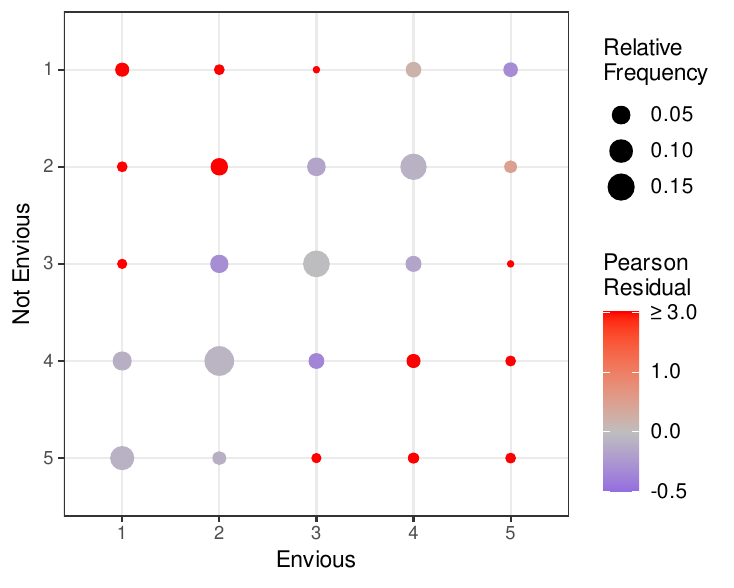}
\caption{Dot plot of cells for the \emph{neuroticism} item adjective pair ``not envious'' and ``envious'' in the data of \citet{arias2020}, where each item has five Likert-type response options, anchored by ``very inaccurate'' (= 1) and ``very accurate'' (= 5). Each dot's size is proportional to the relative empirical frequency of its associated contingency table cell,~$\fhatxy$, whereas its color varies by the value of the cell's PR,~$\fhatxy / \pxy{\thetahat}-1$, at robust parameter estimate with tuning constant~$c=0.6$. Note that some cells have an extremely large PR (substantially larger than the ideal value~0; see also Table~\ref{tab:arias2020-DPR-complete}), indicating a poor model fit for those cells. We therefore censored the color scale so that a fully saturated red is assigned to all cells with PR $\geq 3$, i.e., the empirical frequency $\fhatxy$ being at least four times as large as the model frequency $\pxy{\thetahat}$.
}
\label{fig:application-neuroticism-dotplot}
\end{figure}

This section contains additional results of the empirical application from Section~\ref{sec:application}. Table~\ref{tab:big5scaleadjectives} lists the unipolar markers of the three Big Five scales used by \citet{arias2020}, namely \emph{extroversion, conscientiousness,} and \emph{neuroticism}. Tables~\ref{tab:corrmatN}--\ref{tab:corrmatC} contain the (polychoric) correlation matrices of the items in each scale, estimated by maximum likelihood and our robust estimator, while Figures~\ref{fig:application-neuroticism-heatmapAPP}--\ref{fig:application-conscientiousness-heatmap} visualize the absolute difference between the two estimators for each pairwise correlation. Furthermore, Table~\ref{tab:arias2020-DPR-complete} contains the cellwise Pearson residuals for the item pair ``not envious'' and ``envious'' in the \emph{neuroticism} scale. Figure~\ref{fig:application-neuroticism-dotplot} provides a corresponding visualization.

\begin{table}[!b]
\centering
\begin{tabular}{l c l l}
& \multicolumn{3}{c}{Adjective marker pairs} \\
\noalign{\smallskip}\cline{2-4}\noalign{\smallskip}
Construct & Index & Positive (P) & Negative (N) \\
\noalign{\smallskip}\hline\noalign{\smallskip}
\multirow{6}{*}{Extroversion (E)} & 1 & extraverted & introverted \\
& 2 & energetic & unenergetic \\
& 3 & talkative & silent \\
& 4 & bold & timid \\
& 5 & assertive & unassertive \\
& 6 & adventurous & unadventurous \\
\noalign{\smallskip}\hline\noalign{\smallskip}
\multirow{6}{*}{Conscientiousness (C)} & 1 & organized & disorganized \\
& 2 & responsible & irresponsible \\
& 3 & conscientious & negligent \\
& 4 & practical & impractical \\
& 5 & thorough & careless \\
& 6 & hardworking & lazy \\
\noalign{\smallskip}\hline\noalign{\smallskip}
\multirow{6}{*}{Neuroticism (N)} & 1 & calm & angry \\
& 2 & relaxed & tense \\
& 3 & at ease & nervous \\
& 4 & not envious & envious \\
& 5 & stable & unstable \\
& 6 & contented & discontented \\
\noalign{\smallskip}\hline
\end{tabular}
\caption{Unipolar markers of three Big Five personality traits \citep{goldberg1992}. Each trait is measured by six pairs of items, where each item is a single English adjective. Each item pair consists of a positive and negative item. We explain item identifiers by means of the following example. Item ``C3\_P'' refers to the positive (P) item in the 3rd pair of the conscientiousness (C) scale, that is, adjective ``conscientious'', whereas ``N1\_N'' would refer to ``angry''.}
\label{tab:big5scaleadjectives}
\end{table}

\begin{table}
\centering
\footnotesize
\begin{subtable}{0.99\textwidth}
\centering
\begin{tabular}{r|*{12}{r}}
& \text{N1\_P} & \text{N1\_N} & \text{N2\_P} & \text{N2\_N} & \text{N3\_P} & \text{N3\_N} & \text{N4\_P} & \text{N4\_N} & \text{N5\_P} & \text{N5\_N} & \text{N6\_P} & \text{N6\_N} \\
\hline
N1\_P & 1.00 & -0.47 & 0.80 & -0.58 & 0.79 & -0.56 & 0.30 & -0.26 & 0.63 & -0.54 & 0.49 & -0.39 \\ 
  N1\_N & -0.47 & 1.00 & -0.48 & 0.58 & -0.49 & 0.54 & -0.26 & 0.45 & -0.47 & 0.68 & -0.43 & 0.63 \\ 
  N2\_P & 0.80 & -0.48 & 1.00 & -0.66 & 0.85 & -0.60 & 0.32 & -0.32 & 0.64 & -0.50 & 0.60 & -0.56 \\ 
  N2\_N & -0.58 & 0.58 & -0.66 & 1.00 & -0.70 & 0.76 & -0.37 & 0.49 & -0.48 & 0.60 & -0.35 & 0.55 \\ 
  N3\_P & 0.79 & -0.49 & 0.85 & -0.70 & 1.00 & -0.62 & 0.35 & -0.39 & 0.66 & -0.52 & 0.59 & -0.57 \\ 
  N3\_N & -0.56 & 0.54 & -0.60 & 0.76 & -0.62 & 1.00 & -0.42 & 0.49 & -0.52 & 0.58 & -0.37 & 0.53 \\ 
  N4\_P & 0.30 & -0.26 & 0.32 & -0.37 & 0.35 & -0.42 & 1.00 & -0.92 & 0.35 & -0.30 & 0.30 & -0.33 \\ 
  N4\_N & -0.26 & 0.45 & -0.32 & 0.49 & -0.39 & 0.49 & -0.92 & 1.00 & -0.39 & 0.50 & -0.33 & 0.53 \\ 
  N5\_P & 0.63 & -0.47 & 0.64 & -0.48 & 0.66 & -0.52 & 0.35 & -0.39 & 1.00 & -0.82 & 0.59 & -0.55 \\ 
  N5\_N & -0.54 & 0.68 & -0.50 & 0.60 & -0.52 & 0.58 & -0.30 & 0.50 & -0.82 & 1.00 & -0.44 & 0.61 \\ 
  N6\_P & 0.49 & -0.43 & 0.60 & -0.35 & 0.59 & -0.37 & 0.30 & -0.33 & 0.59 & -0.44 & 1.00 & -0.75 \\ 
  N6\_N & -0.39 & 0.63 & -0.56 & 0.55 & -0.57 & 0.53 & -0.33 & 0.53 & -0.55 & 0.61 & -0.75 & 1.00 \\ 
\end{tabular}
\caption{Robust estimates}
\end{subtable}
\begin{subtable}{0.99\textwidth}
\bigskip
\centering
\begin{tabular}{r|*{12}{r}}
& \text{N1\_P} & \text{N1\_N} & \text{N2\_P} & \text{N2\_N} & \text{N3\_P} & \text{N3\_N} & \text{N4\_P} & \text{N4\_N} & \text{N5\_P} & \text{N5\_N} & \text{N6\_P} & \text{N6\_N} \\
\hline
N1\_P & 1.00 & -0.37 & 0.71 & -0.50 & 0.69 & -0.49 & 0.27 & -0.24 & 0.58 & -0.47 & 0.42 & -0.32 \\ 
  N1\_N & -0.37 & 1.00 & -0.40 & 0.55 & -0.39 & 0.48 & -0.19 & 0.40 & -0.39 & 0.60 & -0.32 & 0.57 \\ 
  N2\_P & 0.71 & -0.40 & 1.00 & -0.55 & 0.75 & -0.54 & 0.26 & -0.26 & 0.55 & -0.41 & 0.53 & -0.47 \\ 
  N2\_N & -0.50 & 0.55 & -0.55 & 1.00 & -0.54 & 0.65 & -0.24 & 0.42 & -0.41 & 0.57 & -0.31 & 0.52 \\ 
  N3\_P & 0.69 & -0.39 & 0.75 & -0.54 & 1.00 & -0.53 & 0.29 & -0.28 & 0.63 & -0.44 & 0.52 & -0.48 \\ 
  N3\_N & -0.49 & 0.48 & -0.54 & 0.65 & -0.53 & 1.00 & -0.29 & 0.43 & -0.45 & 0.58 & -0.29 & 0.47 \\ 
  N4\_P & 0.27 & -0.19 & 0.26 & -0.24 & 0.29 & -0.29 & 1.00 & -0.62 & 0.26 & -0.20 & 0.18 & -0.20 \\ 
  N4\_N & -0.24 & 0.40 & -0.26 & 0.42 & -0.28 & 0.43 & -0.62 & 1.00 & -0.33 & 0.46 & -0.22 & 0.44 \\ 
  N5\_P & 0.58 & -0.39 & 0.55 & -0.41 & 0.63 & -0.45 & 0.26 & -0.33 & 1.00 & -0.70 & 0.53 & -0.46 \\ 
  N5\_N & -0.47 & 0.60 & -0.41 & 0.57 & -0.44 & 0.58 & -0.20 & 0.46 & -0.70 & 1.00 & -0.35 & 0.57 \\ 
  N6\_P & 0.42 & -0.32 & 0.53 & -0.31 & 0.52 & -0.29 & 0.18 & -0.22 & 0.53 & -0.35 & 1.00 & -0.58 \\ 
  N6\_N & -0.32 & 0.57 & -0.47 & 0.52 & -0.48 & 0.47 & -0.20 & 0.44 & -0.46 & 0.57 & -0.58 & 1.00 \\ 
\end{tabular}
\caption{Maximum likelihood estimates\medskip}
\end{subtable}
\caption{Estimated correlation matrices of the items in the \emph{neuroticism} scale from the data in \citet[][Sample~1, $N=725$]{arias2020} using %our
the robust estimator with %tuning constant
$c=0.6$ (top) and the MLE (bottom). The items are ``calm'' (N1\_P), ``angry'' (N1\_N), ``relaxed'' (N2\_P), ``tense'' (N2\_N), ``at ease'' (N3\_P), ``nervous'' (N3\_N), ``not envious'' (N4\_P), ``envious'' (N4\_N), ``stable'' (N5\_P), ``unstable'' (N5\_N), ``contented'' (N6\_P), and ``discontented'' (N6\_N). For the item naming given in parentheses, items with identical identifier (the integer after the first~``N'') are polar opposites, where a last character~``P'' refers to the positive opposite and~``N'' to the negative opposite.}
\label{tab:corrmatN}
\end{table}

\begin{table}
\centering
\footnotesize
\begin{subtable}{0.99\textwidth}
\centering
\begin{tabular}{r|*{12}{r}}
& \text{E1\_P} & \text{E1\_N} & \text{E2\_P} & \text{E2\_N} & \text{E3\_P} & \text{E3\_N} & \text{E4\_P} & \text{E4\_N} & \text{E5\_P} & \text{E5\_N} & \text{E6\_P} & \text{E6\_N} \\
\hline
E1\_P & 1.00 & -0.87 & 0.55 & -0.34 & 0.75 & -0.62 & 0.58 & -0.58 & 0.54 & -0.45 & 0.55 & -0.39 \\ 
  E1\_N & -0.87 & 1.00 & -0.40 & 0.36 & -0.67 & 0.63 & -0.52 & 0.62 & -0.52 & 0.51 & -0.36 & 0.37 \\ 
  E2\_P & 0.55 & -0.40 & 1.00 & -0.84 & 0.50 & -0.32 & 0.56 & -0.38 & 0.55 & -0.43 & 0.57 & -0.44 \\ 
  E2\_N & -0.34 & 0.36 & -0.84 & 1.00 & -0.30 & 0.35 & -0.40 & 0.43 & -0.41 & 0.54 & -0.45 & 0.53 \\ 
  E3\_P & 0.75 & -0.67 & 0.50 & -0.30 & 1.00 & -0.71 & 0.50 & -0.51 & 0.52 & -0.50 & 0.42 & -0.28 \\ 
  E3\_N & -0.62 & 0.63 & -0.32 & 0.35 & -0.71 & 1.00 & -0.38 & 0.62 & -0.47 & 0.47 & -0.30 & 0.37 \\ 
  E4\_P & 0.58 & -0.52 & 0.56 & -0.40 & 0.50 & -0.38 & 1.00 & -0.55 & 0.72 & -0.64 & 0.61 & -0.48 \\ 
  E4\_N & -0.58 & 0.62 & -0.38 & 0.43 & -0.51 & 0.62 & -0.55 & 1.00 & -0.61 & 0.66 & -0.33 & 0.44 \\ 
  E5\_P & 0.54 & -0.52 & 0.55 & -0.41 & 0.52 & -0.47 & 0.72 & -0.61 & 1.00 & -0.85 & 0.44 & -0.29 \\ 
  E5\_N & -0.45 & 0.51 & -0.43 & 0.54 & -0.50 & 0.47 & -0.64 & 0.66 & -0.85 & 1.00 & -0.41 & 0.47 \\ 
  E6\_P & 0.55 & -0.36 & 0.57 & -0.45 & 0.42 & -0.30 & 0.61 & -0.33 & 0.44 & -0.41 & 1.00 & -0.83 \\ 
  E6\_N & -0.39 & 0.37 & -0.44 & 0.53 & -0.28 & 0.37 & -0.48 & 0.44 & -0.29 & 0.47 & -0.83 & 1.00 \\ 
\end{tabular}
\caption{Robust estimates}
\end{subtable}
\begin{subtable}{0.99\textwidth}
\bigskip
\centering
\begin{tabular}{r|*{12}{r}}
& \text{E1\_P} & \text{E1\_N} & \text{E2\_P} & \text{E2\_N} & \text{E3\_P} & \text{E3\_N} & \text{E4\_P} & \text{E4\_N} & \text{E5\_P} & \text{E5\_N} & \text{E6\_P} & \text{E6\_N} \\
\hline
E1\_P & 1.00 & -0.78 & 0.50 & -0.26 & 0.70 & -0.50 & 0.56 & -0.42 & 0.51 & -0.40 & 0.52 & -0.32 \\ 
  E1\_N & -0.78 & 1.00 & -0.38 & 0.34 & -0.59 & 0.61 & -0.45 & 0.54 & -0.47 & 0.50 & -0.35 & 0.37 \\ 
  E2\_P & 0.50 & -0.38 & 1.00 & -0.65 & 0.43 & -0.27 & 0.49 & -0.28 & 0.47 & -0.38 & 0.55 & -0.39 \\ 
  E2\_N & -0.26 & 0.34 & -0.65 & 1.00 & -0.24 & 0.34 & -0.30 & 0.40 & -0.32 & 0.48 & -0.38 & 0.50 \\ 
  E3\_P & 0.70 & -0.59 & 0.43 & -0.24 & 1.00 & -0.59 & 0.44 & -0.36 & 0.46 & -0.40 & 0.41 & -0.25 \\ 
  E3\_N & -0.50 & 0.61 & -0.27 & 0.34 & -0.59 & 1.00 & -0.27 & 0.56 & -0.35 & 0.45 & -0.24 & 0.37 \\ 
  E4\_P & 0.56 & -0.45 & 0.49 & -0.30 & 0.44 & -0.27 & 1.00 & -0.41 & 0.64 & -0.49 & 0.54 & -0.34 \\ 
  E4\_N & -0.42 & 0.54 & -0.28 & 0.40 & -0.36 & 0.56 & -0.41 & 1.00 & -0.49 & 0.60 & -0.27 & 0.40 \\ 
  E5\_P & 0.51 & -0.47 & 0.47 & -0.32 & 0.46 & -0.35 & 0.64 & -0.49 & 1.00 & -0.71 & 0.39 & -0.23 \\ 
  E5\_N & -0.40 & 0.50 & -0.38 & 0.48 & -0.40 & 0.45 & -0.49 & 0.60 & -0.71 & 1.00 & -0.34 & 0.45 \\ 
  E6\_P & 0.52 & -0.35 & 0.55 & -0.38 & 0.41 & -0.24 & 0.54 & -0.27 & 0.39 & -0.34 & 1.00 & -0.68 \\ 
  E6\_N & -0.32 & 0.37 & -0.39 & 0.50 & -0.25 & 0.37 & -0.34 & 0.40 & -0.23 & 0.45 & -0.68 & 1.00 \\ 
\end{tabular}
\caption{Maximum likelihood estimates\medskip}
\end{subtable}
\caption{Estimated correlation matrices of the items in the \emph{extroversion} scale from the data in \citet[][Sample~1, $N=725$]{arias2020} using %our
the robust estimator with %tuning constant
$c=0.6$ (top) and the MLE (bottom). The items are ``extraverted'' (E1\_P), ``introverted'' (E1\_N), ``energetic'' (E2\_P), ``unenergetic'' (E2\_N), ``talkative'' (E3\_P), ``silent'' (E3\_N), ``bold'' (E4\_P), ``timid'' (E4\_N), ``assertive'' (E5\_P), ``unassertive'' (E5\_N), ``adventurous'' (E6\_P), and ``unadventurous'' (E6\_N). For the item naming given in parentheses, items with identical identifier (the integer after the first~``N'') are polar opposites, where a last character~``P'' refers to the positive opposite and~``N'' to the negative opposite.}
\label{tab:corrmatE}
\end{table}

\begin{table}
\centering
\footnotesize
\begin{subtable}{0.99\textwidth}
\centering
\begin{tabular}{r|*{12}{r}}
& \text{C1\_P} & \text{C1\_N} & \text{C2\_P} & \text{C2\_N} & \text{C3\_P} & \text{C3\_N} & \text{C4\_P} & \text{C4\_N} & \text{C5\_P} & \text{C5\_N} & \text{C6\_P} & \text{C6\_N} \\
\hline
C1\_P & 1.00 & -0.89 & 0.57 & -0.56 & 0.36 & -0.46 & 0.43 & -0.35 & 0.54 & -0.56 & 0.49 & -0.52 \\ 
  C1\_N & -0.89 & 1.00 & -0.58 & 0.64 & -0.31 & 0.60 & -0.38 & 0.47 & -0.48 & 0.69 & -0.52 & 0.61 \\ 
  C2\_P & 0.57 & -0.58 & 1.00 & -0.87 & 0.45 & -0.68 & 0.62 & -0.54 & 0.55 & -0.65 & 0.69 & -0.64 \\ 
  C2\_N & -0.56 & 0.64 & -0.87 & 1.00 & -0.44 & 0.80 & -0.57 & 0.74 & -0.50 & 0.76 & -0.61 & 0.66 \\ 
  C3\_P & 0.36 & -0.31 & 0.45 & -0.44 & 1.00 & -0.43 & 0.42 & -0.46 & 0.17 & -0.41 & 0.40 & -0.26 \\ 
  C3\_N & -0.46 & 0.60 & -0.68 & 0.80 & -0.43 & 1.00 & -0.48 & 0.70 & -0.52 & 0.78 & -0.55 & 0.59 \\ 
  C4\_P & 0.43 & -0.38 & 0.62 & -0.57 & 0.42 & -0.48 & 1.00 & -0.68 & 0.39 & -0.47 & 0.44 & -0.33 \\ 
  C4\_N & -0.35 & 0.47 & -0.54 & 0.74 & -0.46 & 0.70 & -0.68 & 1.00 & -0.47 & 0.66 & -0.42 & 0.45 \\ 
  C5\_P & 0.54 & -0.48 & 0.55 & -0.50 & 0.17 & -0.52 & 0.39 & -0.47 & 1.00 & -0.54 & 0.60 & -0.45 \\ 
  C5\_N & -0.56 & 0.69 & -0.65 & 0.76 & -0.41 & 0.78 & -0.47 & 0.66 & -0.54 & 1.00 & -0.59 & 0.61 \\ 
  C6\_P & 0.49 & -0.52 & 0.69 & -0.61 & 0.40 & -0.55 & 0.44 & -0.42 & 0.60 & -0.59 & 1.00 & -0.69 \\ 
  C6\_N & -0.52 & 0.61 & -0.64 & 0.66 & -0.26 & 0.59 & -0.33 & 0.45 & -0.45 & 0.61 & -0.69 & 1.00 \\ 
\end{tabular}
\caption{Robust estimates}
\end{subtable}
\begin{subtable}{0.99\textwidth}
\bigskip
\centering
\begin{tabular}{r|*{12}{r}}
& \text{C1\_P} & \text{C1\_N} & \text{C2\_P} & \text{C2\_N} & \text{C3\_P} & \text{C3\_N} & \text{C4\_P} & \text{C4\_N} & \text{C5\_P} & \text{C5\_N} & \text{C6\_P} & \text{C6\_N} \\
\hline
C1\_P & 1.00 & -0.77 & 0.56 & -0.43 & 0.34 & -0.35 & 0.38 & -0.26 & 0.51 & -0.41 & 0.43 & -0.43 \\ 
  C1\_N & -0.77 & 1.00 & -0.51 & 0.59 & -0.24 & 0.55 & -0.32 & 0.44 & -0.43 & 0.61 & -0.44 & 0.55 \\ 
  C2\_P & 0.56 & -0.51 & 1.00 & -0.70 & 0.42 & -0.56 & 0.57 & -0.43 & 0.51 & -0.54 & 0.65 & -0.55 \\ 
  C2\_N & -0.43 & 0.59 & -0.70 & 1.00 & -0.40 & 0.75 & -0.48 & 0.68 & -0.43 & 0.71 & -0.53 & 0.63 \\ 
  C3\_P & 0.34 & -0.24 & 0.42 & -0.40 & 1.00 & -0.32 & 0.39 & -0.34 & 0.44 & -0.34 & 0.38 & -0.25 \\ 
  C3\_N & -0.35 & 0.55 & -0.56 & 0.75 & -0.32 & 1.00 & -0.37 & 0.60 & -0.38 & 0.72 & -0.45 & 0.54 \\ 
  C4\_P & 0.38 & -0.32 & 0.57 & -0.48 & 0.39 & -0.37 & 1.00 & -0.52 & 0.36 & -0.39 & 0.39 & -0.31 \\ 
  C4\_N & -0.26 & 0.44 & -0.43 & 0.68 & -0.34 & 0.60 & -0.52 & 1.00 & -0.38 & 0.59 & -0.31 & 0.43 \\ 
  C5\_P & 0.51 & -0.43 & 0.51 & -0.43 & 0.44 & -0.38 & 0.36 & -0.38 & 1.00 & -0.43 & 0.54 & -0.39 \\ 
  C5\_N & -0.41 & 0.61 & -0.54 & 0.71 & -0.34 & 0.72 & -0.39 & 0.59 & -0.43 & 1.00 & -0.43 & 0.53 \\ 
  C6\_P & 0.43 & -0.44 & 0.65 & -0.53 & 0.38 & -0.45 & 0.39 & -0.31 & 0.54 & -0.43 & 1.00 & -0.61 \\ 
  C6\_N & -0.43 & 0.55 & -0.55 & 0.63 & -0.25 & 0.54 & -0.31 & 0.43 & -0.39 & 0.53 & -0.61 & 1.00 \\ 
\end{tabular}
\caption{Maximum likelihood estimates\medskip}
\end{subtable}
\caption{Estimated correlation matrices of the items in the \emph{conscientiousness} scale from the data in \citet[][Sample~1, $N=725$]{arias2020} using %our
the robust estimator with %tuning constant
$c=0.6$ (top) and the MLE (bottom). The items are ``calm'' (C1\_P), ``angry'' (C1\_N), ``relaxed'' (C2\_P), ``tense'' (C2\_N), ``at ease'' (C3\_P), ``nervous'' (C3\_N), ``not envious'' (C4\_P), ``envious'' (C4\_N), ``stable'' (C5\_P), ``unstable'' (C5\_N), ``contented'' (C6\_P), and ``discontented'' (C6\_N). For the item naming given in parentheses, items with identical identifier (the integer after the first~``N'') are polar opposites, where a last character~``P'' refers to the positive opposite and~``N'' to the negative opposite.}
\label{tab:corrmatC}
\end{table}

 \FloatBarrier
\begin{table}[h]
\centering
\small
\begin{tabular}{c|*{5}{r}}
$X$\textbackslash $Y$     & 1        & 2        & 3        & 4        & 5        \\
\hline
1     & 9,780,473,685.31     & 15,982.37 & 10.81  & 0.14     & $-0.35$  \\
2     & 2,419.15  & 9.06     & $-0.20$ & $-0.10$  &  0.42    \\
3     & 4.48      & $-0.35$  & $-0.01$ & $-0.20$  & 76.11    \\
4     & $-0.12$   & $-0.08$  & $-0.39$ & 11.66    & 222,240.11 \\
5     & $-0.11$   & $-0.12$  & 34.98   & 55,329.21 & 991,790,294,422.36 \\
\end{tabular}
\caption{Pearson residual, $\fhatxy \big/ \pxy{\thetahat}-1$, of each response $(x,y)$ for the item pair ``not envious'' ($X$) and ``envious'' ($Y$) in the measurements of \citet{arias2020} of the \emph{neuroticism} scale. Estimate~$\thetahat$ was computed via the robust estimator with tuning constant $c=0.6$.}
\label{tab:arias2020-DPR-complete}
\end{table}

\FloatBarrier

\end{document}